\documentclass[10pt, a4paper, onecolumn]{article} % 10pt font size (11 and 12 also possible), A4 paper (letterpaper for US letter) and two column layout
\usepackage[utf8]{inputenc}
\usepackage[title]{appendix}
\usepackage{lscape}
\usepackage{hyperref}
\usepackage{amsfonts,amssymb}
\usepackage{graphicx}
\usepackage{float}
\usepackage{amsmath}
\usepackage{cuted}
\usepackage{geometry}
\usepackage{rotating}
\usepackage{pdflscape}
\usepackage{blindtext}

\newcommand{\ket}[1]{\left|#1\right\rangle}
\newcommand{\bra}[1]{\left\langle#1\right|}

 \usepackage{braket}
\usepackage{chngcntr}
 \numberwithin{equation}{section}
\title{Electrostatically interacting Wannier qubits in curved space}
\author{ Krzysztof Pomorski$^{1,2} $ \\ \\ %\textsuperscript{1,2,3}
		\newline\newline % Space before institutions
     1: Cracow University of Technology,\\ Faculty of Computer Science and Telecommunications, \\ Department of Computer Science, Poland %}
     \\ \\
    2: Quantum Hardware Systems (\texttt{www.quantumhardwaresystems.com}) %} % Institution 3
}
\providecommand{\keywords}[1]
{
  \small	
  \textbf{\textit{Keywords---}} #1
}

\begin{document}
\maketitle

\begin{abstract}
Derivation of tight-binding model from Schr\"{o}dinger formalism for various topologies of position-based semiconductor qubits is presented in this work in case of static and time-dependent electric fields. Simplistic tight-binding model allows for description of single-electron devices at large integration scale.
The case of two electrostatically Wannier qubits (that are also known as position based qubits) in Schr\"{o}dinger model is presented with omission spin degrees of freedom.
The concept of programmable quantum matter can be implemented in the chain of coupled semiconductor quantum dots. Indeed highly integrated and developed cryogenic CMOS nanostructures can be mapped to coupled quantum dots, whose connectivity can be controlled by voltage applied across transistor gates as well as external magnetic field. Using anti-correlation principle arising from Coulomb repulsion interaction between electrons one can implement classical and quantum inverter (Classical/Quantum Swap gate) and many other logical gates. This anti-correlation will be weaken due to the fact of quantumness of physical process is bringing coexistence of correlation and anti-correlation at the same time.
%%Justification of tight-binding model from Schroedinger formalism for various topologies of position-based semiconductor qubits is presented in this work. Simplistic tight-binding model allows for description of single-electron devices at large integration scale. However it is due to the fact that tight-binding model omits the integro-differential equations that arise from electron-electron interaction in Schroedinger model. Two approaches are given in derivation of tight-binding model from Schroedinger equation. First approach is conducted by usage of Green functions obtained from Schroedinger equation. Second approach is given by usage of Taylor expansion applied to Schroedinger equation. The obtained results can be extended for the case of many Wannier qubits with more than one electron and can be applied to 2 and 3 dimensional model. Furthermore various correlation functions are proposed in Schroedinger formalism that can account for static and time-dependent electric and magnetic field polarizing given Wannier qubit system.
One of the central results presented in this work relies on the emergence of dissipation processes during smooth bending of semiconductor nanowires both in the case of classical and quantum picture. Presented results give the base for physical description of electrostatic Q-Swap gate of any topology using open loop nanowires, whose functionality can be programmed. We observe strong localization of wavepacket due to nanowire bending. Therefore it is not always necessary to built barrier between two nanowires to obtain two quantum dot system. On another hand the obtained results can be mapped to problem of electron in curved space, so they can be expressed by programmable position-dependent metric embedded in Schr\"{o}dinger equation. Indeed semiconductor quantum dot system is capable of mimicking the curved space what provides bridge between fundamental and applied science present in implementation of single-electron devices.
\end{abstract}

%%%\tableofcontents

\section{Philosophy behind charged based classical and quantum logic}
\begin{figure}
\centering
\includegraphics[scale=0.6]{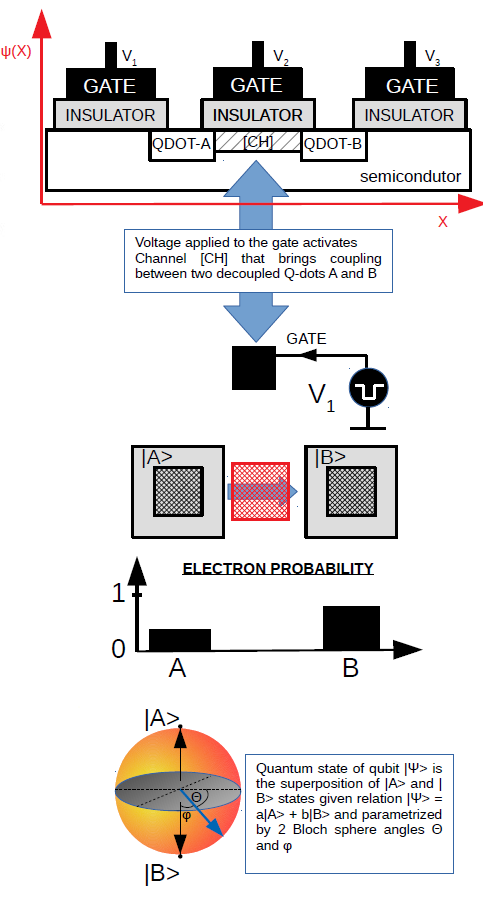}\includegraphics[scale=0.7]{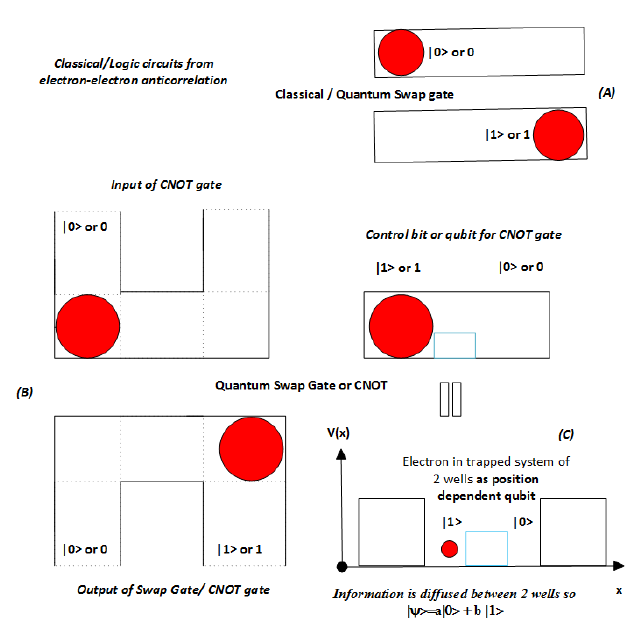}
\caption{Left:Position based qubit also known as Wannier qubit in CMOS circuit implementation as by \cite{SEL}, \cite{Cryogenics}, \cite{Nbodies},\cite{qchip}, Right:(A) Electrostatic inverter (Quantum Swap Gate) made from position based qubits also known as Wannier qubit and (B): Controllable NOT gate. The generalized version of electrostatic double quantum dot system (qubit) is given by Fig.\ref{fiber} and generalized quantum swap gate is depicted in Fig.\ref{2fibers}.}
\label{fig:qubit1}
\end{figure}

\begin{figure}
\centering
\includegraphics[scale=0.7]{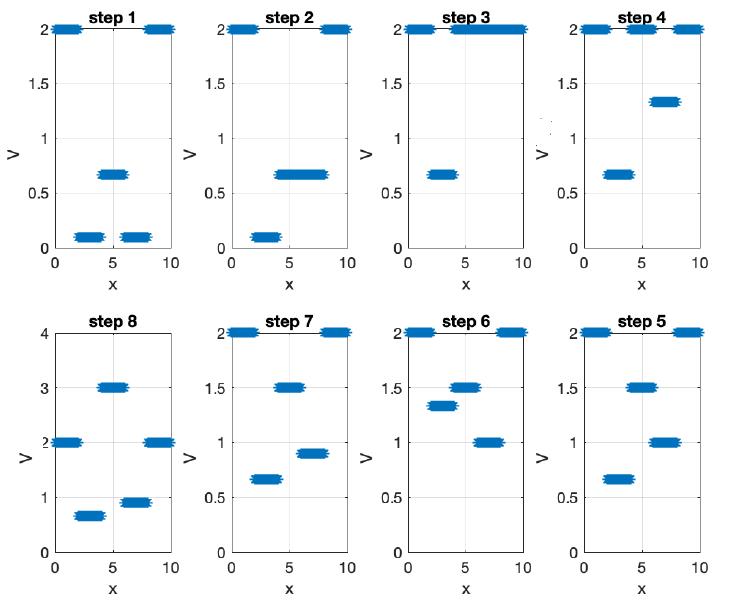} \\
\includegraphics[scale=0.6]{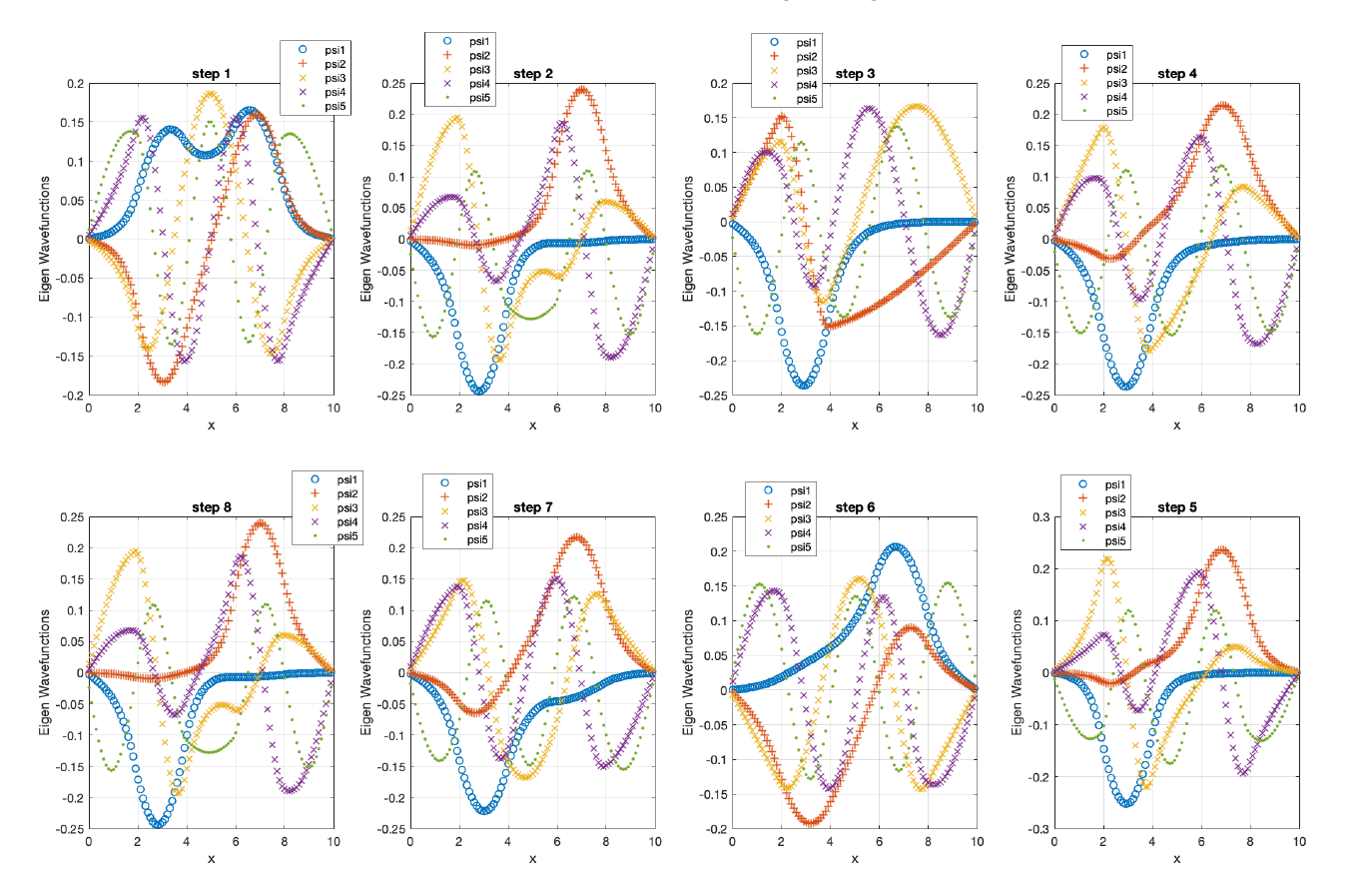}
\caption{ Upper: Effective potentials V(x) for single electron under different voltage biasing circumstances \cite{Noninvasive}.
Lower: Electron wave-functions for subsequent eiengenergies \cite{Noninvasive} obtained by Wannier qubit. One can spot maximum localized functions for various qubit electrostatic biasing potentials expressed by effective potential. }
\label{fig:qubit2}
\end{figure}

Single electron devices in semiconductor quantum dots \cite{Likharev} becomes quite promising way of implementation of qubit and quantum computation as well as quantum communication and it was being confirmed experimentally by \cite{Fujisawa}, \cite{Petta}.
This is particularly attractive perspective in the framework of CMOS technology \cite{Dirk}, \cite{CNOT}, \cite{Limitation}, \cite{QDotArray}, \cite{CryoCMOS}, \cite{Panos}, \cite{Pomorski_spie}, \cite{QHSchannel}, \cite{SEL}, \cite{qchip}, \cite{Cryogenics}. In case of small field effect transistor the source and drain are playing the role of quantum dots, whose connectivity is regulated by voltage applied on the top gate that is in-between as depicted in Fig.\ref{fig:qubit1} and in Fig.\ref{fig:qubit2}. We consider only single-electron devices that corresponds to the fact that one electron occupies one channel that is assumed to be quasi-one dimensional nanowire \cite{Xu}.
It is natural to expect in case of two coupled quantum dot systems the oscillations of occupancy in left and right quantum dots when at least two eigenergy levels are occupied.
In case of time-independent Hamiltonian that corresponds to time-independent magnetic and electric field as well as time independent boundary conditions (that is guaranteed by stiffness of nanostructure) we have the system wave-function that can be written as linear combination of maximum localized left and right wave-function (written as linear combination of two eigenergy wavefunctions that are orthonormal). In condensed matter physics we know Wannier functions in case of elementary cells accounting for crystal lattice and those Wannier functions are maximum localized on particular atoms and orthonormal one to each other as given by the work \cite{WannierCondensates}, \cite{WannierFunctions}, \cite{Spalek}.
%%\begin{figure}
%%\includegraphics[scale=0.3]{qgatesKP.png}
%%\includegraphics[scale=0.3]{qgatesKP.png}
%%\end{figure}

The most simple model describing semiconductor Wannier position based qubit is given by simplistic tight-binding model. We have dynamics of quantum state with time given by
\begin{eqnarray}
\hat{H}_t \ket{\psi(t)}=
\begin{pmatrix}
E_{p1}(t) & t_{s12}(t) \\
t_{s12}^{*}(t) & E_{p2}(t)
\end{pmatrix}
\begin{pmatrix}
\alpha_q(t) \\
\beta_q(t)
\end{pmatrix}
=i\hbar \frac{d}{dt}
\begin{pmatrix}
\alpha_q(t) \\
\beta_q(t)
\end{pmatrix}=i\hbar \frac{d}{dt}\ket{\psi(t)}, |\alpha_q(t)|^2+|\beta_q(t)|^2=1, \nonumber \\
\label{eqn:fundamentalbasic}
\end{eqnarray}
where $E_{p1}$ denotes maximum localized energy due to presence of electron on the left quantum dot and
$E_{p2}$ denotes maximum localized energy due to presence of electron on the right quantum dot and $|t_{s12}|$ is hopping energy due to electron movement from left to right quantum dot. We can refer tight-binding model to Schrodinger equation
\begin{eqnarray}
\psi(x,t)=\alpha_q(t)w_L(x)+\beta_q(t)w_R(x), \int_{-\infty}^{+\infty} dx|w_{L(R)}(x)|^2=1, \int_{-\infty}^{+\infty} dx w_{R}^{*}(x) w_{L}(x)=\bra{w_R}\ket{w_L}=0, \nonumber \\
\alpha_q(t)=\int_{-\infty}^{+\infty}dxw_L^{*}(x)\psi(x,t)=\bra{w_L}\ket{\psi},\beta_q(t)=\int_{-\infty}^{+\infty}dxw_R^{*}(x)\psi(x,t)=\bra{w_L}\ket{\psi},
\end{eqnarray}
where $w_L(x)$ and $w_R(x)$ are maximum localized orthonormal wavefunctions (Wannier functions) of single electron on left and right quantum dot.
They are defined by wavefunction distribution implementing maximum occupancy of electron on left or right side.
Prescription for tight-binding model in terms of Wannier functions is given by
\begin{eqnarray}
E_{p1(p2)}=\int_{-\infty}^{+\infty}dxw_{L(R)}(x)^{*}(\frac{-\hbar^2}{2m}\frac{d^2}{dx^2}+V(x))w_{L(R)}(x)=\int_{-\infty}^{+\infty}dxw_{L(R)}(x)^{*}\hat{H}(x)w_{L(R)}(x), \nonumber \\
t_{s12(s21)}=\int_{-\infty}^{+\infty}dxw_{L(R)}(x)^{*}(\frac{-\hbar^2}{2m}\frac{d^2}{dx^2}+V(x))w_{R(L)}(x)=\int_{-\infty}^{+\infty}dxw_{L(R)}(x)^{*}\hat{H}(x)w_{R(L)}(x),
\label{basicexp}
\end{eqnarray}
where $H$ is Hamiltonian of double quantum dot system. In this work we justify the formulas for $E_{p1}$, $E_{p2}$, $t_{s12}$ and for $t_{s21}$ for most general case \cite{Extended}.
% in terms of Green functions and Taylor wavefunction expansion for any physical situation \cite{Extended}.
Physics is relying on mass and electric or topological charge conservation principles and both quantities are quasi-continuous on macroscale and on nanoscale become integer multiplicity of elementary values. The charge flows in such a way as the energy of the electric/magnetic field tends to be minimized. One of its consequences is the repulsion of two charges of the same sign and attraction of two charges of opposite signs that is commonly known as Coulomb law. The electric currents flow also tends to minimize the energy of magnetic field, so the most equilibrium state of isolated capacitor is discharged device. Due to electron and hole mobility charge can be used for information or energy transfer across metallic or semiconductor nanowires. One can use the electric and magnetic fields as parameters controlling the evolution of the given physical system with time, so desired final state can be achieved upon previous setting the system with initial configuration that is formally expressed by circuit theory both in classical and in quantum regime. Furthermore the simple rules of dynamics of charged billiard balls confined in boxes can lead to a simple scheme for implementation of logical operations as
logical inverter (Quantum Swap Gate) or controllable inverter (CNOT gate or Controllable Quantum Swap Gate) as depicted in Fig.1 and by \cite{Likharev}, \cite{Dirk}, \cite{Panos}, \cite{Pomorski_spie}, \cite{SEL}, \cite{qchip}. However the electric charge is confirmed to be quantized by experiments (except fractional quantum Hall effect where fractionation of electric charge is being observed) and expressed by electron, proton or hole charge in condensed matter systems. The quantization and control of single electron flow by distinct integer values can be achieved in nanotechnological experiments as in the chain of coupled quantum dots that can have particularly small diameters in semiconductors and in most recent CMOS technology, so even size of 3nm can be achieved for very highly integrated circuits. In such types of structures the usage of magnetic fields is less practical since waveguides and solenoids are very hardly scalable. Therefore it is favourable to use only electric fields as a controlling factor so it promotes Wannier qubits that are also known as position-based qubits. Wannier qubits are using maximum localized wavefunctions as present in two coupled quantum dots in order to encode quantum information in qubit what makes such qubit different from eigenergy based qubit using two eigenergies to span the qubit state. However, it shall be underlined that even in cryogenic conditions the semiconductors are having the intrinsic noise that is significantly higher than in case of superconductors. %Since IBM quantum experience is using low temperature superconductors and transmon low noise Josephson junction qubit one can think about the possibility of merging those two quantum computer architectures into one chip. This leads to the necessity of constructing an interface between Josephson junction and electrostatic Wannier based qubit. In this consideration we exclude the role of electron spin that becomes important in mK regime, while charged based qubits shall be operational at 4 K. The minimum necessity for qubit operation is the partial occupation of two energy levels that will create the oscillations of occupancy between two coupled quantum dots with frequency proportional to energy difference between two energy levels.
%Such a problem of placing electrons in effective potential is easily solvable by Schroedinger equation. However when one considers two or more electrostatically interacting qubits then integro-differential equations emerge and are not solvable by analytical methods. Electron movement between two quantum dots is represented by tight-binding approach that
%essentially describes energy flow between neighbouring quantum dots and describes localized energy in each quantum dot. Such an approach is simplistic but quite powerful and will be conducted in the presented paper. This approach is caching the essence of electrostatic entanglement between coupled position based qubits. Furthermore, the tight-binding model can be used in the context of Bogoliubov-de Gennes equations describing Andreev bound states being the essence of Josephson effect. Such Andreev bound state of Josephson junction is being modified by the presence of position-based qubit implemented in semiconductor. This modification of the Andreev bound state by a single electron in Wannier semiconductor qubit (position-based qubit) is the subject of this work. It shall have its scientific, technological and didactic value. The results of this work might be important for creation hybrid semiconductor-superconducting quantum computer (hybrid semiconductor-superconducting quantum chip) and for quantum information processing in such structures.
\section{From Schr\"{o}dinger to Wannier functions}
Let us consider the system of 2 coupled quantum dots and let us assign the occupancy of left quantum dot by electron as wavepacket presence in $x \in (-\infty,0)$ and wavepacket occupancy of right quantum dot as wavepacket presence in $x \in (0,+\infty)$. We can assume that Wannier wavefunctions are linear transformation of system eigenenergy wavefunctions.

We propose maximum localized orthonormal Wannier functions of the form
\begin{eqnarray}
w_L(x)=(+\alpha \psi_{E1}(x)+\beta \psi_{E2}(x))=w_{1,1}\psi_{E1}(x)+w_{1,2}\psi_{E2}(x), \nonumber \\
w_R(x)=(-\beta \psi_{E1}(x)+\alpha\psi_{E2}(x))=w_{2,1}\psi_{E1}(x)+w_{2,2}\psi_{E2}(x), \nonumber \\
\end{eqnarray}
and formally we have
\begin{eqnarray}
\begin{pmatrix}
w_L(x) \\
w_R(x)
\end{pmatrix}= \hat{W}
\begin{pmatrix}
\psi_{E1}(x) \\
\psi_{E2}(x)
\end{pmatrix}
=
\begin{pmatrix}
w_{1,1} & w_{1,2} \\
w_{2,1} & w_{2,2}
\end{pmatrix}
\begin{pmatrix}
\psi_{E1}(x) \\
\psi_{E2}(x)
\end{pmatrix},
%% \hat{W}^{-1}=\frac{1}{Det(\hat{W})}
%% \begin{pmatrix}
%%w_{2,2} & -w_{1,2} \\
%%-w_{2,1} & w_{1,1}
%%\end{pmatrix}
\end{eqnarray}
%%\begin{eqnarray}
%%1=\int dx w_L^{*}(x)w_L(x)=\int dx \alpha_L^{\dag}(\psi(x)_{E1}^{\dag}c_{E1}^{\dag}+\beta^{\dag}c_{E2}^{\dag} \psi_{E2}(x)^{\dag})\alpha_L(c_{E1}\psi_{E1}(x)+\beta c_{E2}\psi_{E2}(x))= \nonumber \\
%%|\alpha_L|^2[\int dx [(|c_{E2}|^2|\beta|^2|\psi_{E2}(x)|^2+|c_{E1}|^2|\psi(x)_{E1}|^2)]+\int dx [ \beta c_{E1}^{\dag}c_{E2}\psi(x)_{E1}^{\dag} \psi_{E2}(x) + c_{E1}c_{E2}^{\dag}\beta^{\dag} \psi_{E2}(x)^{\dag} \psi_{E1}(x)]]=\nonumber \\
%%|\alpha_L|^2[ [(|c_{E2}|^2|\beta|^2+|c_{E1}|^2)]=|\alpha_L|^2[ [(1-|c_{E1}|^2|\beta|^2+|c_{E1}|^2)].
%%%|\frac{1}{\sqrt{1+|\beta|^2+\beta c_1 +\beta^{\dag} c_2}}|^2[ [(|\beta|^2+1)]+ [ \beta \int dx \psi(x)^{\dag} \psi(-x) + \beta^{\dag} \int dx \psi(-x)^{\dag} \psi(x)]]=1.
%%\end{eqnarray}

We have 4 conditions to be fulfilled:
\begin{eqnarray} \label{eqnset}
1=\int dx w_L^{*}(x)w_L(x)=\int dx (+\alpha^{\dag} \psi_{E1}(x)^{\dag}+\beta^{\dag} \psi_{E2}(x)^{\dag})(\alpha\psi_{E1}(x)+\beta \psi_{E2}(x)), \nonumber \\
1=\int dx w_R^{*}(x)w_R(x)=\int dx (-\beta^{\dag} \psi_{E1}(x)+\alpha^{\dag}\psi_{E2}(x)^{\dag})(-\beta\psi_{E1}(x)+\alpha \psi_{E2}(x)), \nonumber \\
0=\int dx w_R^{*}(x)w_L(x)=\int dx (-\beta^{\dag} \psi_{E1}(x)^{\dag}+\alpha^{\dag} \psi_{E2}(x)^{\dag})(\alpha\psi_{E1}(x)+\beta \psi_{E2}(x)), \nonumber \\
0=\int dx w_L^{*}(x)w_R(x)=\int dx (\alpha^{\dag}\psi_{E1}(x)^{\dag}+\beta^{\dag} \psi_{E2}(x)^{\dag})(-\beta \psi_{E1}(x)+\alpha \psi_{E2}(x)). %%% \nonumber \\
\end{eqnarray}
Due to orthogonality of 2 wavefunctions $\psi_{E1}$ and $\psi_{E2}$ we have from first two equations $|\alpha|^2+|\beta|^2=1$, so $|\beta|^2=1-|\alpha|^2$ and hence $|\alpha|=cos(\gamma)$ and $|\beta|=sin(\gamma)$.
From 3rd and 4th equation we have $\alpha^{\dag}\beta=\alpha\beta^{\dag}$ that is fulfilled when $\alpha=|\alpha|e^{i\delta}, \beta=|\beta|e^{+i\delta}=\sqrt{1-|\alpha|^2}e^{+i\delta}$.

%%Therefore we have

We propose maximum localized orthonormal Wannier functions of the form
\begin{eqnarray}
w_L(x)=(+|\alpha|e^{i\delta} \psi_{E1}(x)+\sqrt{1-|\alpha|^2}e^{+i\delta} \psi_{E2}(x)), \nonumber \\
w_R(x)=(-\sqrt{1-|\alpha|^2}e^{+i\delta} \psi_{E1}(x)+|\alpha|e^{i\delta}\psi_{E2}(x)), \nonumber \\
\end{eqnarray}
%or
%\begin{eqnarray}
%w_L(x)=(+|\alpha|e^{i\delta} \psi_{E1}(x)+\sqrt{1-|\alpha|^2}e^{-i\delta} \psi_{E2}(x)), \nonumber \\
%w_R(x)=(+\sqrt{1-|\alpha|^2}e^{-i\delta} \psi_{E1}(x)-|\alpha|e^{i\delta}\psi_{E2}(x)), \nonumber \\
%\end{eqnarray}

The last criteria to be matched is that $w_L(x)$ is maximum localized on the left quantum dot that geometric position is given by $x \in (-\infty,0)$ and that $w_R(x)$ is maximum localized on the right quantum dot that is denoted by $x \in (0,+\infty)$. Formally we can define
\begin{eqnarray}
S_L(\alpha)[\psi_{E1}(x),\psi_{E2}(x)]=S_L(\gamma)[\psi_{E1}(x),\psi_{E2}(x)]=\int_{-\infty}^{0}w_L(x)^{*}w_L(x)dx= \nonumber \\
=\int_{-\infty}^{0}dx[+|\alpha| \psi_{E1}^{\dag}(x)+\sqrt{1-|\alpha|^2}\psi_{E2}^{\dag}(x))][+|\alpha| \psi_{E1}(x)+\sqrt{1-|\alpha|^2} \psi_{E2}(x))]= \nonumber \\
=\int_{-\infty}^{0}dx [ (1-|\alpha|^2)|\psi_{E2}|^2+|\alpha|^2|\psi_{E1}(x)|^2+|\alpha|\sqrt{1-|\alpha|^2}(\psi_{E1}\psi_{E2}^{\dag}(x)+\psi_{E1}^{\dag}\psi_{E2}(x)) ] = \nonumber \\
=\int_{-\infty}^{0}dx [ (1-cos(\gamma)^2)|\psi_{E2}(x)|^2+cos(\gamma)^2|\psi_{E1}(x)|^2+sin(\gamma)cos(\gamma)(\psi_{E1}(x)\psi_{E2}^{\dag}(x)+\psi_{E1}(x)^{\dag}\psi_{E2}(x)) ]. %=
%%S_L(\gamma)[\psi_{E1}(x),\psi_{E2}(x)].
\end{eqnarray}
Since $S_L(\gamma)[\psi_{E1}(x),\psi_{E2}(x)]$ reaches maximum with respect to $\gamma$ what implies $\frac{d}{d \gamma}S_L(\gamma)[\psi_{E1}(x),\psi_{E2}(x)]=0$.
\begin{eqnarray}
0=\int_{-\infty}^{0}dx [-2 sin(\gamma) cos(\gamma)(|\psi_{E1}(x)|^2-|\psi_{E2}(x)|^2)+(cos(\gamma)^2-sin(\gamma)^2)(\psi_{E1}(x)\psi_{E2}^{\dag}(x)+\psi_{E1}(x)^{\dag}\psi_{E2}(x)) ].
%1+4|\alpha|^4-4|\alpha|^2=r^2|\alpha|^2-r^4|\alpha|^4.
\end{eqnarray}
that can be summarized as
\begin{eqnarray}
0=\int_{-\infty}^{0}dx [-sin(2\gamma)(|\psi_{E1}(x)|^2-|\psi_{E2}(x)|^2)+(cos(2\gamma))(\psi_{E1}(x)\psi_{E2}^{\dag}(x)+\psi_{E1}(x)^{\dag}\psi_{E2}(x)) ].
%1+4|\alpha|^4-4|\alpha|^2=r^2|\alpha|^2-r^4|\alpha|^4.
\end{eqnarray}
%\begin{eqnarray}
%(r^4+4)|\alpha|^4-(4+r^2)|\alpha|^2+1=0.
%\end{eqnarray}
and finally we have
\begin{eqnarray}
\gamma=\frac{1}{2}ArcTan [ \frac{\int_{-\infty}^{0}dx (\psi_{E1}(x)\psi_{E2}^{\dag}(x)+\psi_{E1}(x)^{\dag}\psi_{E2}(x))}{\int_{-\infty}^{0}dx (|\psi_{E1}(x)|^2-|\psi_{E2}(x)|^2)}]=\frac{1}{2}ArcTan [r],
%% \frac{ \int_{-\infty}^{0}dx (|\psi_{E1}(x)|^2-|\psi_{E2}(x)|^2 }{ \int_{-\infty}^{0}dx (\psi_{E1}(x)\psi_{E2}^{\dag}(x)+\psi_{E1}(x)^{\dag}\psi_{E2}(x)) } ].
%1+4|\alpha|^4-4|\alpha|^2=r^2|\alpha|^2-r^4|\alpha|^4.
\end{eqnarray}
where
\begin{eqnarray}
\label{special}
r=\frac{\int_{-\infty}^{0}dx (\psi_{E1}(x)\psi_{E2}^{\dag}(x)+\psi_{E1}(x)^{\dag}\psi_{E2}(x))}{\int_{-\infty}^{0}dx (|\psi_{E1}(x)|^2-|\psi_{E2}(x)|^2)}.
\end{eqnarray}

% $\Delta=(4+r^2)^2-4(r^4+4)=16+r^4+8r^2-4r^4-16=-3r^4+8r^2=4r^2(2-\frac{3}{4}r^2)$.
Consequently we have
\begin{eqnarray}
|\alpha|=
cos(\frac{1}{2} ArcTan [ \frac{\int_{-\infty}^{0}dx (\psi_{E1}(x)\psi_{E2}^{\dag}(x)+\psi_{E1}(x)^{\dag}\psi_{E2}(x))}{\int_{-\infty}^{0}dx (|\psi_{E1}(x)|^2-|\psi_{E2}(x)|^2)}]), \nonumber \\
|\beta|=
sin(\frac{1}{2} ArcTan [ \frac{\int_{-\infty}^{0}dx (\psi_{E1}(x)\psi_{E2}^{\dag}(x)+\psi_{E1}(x)^{\dag}\psi_{E2}(x))}{\int_{-\infty}^{0}dx (|\psi_{E1}(x)|^2-|\psi_{E2}(x)|^2)}]), \nonumber
\end{eqnarray}
%%%In order to have $|\alpha|>=0$ we have the condition $2>\frac{3}{4}r^2$ that is $2.666(6)=\sqrt{\frac{8}{3}}>|r|$.
Finally one can write

\begin{eqnarray}
\begin{pmatrix}
w_L(x) \\
w_R(x)
\end{pmatrix}
=
\begin{pmatrix}
+cos(\frac{1}{2}ArcTan [\frac{\int_{-\infty}^{0}dx (\psi_{E1}(x)\psi_{E2}^{\dag}(x)+\psi_{E1}(x)^{\dag}\psi_{E2}(x))}{\int_{-\infty}^{0}dx (|\psi_{E1}(x)|^2-|\psi_{E2}(x)|^2)}]) & sin(\frac{1}{2}ArcTan [\frac{\int_{-\infty}^{0}dx (\psi_{E1}(x)\psi_{E2}^{\dag}(x)+\psi_{E1}(x)^{\dag}\psi_{E2}(x))}{\int_{-\infty}^{0}dx (|\psi_{E1}(x)|^2-|\psi_{E2}(x)|^2)}]) \\
-sin(\frac{1}{2}ArcTan [ \frac{\int_{-\infty}^{0}dx (\psi_{E1}(x)\psi_{E2}^{\dag}(x)+\psi_{E1}(x)^{\dag}\psi_{E2}(x))}{\int_{-\infty}^{0}dx (|\psi_{E1}(x)|^2-|\psi_{E2}(x)|^2)}]) & cos(\frac{1}{2}ArcTan [ \frac{\int_{-\infty}^{0}dx (\psi_{E1}(x)\psi_{E2}^{\dag}(x)+\psi_{E1}(x)^{\dag}\psi_{E2}(x))}{\int_{-\infty}^{0}dx (|\psi_{E1}(x)|^2-|\psi_{E2}(x)|^2)}])
\end{pmatrix}
\begin{pmatrix}
\psi_{E1}(x) \\
\psi_{E2}(x)
\end{pmatrix}. \nonumber \\
\end{eqnarray}
 %with $r=\frac{\int_{-\infty}^{0}dx(|\psi_{E2}(x)|^2-|\psi_{E1}(x)|^2)}{\int_{-\infty}^{0}dx\psi_{E1}(x)\psi_{E2}(x)}$.
 Such reasoning can be conducted for any 2 different energy levels as well as for N different energetic levels.
 If quantum state is given as
 \begin{eqnarray}
 |\psi>=e^{\frac{E_1 (t-t_0)}{i \hbar}}e^{i\gamma_{E1}}\sqrt{p_{E1}}|E_1>+e^{\frac{E_2 (t-t_0)}{i \hbar}}e^{i\gamma_{E2}}\sqrt{p_{E2}}|E_2>
 \end{eqnarray}
 then

\begin{eqnarray}
\begin{pmatrix}
\alpha(t) w_L(x) \\
\beta(t) w_R(x)
\end{pmatrix}
=
\begin{pmatrix}
+cos(\frac{1}{2}ArcTan(r)) & +sin(\frac{1}{2}ArcTan(r)) \\
-sin(\frac{1}{2}ArcTan(r)) & +cos(\frac{1}{2}ArcTan(r))
\end{pmatrix}
\begin{pmatrix}
e^{\frac{E_1 (t-t_0)}{i \hbar}}e^{i\gamma_{E1}}\sqrt{p_{E1}} \psi_{E1}(x) \\
e^{\frac{E_2 (t-t_0)}{i \hbar}}e^{i\gamma_{E2}}\sqrt{p_{E2}} \psi_{E2}(x)
\end{pmatrix},
\end{eqnarray}
The last implies that
\begin{eqnarray}
\alpha_c(t)= \int_{-\infty}^{+\infty}dx
\begin{pmatrix}
[cos(\frac{1}{2}ArcTan(r))\psi_{E1}^{\dag}(x), & sin(\frac{1}{2}ArcTan(r))\psi_{E2}^{\dag}(x)
\end{pmatrix} \times \nonumber \\
\begin{pmatrix}
+cos(\frac{1}{2}ArcTan(r)), & +sin(\frac{1}{2}ArcTan(r)) \\
-sin(\frac{1}{2}ArcTan(r)), & +cos(\frac{1}{2}ArcTan(r))
\end{pmatrix}
\begin{pmatrix}
e^{\frac{E_1 (t-t_0)}{i \hbar}}e^{i\gamma_{E1}}\sqrt{p_{E1}} \psi_{E1}(x) \\
e^{\frac{E_2 (t-t_0)}{i \hbar}}e^{i\gamma_{E2}}\sqrt{p_{E2}} \psi_{E2}(x)
\end{pmatrix}= \nonumber \\
=\int_{-\infty}^{+\infty}dx
\begin{pmatrix}
w_{1,1}^{*}\psi_{E1}^{\dag}(x), & w_{1,2}^{*}\psi_{E2}^{\dag}(x)
\end{pmatrix} \times %\nonumber \\
\begin{pmatrix}
w_{1,1} & w_{1,2} \\
w_{2,1} & w_{2,2}
\end{pmatrix}
\begin{pmatrix}
e^{\frac{E_1 (t-t_0)}{i \hbar}}e^{i\gamma_{E1}}\sqrt{p_{E1}} \psi_{E1}(x) \\
e^{\frac{E_2 (t-t_0)}{i \hbar}}e^{i\gamma_{E2}}\sqrt{p_{E2}} \psi_{E2}(x)
\end{pmatrix}= \nonumber \\
=\int_{-\infty}^{+\infty}dx
\begin{pmatrix}
w_{1,1}^{*}\psi_{E1}^{\dag}(x), & w_{1,2}^{*}\psi_{E2}^{\dag}(x)
\end{pmatrix}
\begin{pmatrix}
w_{1,1}e^{\frac{E_1 (t-t_0)}{i \hbar}}e^{i\gamma_{E1}}\sqrt{p_{E1}} \psi_{E1}(x)+w_{1,2}e^{\frac{E_2 (t-t_0)}{i \hbar}}e^{i\gamma_{E2}}\sqrt{p_{E2}} \psi_{E2}(x) \\
w_{2,1}e^{\frac{E_1 (t-t_0)}{i \hbar}}e^{i\gamma_{E1}}\sqrt{p_{E1}} \psi_{E1}(x)+w_{2,2}e^{\frac{E_2 (t-t_0)}{i \hbar}}e^{i\gamma_{E2}}\sqrt{p_{E2}} \psi_{E2}(x)
\end{pmatrix}= \nonumber \\
=w_{1,1}^{*}w_{1,1}e^{\frac{E_1 (t-t_0)}{i \hbar}}e^{i\gamma_{E1}}\sqrt{p_{E1}} +w_{1,2}^{*}w_{2,2}e^{\frac{E_2 (t-t_0)}{i \hbar}}e^{i\gamma_{E2}}\sqrt{p_{E2}}= \nonumber \\
=cos(\frac{1}{2}ArcTan(r))^2e^{\frac{E_1 (t-t_0)}{i \hbar}}e^{i\gamma_{E1}}\sqrt{p_{E1}} + %%\nonumber \\
sin(\frac{1}{2}ArcTan(r))cos(\frac{1}{2}ArcTan(r)) e^{\frac{E_2 (t-t_0)}{i \hbar}}e^{i\gamma_{E2}}\sqrt{p_{E2}}= \nonumber \\
=cos(\frac{1}{2}ArcTan(r)) [ cos(2ArcTan(r)) e^{\frac{E_1 (t-t_0)}{i \hbar}}e^{i\gamma_{E1}}\sqrt{p_{E1}}+ sin(\frac{1}{2}ArcTan(r)) e^{\frac{E_2 (t-t_0)}{i \hbar}}e^{i\gamma_{E2}}\sqrt{p_{E2}} ]
= \alpha(t). \nonumber \\
%%[+\sqrt{\frac{(4+r^2)\pm 2r\sqrt{2-\frac{3}{4}r^2}}{2(r^4+4)}}]^{\dag}[\sqrt{\frac{(4+r^2)\pm 2r\sqrt{2-\frac{3}{4}r^2}}{2(r^4+4)}}]e^{\frac{E_2 (t-t_0)}{i \hbar}}e^{i\gamma_{E2}}\sqrt{p_{E2}}. \nonumber \\
\end{eqnarray}
and
 \begin{eqnarray}
\beta_c(t)= \int_{-\infty}^{+\infty}dx
\begin{pmatrix}
 -sin(\frac{1}{2}ArcTan(r))\psi_{E1}^{\dag}(x), & cos(\frac{1}{2}ArcTan(r))\psi_{E2}^{\dag}(x)
\end{pmatrix} \times \nonumber \\
\begin{pmatrix}
+cos(\frac{1}{2}ArcTan(r)), & +sin(\frac{1}{2}ArcTan(r)) \\
-sin(\frac{1}{2}ArcTan(r)), & +cos(\frac{1}{2}ArcTan(r))
\end{pmatrix}
\begin{pmatrix}
e^{\frac{E_1 (t-t_0)}{i \hbar}}e^{i\gamma_{E1}}\sqrt{p_{E1}} \psi_{E1}(x) \\
e^{\frac{E_2 (t-t_0)}{i \hbar}}e^{i\gamma_{E2}}\sqrt{p_{E2}} \psi_{E2}(x)
\end{pmatrix}= \nonumber \\
=\int_{-\infty}^{+\infty}dx
\begin{pmatrix}
w_{2,1}^{*}\psi_{E1}^{\dag}(x), & w_{2,2}^{*}\psi_{E2}^{\dag}(x)
\end{pmatrix} \times %\nonumber \\
\begin{pmatrix}
w_{1,1} & w_{1,2} \\
w_{2,1} & w_{2,2}
\end{pmatrix}
\begin{pmatrix}
e^{\frac{E_1 (t-t_0)}{i \hbar}}e^{i\gamma_{E1}}\sqrt{p_{E1}} \psi_{E1}(x) \\
e^{\frac{E_2 (t-t_0)}{i \hbar}}e^{i\gamma_{E2}}\sqrt{p_{E2}} \psi_{E2}(x)
\end{pmatrix}= \nonumber \\
=\int_{-\infty}^{+\infty}dx
\begin{pmatrix}
w_{2,1}^{*}\psi_{E1}^{\dag}(x), & w_{2,2}^{*}\psi_{E2}^{\dag}(x)
\end{pmatrix}
\begin{pmatrix}
w_{1,1}e^{\frac{E_1 (t-t_0)}{i \hbar}}e^{i\gamma_{E1}}\sqrt{p_{E1}} \psi_{E1}(x)+w_{1,2}e^{\frac{E_2 (t-t_0)}{i \hbar}}e^{i\gamma_{E2}}\sqrt{p_{E2}} \psi_{E2}(x) \\
w_{2,1}e^{\frac{E_1 (t-t_0)}{i \hbar}}e^{i\gamma_{E1}}\sqrt{p_{E1}} \psi_{E1}(x)+w_{2,2}e^{\frac{E_2 (t-t_0)}{i \hbar}}e^{i\gamma_{E2}}\sqrt{p_{E2}} \psi_{E2}(x)
\end{pmatrix}= \nonumber \\
=w_{2,1}^{*}w_{1,1}e^{\frac{E_1 (t-t_0)}{i \hbar}}e^{i\gamma_{E1}}\sqrt{p_{E1}} +w_{2,2}^{*}w_{2,2}e^{\frac{E_2 (t-t_0)}{i \hbar}}e^{i\gamma_{E2}}\sqrt{p_{E2}}= \nonumber \\
=-sin(\frac{1}{2}ArcTan(r))cos(\frac{1}{2}ArcTan(r))e^{\frac{E_1 (t-t_0)}{i \hbar}}e^{i\gamma_{E1}}\sqrt{p_{E1}} %%%+ \nonumber \\
+cos(\frac{1}{2}ArcTan(r))^2e^{\frac{E_2 (t-t_0)}{i \hbar}}e^{i\gamma_{E2}}\sqrt{p_{E2}}= \nonumber \\
=cos(\frac{1}{2}ArcTan(r))[-sin(\frac{1}{2}ArcTan(r))e^{\frac{E_1 (t-t_0)}{i \hbar}}e^{i\gamma_{E1}}\sqrt{p_{E1}} %%%+ \nonumber \\
+cos(\frac{1}{2}ArcTan(r))e^{\frac{E_2 (t-t_0)}{i \hbar}}e^{i\gamma_{E2}}\sqrt{p_{E2}}]=\beta(t)
\end{eqnarray}

\normalsize
In such way it was shown how to convert the quantum information represented by eigenergy qubits (as mostly used with formula $|\psi>=\sqrt{p_{E1}}e^{i\gamma_{E1}}|\psi>_{E1}+\sqrt{p_{E2}}e^{i\gamma_{E2}}|\psi>_{E2}$) in position based format $|\psi>=\alpha(t)|w>_{1}+\beta(t)|w>_{2}$ (Wannier qubit format).
We notice that
\begin{eqnarray}
\frac{\alpha_c(t)}{\beta_c(t)}=\frac{[ cos(\frac{1}{2}ArcTan(r)) e^{\frac{E_1 (t-t_0)}{i \hbar}}e^{i\gamma_{E1}}\sqrt{p_{E1}}+ sin(\frac{1}{2}ArcTan(r)) e^{\frac{E_2 (t-t_0)}{i \hbar}}e^{i\gamma_{E2}}\sqrt{p_{E2}} ]}{[-sin(\frac{1}{2}ArcTan(r))e^{\frac{E_1 (t-t_0)}{i \hbar}}e^{i\gamma_{E1}}\sqrt{p_{E1}} %%%+ \nonumber \\
+cos(\frac{1}{2}ArcTan(r))e^{\frac{E_2 (t-t_0)}{i \hbar}}e^{i\gamma_{E2}}\sqrt{p_{E2}}]}= \nonumber \\
=\frac{[ cos(\frac{1}{2}ArcTan(r)) \sqrt{p_{E1}}+ sin(\frac{1}{2}ArcTan(r)) e^{\frac{(E_2-E_1)(t-t_0)}{i \hbar}}e^{i(\gamma_{E2}-\gamma_{E1})}\sqrt{p_{E2}} ]}{[-sin(\frac{1}{2}ArcTan(r))\sqrt{p_{E1}} %%%+ \nonumber \\
+cos(\frac{1}{2}ArcTan(r))e^{\frac{(E_2-E_1) (t-t_0)}{i \hbar}}e^{i(\gamma_{E2}-\gamma_{E1})}\sqrt{p_{E2}}]}= \nonumber \\
=\frac{[ \frac{\sqrt{p_{E1}}}{\sqrt{p_{E2}}}+ Tan(\frac{1}{2}ArcTan(r)) e^{-i\frac{(E_2-E_1)(t-t_0)}{\hbar}}e^{i(\gamma_{E2}-\gamma_{E1})} ]}{[-Tan(\frac{1}{2}ArcTan(r))\frac{\sqrt{p_{E1}}}{\sqrt{p_{E2}}}] %%%+ \nonumber \\
+e^{-i\frac{(E_2-E_1) (t-t_0)}{\hbar}}e^{i(\gamma_{E2}-\gamma_{E1})}},
\end{eqnarray}
what implies that occupancy of full Bloch sphere is not achievable by single Wannier qubit in static electric and magnetic field. Still we can approach arbitrary close to South and North pole of Bloch sphere by regulating $r$ \newline ( achieved from different effective potential generated by biasing electrodes) and by setting arbitrary ratio $\frac{\sqrt{p_{E1}}}{\sqrt{p_{E2}}}$. Furthermore we immediately obtain
\begin{eqnarray} \label{formts11}
E_{p1}=\int_{-\infty}^{+\infty}dx[w_L^{*}(x)\hat{H}w_L(x)]=\int_{-\infty}^{+\infty}dx[(\alpha^{\dag} \psi_{E1}^{\dag}(x) +\beta^{\dag} \psi_{E2}^{\dag}(x)) %\times \nonumber \\
 \hat{H}(\alpha \psi_{E1}(x)+\beta \psi_{E2}(x))]= \nonumber \\
=|\alpha|^2 E_1+|\beta|^2 E_2 %%%=|\alpha|^2 E_1+(1-|\alpha|^2) E_2 %=E_2-|\alpha|^2(E_2-E_1)= \nonumber \\
=(1-|\beta|^2) E_1+|\beta|^2 E_2=E_1+|\beta|^2(E_2-E_1)= \nonumber \\
=E_1+(E_2-E_1)|sin(\frac{1}{2}ArcTan [\frac{\int_{-\infty}^{0}dx (\psi_{E1}(x)\psi_{E2}^{\dag}(x)+\psi_{E1}(x)^{\dag}\psi_{E2}(x))}{\int_{-\infty}^{0}dx (|\psi_{E1}(x)|^2-|\psi_{E2}(x)|^2)}])|^2
 %=p_{E1}(E_1+E_2) -(E_2-E_1)|\alpha|^2
, \nonumber \\
\end{eqnarray}
\begin{eqnarray} \label{formts22}
E_{p2}=\int_{-\infty}^{+\infty}dx[w_R^{*}(x)\hat{H}w_R(x)]=\int_{-\infty}^{+\infty}dx(-\beta^{\dag} \psi_{E1}^{\dag}(x)+\alpha^{\dag} \psi_{E2}^{\dag}(x) ) \times \nonumber \\
\times \hat{H}(-\beta\psi_{E1}(x)+\alpha \psi_{E2}(x))=|\beta|^2 E_1+|\alpha|^2 E_2=E_1+(E_2-E_1)|\alpha|^2= \nonumber \\
%%%=-E_1+(E_2+E_1)|\alpha|^2
=E_1+(E_2-E_1)|cos(\frac{1}{2}ArcTan [\frac{\int_{-\infty}^{0}dx (\psi_{E1}(x)\psi_{E2}^{\dag}(x)+\psi_{E1}(x)^{\dag}\psi_{E2}(x))}{\int_{-\infty}^{0}dx (|\psi_{E1}(x)|^2-|\psi_{E2}(x)|^2)}])|^2,
\end{eqnarray}

\begin{eqnarray} \label{formts21}
t_{s,2 \rightarrow 1}=\int_{-\infty}^{+\infty}dx[w_R^{*}(x)\hat{H}w_L(x)]=\int_{-\infty}^{+\infty}dx(-\beta^{\dag} \psi_{E1}^{\dag}(x) +\alpha^{\dag} \psi_{E2}^{\dag}(x) )\times \nonumber \\
\times \hat{H}(\alpha\psi_{E1}(x)+\beta \psi_{E2}(x))=-\alpha\beta^{*} E_1+\alpha^{*}\beta E_2=(E_2-E_1)\alpha \beta = \nonumber \\
= \frac{1}{2}sin(2 \frac{\int_{-\infty}^{0}dx (\psi_{E1}(x)\psi_{E2}^{\dag}(x)+\psi_{E1}(x)^{\dag}\psi_{E2}(x))}{\int_{-\infty}^{0}dx (|\psi_{E1}(x)|^2-|\psi_{E2}(x)|^2)})(E_2-E_1)= \nonumber \\
=\frac{(E_2-E_1)}{2}\frac{\frac{\int_{-\infty}^{0}dx (\psi_{E1}(x)\psi_{E2}^{\dag}(x)+\psi_{E1}(x)^{\dag}\psi_{E2}(x))}{\int_{-\infty}^{0}dx (|\psi_{E1}(x)|^2-|\psi_{E2}(x)|^2)}}{\sqrt{1+[\frac{\int_{-\infty}^{0}dx (\psi_{E1}(x)\psi_{E2}^{\dag}(x)+\psi_{E1}(x)^{\dag}\psi_{E2}(x))}{\int_{-\infty}^{0}dx (|\psi_{E1}(x)|^2-|\psi_{E2}(x)|^2)}]^2}},
\end{eqnarray}

\begin{eqnarray} \label{formts12}
t_{s,1 \rightarrow 2}=\int_{-\infty}^{+\infty}dx[w_L^{*}(x)\hat{H}w_R(x)]=\int_{-\infty}^{+\infty}dx(\alpha^{\dag} \psi_{E1}^{\dag}(x) +\beta^{\dag} \psi_{E2}^{\dag}(x) )\times \nonumber \\
\times \hat{H}(-\beta\psi_{E1}(x)+\alpha \psi_{E2}(x))=-\alpha^{*}\beta E_1+\alpha\beta^{*} E_2=(E_2-E_1)\alpha \beta= \nonumber \\ =(E_2-E_1)\frac{1}{2}2cos(\frac{1}{2}ArcTan(r))sin(\frac{1}{2}ArcTan(r))=(E_2-E_1)\frac{1}{2}sin(ArcTan(r))= \nonumber \\
=(E_2-E_1)\frac{1}{2}sin(ArcTan(r))=\frac{(E_2-E_1)}{2}\frac{r}{\sqrt{1+r^2}}=\nonumber \\
=\frac{(E_2-E_1)}{2}\frac{\frac{\int_{-\infty}^{0}dx (\psi_{E1}(x)\psi_{E2}^{\dag}(x)+\psi_{E1}(x)^{\dag}\psi_{E2}(x))}{\int_{-\infty}^{0}dx (|\psi_{E1}(x)|^2-|\psi_{E2}(x)|^2)}}{\sqrt{1+[\frac{\int_{-\infty}^{0}dx (\psi_{E1}(x)\psi_{E2}^{\dag}(x)+\psi_{E1}(x)^{\dag}\psi_{E2}(x))}{\int_{-\infty}^{0}dx (|\psi_{E1}(x)|^2-|\psi_{E2}(x)|^2)}]^2}}
%
%ArcTan [ \frac{\int_{-\infty}^{0}dx (\psi_{E1}(x)\psi_{E2}^{\dag}(x)+\psi_{E1}(x)^{\dag}\psi_{E2}(x))}{\int_{-\infty}^{0}dx (|\psi_{E1}(x)|^2-|\psi_{E2}(x)|^2)}].
\end{eqnarray}
Therefore tight-binding model useful for description of Wannier qubits (position based qubits) was fundamentally derived from Schr\"{o}dinger formalism.
The obtained results can be summarized by tight-binding model being functional of eigenenergies of Schr\"{o}dinger Hamiltonian in the form as
\begin{eqnarray}
\label{MainFormula}
\begin{pmatrix}
E_{p1} & t_{s21}  \\
t_{s12} & E_{p2}
\end{pmatrix}=
\begin{pmatrix}
E_1+|sin(\frac{1}{2}ArcTan(r))|^2(E_2-E_1) & (E_2-E_1)\frac{1}{2}sin(ArcTan(r))  \\
(E_2-E_1)\frac{1}{2}sin(ArcTan(r)) & E_1+(E_2-E_1)|cos(\frac{1}{2}ArcTan(r))|^2
\end{pmatrix},
\end{eqnarray}
% cos(\frac{1}{2}ArcTan(r))
with r given by \ref{special} and use of formulas as \ref{formts11} ,\ref{formts22}, \ref{formts12}, \ref{formts21} basing on formula \ref{eqn:fundamentalbasic}.
We will prove the equation \ref{eqn:fundamentalbasic} using Schr\"{o}dinger equation as given by the next section.
If we assume the possible escape of electron from 2 coupled quantum dot system the wavefunction is no longer normalized to one and we can replace real value eigenenergies with
complex value energies, so $E_1 \rightarrow E_{1r}+iE_{1i}$ and $E_2 \rightarrow E_{2r}+iE_{2i}$. In such case the effective tight-binding model corresponding to complex value eigenenergies can be expressed as
\begin{eqnarray}
\label{MainFormulaD}
\begin{pmatrix}
E_{p1D} & t_{s21D}  \\
t_{s12D} & E_{p2D}
\end{pmatrix}=
\begin{pmatrix}
E_{1r}+|sin(\frac{1}{2}ArcTan(r))|^2(E_{2r}-E_{1r}) & (E_{2r}-E_{1r})\frac{1}{2}sin(ArcTan(r))  \\
(E_{2r}-E_{1r})\frac{1}{2}sin(ArcTan(r)) & E_{1r}+(E_{2r}-E_{1r})|cos(\frac{1}{2}ArcTan(r))|^2
\end{pmatrix}+ \nonumber \\
+\sqrt{-1}
\begin{pmatrix}
E_{1i}+|sin(\frac{1}{2}ArcTan(r))|^2(E_{2i}-E_{1i}) & (E_{2i}-E_{1i})\frac{1}{2}sin(ArcTan(r))  \\
(E_{2i}-E_{1i})\frac{1}{2}sin(ArcTan(r)) & E_{1i}+(E_{2i}-E_{1i})|cos(\frac{1}{2}ArcTan(r))|^2
\end{pmatrix}
\end{eqnarray}
The dissipative version of tight-binding model accounting for electron escape from 2 quantum dot system due to tunneling is non-Hermitian, while non-dissipative version of tight-binding model is Hermitian.
\subsection{Equivalence of Wannier and Schr\"{o}dinger formalism}
Let us start from static case of electric and magnetic time-independent fields so
\begin{eqnarray}
\hat{H}|\psi>= (-\frac{\hbar^2}{2m}\frac{d^2}{dx^2}+V(x))
\begin{pmatrix}
e^{i\gamma_{E1}(t)}\sqrt{p_{E1}}\psi_{E1}(x) \\
e^{i\gamma_{E2}(t)}\sqrt{p_{E2}}\psi_{E2}(x)
\end{pmatrix}
=
\begin{pmatrix}
E_1 & 0 \\
0   & E_2
\end{pmatrix}
\begin{pmatrix}
e^{i\gamma_{E1}(t)}\sqrt{p_{E1}}\psi_{E1}(x) \\
e^{i\gamma_{E2}(t)}\sqrt{p_{E2}}\psi_{E2}(x)
\end{pmatrix}
= i \hbar \frac{d}{dt}
\begin{pmatrix}
e^{i\gamma_{E1}(t)}\sqrt{p_{E1}}\psi_{E1}(x) \\
e^{i\gamma_{E2}(t)}\sqrt{p_{E2}}\psi_{E2}(x)
\end{pmatrix}
%%%\sqrt{p_{E1}}|\psi>_{E1}+\sqrt{p_{E2}}|\psi>_{E2}
\end{eqnarray} that is equivalent to
\begin{eqnarray}
\begin{pmatrix}
E_1 & 0 \\
0   & E_2
\end{pmatrix}
\begin{pmatrix}
e^{i\gamma_{E1}(t)}\sqrt{p_{E1}} \\
e^{i\gamma_{E2}(t)}\sqrt{p_{E2}}
\end{pmatrix}
= i \hbar \frac{d}{dt}
\begin{pmatrix}
e^{i\gamma_{E1}(t)}\sqrt{p_{E1}} \\
e^{i\gamma_{E2}(t)}\sqrt{p_{E2}}
\end{pmatrix}
%%%\sqrt{p_{E1}}|\psi>_{E1}+\sqrt{p_{E2}}|\psi>_{E2}
\end{eqnarray}
and we have
\begin{eqnarray}
\begin{pmatrix}
w_{1,1} & w_{1,2} \\
w_{2,1} & w_{2,2}
\end{pmatrix}
\begin{pmatrix}
E_1 & 0 \\
0   & E_2
\end{pmatrix}
\begin{pmatrix}
e^{i\gamma_{E1}(t)}\sqrt{p_{E1}}\psi_{E1}(x) \\
e^{i\gamma_{E2}(t)}\sqrt{p_{E2}}\psi_{E2}(x)
\end{pmatrix}
= i \hbar \frac{d}{dt}
\begin{pmatrix}
w_{1,1} & w_{1,2} \\
w_{2,1} & w_{2,2}
\end{pmatrix}
\begin{pmatrix}
e^{i\gamma_{E1}(t)}\sqrt{p_{E1}}\psi_{E1}(x) \\
e^{i\gamma_{E2}(t)}\sqrt{p_{E2}}\psi_{E2}(x)
\end{pmatrix}
%%%\sqrt{p_{E1}}|\psi>_{E1}+\sqrt{p_{E2}}|\psi>_{E2}
\end{eqnarray}
that yields

\begin{eqnarray}
\begin{pmatrix}
w_{1,1} & w_{1,2} \\
w_{2,1} & w_{2,2}
\end{pmatrix}
\begin{pmatrix}
E_1 & 0 \\
0   & E_2
\end{pmatrix}
\begin{pmatrix}
e^{i\gamma_{E1}(t)}\sqrt{p_{E1}}\psi_{E1}(x) \\
e^{i\gamma_{E2}(t)}\sqrt{p_{E2}}\psi_{E2}(x)
\end{pmatrix}
= i \hbar \frac{d}{dt}
\begin{pmatrix}
w_{1,1} & w_{1,2} \\
w_{2,1} & w_{2,2}
\end{pmatrix}
\begin{pmatrix}
e^{i\gamma_{E1}(t)}\sqrt{p_{E1}}w_{E1}(x) \\
e^{i\gamma_{E2}(t)}\sqrt{p_{E2}}w_{E2}(x)
\end{pmatrix}
%%%\sqrt{p_{E1}}|\psi>_{E1}+\sqrt{p_{E2}}|\psi>_{E2}
\end{eqnarray}
what implies
\small
\begin{eqnarray}
\begin{pmatrix}
w_{1,1} & w_{1,2} \\
w_{2,1} & w_{2,2}
\end{pmatrix}
\begin{pmatrix}
E_1 & 0 \\
0   & E_2
\end{pmatrix}
\begin{pmatrix}
w_{1,1} & w_{1,2} \\
w_{2,1} & w_{2,2}
\end{pmatrix}^{-1}
\begin{pmatrix}
w_{1,1} & w_{1,2} \\
w_{2,1} & w_{2,2}
\end{pmatrix}
\begin{pmatrix}
e^{i\gamma_{E1}(t)}\sqrt{p_{E1}}\psi_{E1}(x) \\
e^{i\gamma_{E2}(t)}\sqrt{p_{E2}}\psi_{E2}(x)
\end{pmatrix}
= i \hbar \frac{d}{dt}
\begin{pmatrix}
w_{1,1} & w_{1,2} \\
w_{2,1} & w_{2,2}
\end{pmatrix}
\begin{pmatrix}
e^{i\gamma_{E1}(t)}\sqrt{p_{E1}}w_{E1}(x) \\
e^{i\gamma_{E2}(t)}\sqrt{p_{E2}}w_{E2}(x)
\end{pmatrix}
%%%\sqrt{p_{E1}}|\psi>_{E1}+\sqrt{p_{E2}}|\psi>_{E2}
\end{eqnarray}
\normalsize
%\begin{pmatrix}
%w_{1,1} & w_{1,2} \\
%w_{2,1} & w_{2,2}
%\end{pmatrix}
Having
\begin{eqnarray}
\alpha_c(t)|w_{L}(x)>+\beta_c(t)|w_{R}(x)>=
\begin{pmatrix}
\alpha_c(t) w_{L}(x) \\
\beta_c(t) w_{R}(x)
\end{pmatrix}
=
\begin{pmatrix}
w_{1,1} & w_{1,2} \\
w_{2,1} & w_{2,2}
\end{pmatrix}
\begin{pmatrix}
e^{i\gamma_{E1}(t)}\sqrt{p_{E1}}w_{E1}(x) \\
e^{i\gamma_{E2}(t)}\sqrt{p_{E2}}w_{E2}(x)
\end{pmatrix}
%%%\sqrt{p_{E1}}|\psi>_{E1}+\sqrt{p_{E2}}|\psi>_{E2}
\end{eqnarray}
we obtain
\small
\begin{eqnarray}
\Bigg[
\begin{pmatrix}
w_{1,1} & w_{1,2} \\
w_{2,1} & w_{2,2}
\end{pmatrix}
\begin{pmatrix}
E_1 & 0 \\
0   & E_2
\end{pmatrix}
\begin{pmatrix}
w_{2,2} & -w_{1,2} \\
-w_{2,1} & w_{1,1}
\end{pmatrix}
\Bigg]
\begin{pmatrix}
\alpha_c(t)w_{L}(x) \\
\beta_c(t) w_{R}(x)
\end{pmatrix}
= i \hbar \frac{d}{dt}
\begin{pmatrix}
\alpha_c(t)w_{L}(x) \\
\beta_c(t) w_{R}(x)
\end{pmatrix}
%%%\sqrt{p_{E1}}|\psi>_{E1}+\sqrt{p_{E2}}|\psi>_{E2}
\end{eqnarray}
\normalsize
and
\small
%
%\begin{eqnarray}
%\Bigg[
%\begin{pmatrix}
%w_{1,1} & w_{1,2} \\
%w_{2,1} & w_{2,2}
%\end{pmatrix}
%\begin{pmatrix}
%w_{2,2}E_1 & -w_{1,2}E_1 \\
%-w_{2,1}E_2 & w_{1,1}E_2
%\end{pmatrix}
%\Bigg]
%\begin{pmatrix}
%\alpha_c(t)w_{L}(x) \\
%\beta_c(t) w_{R}(x)
%\end{pmatrix}
%= i \hbar \frac{d}{dt}
%\begin{pmatrix}
%\alpha_c(t)w_{L}(x) \\
%\beta_c(t) w_{R}(x)
%\end{pmatrix}
%%%%\sqrt{p_{E1}}|\psi>_{E1}+\sqrt{p_{E2}}|\psi>_{E2}
%\end{eqnarray}
\normalsize
%Furthermore we have
%\begin{eqnarray}
%\Bigg[
%\begin{pmatrix}
%w_{1,1} & w_{1,2} \\
%w_{2,1} & w_{2,2}
%\end{pmatrix}
%\begin{pmatrix}
%w_{2,2}E_1 & -w_{1,2}E_1 \\
%-w_{2,1}E_2 & w_{1,1}E_2
%\end{pmatrix}
%\Bigg]
%\begin{pmatrix}
%\alpha_c(t)w_{L}(x) \\
%\beta_c(t) w_{R}(x)
%\end{pmatrix}
%= i \hbar \frac{d}{dt}
%\begin{pmatrix}
%\alpha_c(t)w_{L}(x) \\
%\beta_c(t) w_{R}(x)
%\end{pmatrix}
%%%%\sqrt{p_{E1}}|\psi>_{E1}+\sqrt{p_{E2}}|\psi>_{E2}
%\end{eqnarray}
that gives
\begin{eqnarray}
\Bigg[
\begin{pmatrix}
w_{1,1}w_{2,2}E_1-w_{1,2}w_{2,1}E_2 & -w_{1,1}w_{1,2}E_1+w_{1,2}w_{1,1}E_2 \\
w_{2,1}w_{2,2}E_1-w_{2,1}w_{2,2}E_2 & -w_{1,2}w_{2,1}E_1+w_{1,1}w_{2,2}E_2
\end{pmatrix}
\Bigg]
\begin{pmatrix}
\alpha_c(t)w_{L}(x) \\
\beta_c(t) w_{R}(x)
\end{pmatrix}
= i \hbar \frac{d}{dt}
\begin{pmatrix}
\alpha_c(t)w_{L}(x) \\
\beta_c(t) w_{R}(x)
\end{pmatrix}
%%%\sqrt{p_{E1}}|\psi>_{E1}+\sqrt{p_{E2}}|\psi>_{E2}
\end{eqnarray}
%%%%Last equation can be rewritten as
%\small
%\begin{eqnarray}
%\Bigg[
%\begin{pmatrix}
%w_{1,1}w_{2,2}E_1-w_{1,2}w_{2,1}E_2 & -w_{1,1}w_{1,2}E_1+w_{1,2}w_{1,1}E_2 \\
%w_{2,1}w_{2,2}E_1-w_{2,1}w_{2,2}E_2 & -w_{1,2}w_{2,1}E_1+w_{1,1}w_{2,2}E_2
%\end{pmatrix}
%\begin{pmatrix}
%w_{L}(x) & 0 \\
%0 & w_{R}(x)
%\end{pmatrix}
%\Bigg]
%\begin{pmatrix}
%\alpha_c(t) \\
%\beta_c(t)
%\end{pmatrix}
%= i \hbar \frac{d}{dt} \Bigg[
%\begin{pmatrix}
%w_{L}(x) & 0 \\
%0 & w_{R}(x)
%\end{pmatrix}
%\begin{pmatrix}
%\alpha_c(t) \\
%\beta_c(t)
%\end{pmatrix}
%\Bigg]
%%%%\sqrt{p_{E1}}|\psi>_{E1}+\sqrt{p_{E2}}|\psi>_{E2}
%\end{eqnarray}
%so we have
\small
%\begin{eqnarray}
%\Bigg[
%\begin{pmatrix}
%(w_{1,1}w_{2,2}E_1-w_{1,2}w_{2,1}E_2)w_L(x) & (-w_{1,1}w_{1,2}E_1+w_{1,2}w_{1,1}E_2)w_R(x) \\
%(w_{2,1}w_{2,2}E_1-w_{2,1}w_{2,2}E_2)w_L(x) & (-w_{1,2}w_{2,1}E_1+w_{1,1}w_{2,2}E_2)w_R(x)
%\end{pmatrix}
%\Bigg]
%\begin{pmatrix}
%\alpha_c(t) \\
%\beta_c(t)
%\end{pmatrix}
%= i \hbar \frac{d}{dt} \Bigg[
%\begin{pmatrix}
%w_{L}(x) & 0 \\
%0 & w_{R}(x)
%\end{pmatrix}
%\begin{pmatrix}
%\alpha_c(t) \\
%\beta_c(t)
%\end{pmatrix}
%\Bigg]
%%%%\sqrt{p_{E1}}|\psi>_{E1}+\sqrt{p_{E2}}|\psi>_{E2}
%\end{eqnarray}
%\normalsize
%Applying
%$ \int_{-\infty}^{+\infty} dx
%\begin{pmatrix}
%w_{L}(x)^{*} & 0 \\
%0 & w_{R}(x)^{*}
%\end{pmatrix}
%$
%to both sides we obtain
%\small
%\begin{eqnarray}
%\Bigg[
%\int_{-\infty}^{+\infty} dx
%\begin{pmatrix}
%w_{L}(x){*} & 0 \\
%0 & w_{R}(x)^{*}
%\end{pmatrix}
%\begin{pmatrix}
%(w_{1,1}w_{2,2}E_1-w_{1,2}w_{2,1}E_2)w_L(x) & (-w_{1,1}w_{1,2}E_1+w_{1,2}w_{1,1}E_2)w_R(x) \\
%(w_{2,1}w_{2,2}E_1-w_{2,1}w_{2,2}E_2)w_L(x) & (-w_{1,2}w_{2,1}E_1+w_{1,1}w_{2,2}E_2)w_R(x)
%\end{pmatrix}
%\Bigg]
%\begin{pmatrix}
%\alpha_c(t) \\
%\beta_c(t)
%\end{pmatrix}
%= i \hbar \frac{d}{dt}
%\begin{pmatrix}
%\alpha_c(t) \\
%\beta_c(t)
%\end{pmatrix}
%%%%\sqrt{p_{E1}}|\psi>_{E1}+\sqrt{p_{E2}}|\psi>_{E2}
%\end{eqnarray}
%%%%and thus we obtain
\small
%\begin{eqnarray}
%\Bigg[
%\begin{pmatrix}
%\int_{-\infty}^{+\infty} dxw_{L}(x)^{*}(w_{1,1}w_{2,2}E_1-w_{1,2}w_{2,1}E_2)w_L(x) & \int_{-\infty}^{+\infty} dxw_{L}(x)^{*}(-w_{1,1}w_{1,2}E_1+w_{1,2}w_{1,1}E_2)w_R(x) \\
%\int_{-\infty}^{+\infty} dxw_{R}(x)^{*}(w_{2,1}w_{2,2}E_1-w_{2,1}w_{2,2}E_2)w_L(x) & \int_{-\infty}^{+\infty} dxw_{R}(x)^{*}(-w_{1,2}w_{2,1}E_1+w_{1,1}w_{2,2}E_2)w_R(x)
%\end{pmatrix}
%\Bigg]
%\begin{pmatrix}
%\alpha_c(t) \\
%\beta_c(t)
%\end{pmatrix}
%= i \hbar \frac{d}{dt}
%\begin{pmatrix}
%\alpha_c(t) \\
%\beta_c(t)
%\end{pmatrix}
%%%%\sqrt{p_{E1}}|\psi>_{E1}+\sqrt{p_{E2}}|\psi>_{E2}
%\end{eqnarray}
%\begin{eqnarray}
%\begin{pmatrix}
%E_{p1} & t_{s21} \\
%t_{s12} & E_{p2}
%\end{pmatrix}
%\begin{pmatrix}
%\alpha_c(t) \\
%\beta_c(t)
%\end{pmatrix}
%=
%\Bigg[
%\begin{pmatrix}
%w_{1,1}w_{2,2}E_1-w_{1,2}w_{2,1}E_2 & -w_{1,1}w_{1,2}E_1+w_{1,2}w_{1,1}E_2 \\
%w_{2,1}w_{2,2}E_1-w_{2,1}w_{2,2}E_2 & -w_{1,2}w_{2,1}E_1+w_{1,1}w_{2,2}E_2
%\end{pmatrix}
%\Bigg]
%\begin{pmatrix}
%\alpha_c(t) \\
%\beta_c(t)
%\end{pmatrix}
%=
%i \hbar \frac{d}{dt}
%\begin{pmatrix}
%\alpha_c(t) \\
%\beta_c(t)
%\end{pmatrix}
%%%%\sqrt{p_{E1}}|\psi>_{E1}+\sqrt{p_{E2}}|\psi>_{E2}
%\end{eqnarray}
Last formula can be rewritten to be as
\small
\begin{eqnarray}
\begin{pmatrix}
\int_{-\infty}^{+\infty} dxw_{L}^{*}(x)(-\frac{\hbar^2}{2m}+V(x))w_L(x) & \int_{-\infty}^{+\infty} dxw_{L}^{*}(x)(-\frac{\hbar^2}{2m}+V(x))w_R(x) \\
\int_{-\infty}^{+\infty} dxw_{R}^{*}(x)(-\frac{\hbar^2}{2m}+V(x))w_L(x) & \int_{-\infty}^{+\infty} dxw_{R}^{*}(x)(-\frac{\hbar^2}{2m}+V(x))w_R(x)
\end{pmatrix}
\begin{pmatrix}
\alpha_c(t) \\
\beta_c(t)
\end{pmatrix}
= i \hbar \frac{d}{dt}
\begin{pmatrix}
\alpha_c(t) \\
\beta_c(t)
\end{pmatrix}
%%%\sqrt{p_{E1}}|\psi>_{E1}+\sqrt{p_{E2}}|\psi>_{E2}
\end{eqnarray}
\normalsize
or equivalently
\begin{eqnarray}
\begin{pmatrix}
E_{p1} & t_s \\
t_s^{*} & E_{p2}
\end{pmatrix}
\begin{pmatrix}
\alpha_c(t)w_L(x) \\
\beta_c(t)w_R(x)
\end{pmatrix}
= \nonumber \\
\begin{pmatrix}
\int_{-\infty}^{+\infty} dxw_{L}^{*}(x)(-\frac{\hbar^2}{2m}+V(x))w_L(x) & \int_{-\infty}^{+\infty} dxw_{L}^{*}(x)(-\frac{\hbar^2}{2m}+V(x))w_R(x) \\
\int_{-\infty}^{+\infty} dxw_{R}^{*}(x)(-\frac{\hbar^2}{2m}+V(x))w_L(x) & \int_{-\infty}^{+\infty} dxw_{R}^{*}(x)(-\frac{\hbar^2}{2m}+V(x))w_R(x)
\end{pmatrix}
\begin{pmatrix}
\alpha_c(t)w_L(x) \\
\beta_c(t)w_R(x)
\end{pmatrix}
= i \hbar \frac{d}{dt}
\begin{pmatrix}
\alpha_c(t)w_L(x) \\
\beta_c(t)w_R(x)
\end{pmatrix}
%%%\sqrt{p_{E1}}|\psi>_{E1}+\sqrt{p_{E2}}|\psi>_{E2}
\end{eqnarray}
Therefore we can always obtain tight-binding model from Schr\"{o}dinger model with two eigenergies.
\subsection{Rabi oscillations in tight-binding model}
In time-dependent case we can consider the existence of Rabi oscillations by assuming the effective Hamiltonian to be of the form
$H=E_1\ket{E_1}\bra{E_1}+E_2\ket{E_2}\bra{E_2}+f_1(t)e^{i\xi(t)}\ket{E_2}\bra{E_1}+f_1(t)e^{-i\xi(t)}\ket{E_1}\bra{E_2}$ with $f(t)=f_1(t)e^{i\xi(t)}$ acting on q-state given by
$\ket{\psi}=\sqrt{p_{E1}}e^{i \gamma_1}\ket{E_1}+\sqrt{p_{E2}}e^{i \gamma_2}\ket{E_2}$ and immediately we obtain
%$
%\begin{eqnarray}
%\sqrt{p_{E1}}e^{i \gamma_1} \\
%\sqrt{p_{E2}}e^{i \gamma_2} \\
%\end{eqnarray}
%$
%what results in
%\begin{eqnarray}
%% \nonumber % Remove numbering (before each equation)
%  \sqrt{p_{E1}}e^{i \gamma_1}E_1+f_1\sqrt{p_{E2}}e^{i \gamma_2}e^{i \xi} &=& e^{i\gamma_1}\hbar[i \frac{1}{2 \sqrt{p_{E1}}} \frac{d}{dt}p_{E1}-\sqrt{p_{E1}}\frac{d}{dt}\gamma_1 ] \\
% f_1\sqrt{p_{E1}}e^{i \gamma_1}e^{-i\xi}+ \sqrt{p_{E2}}e^{i \gamma_2}E_2 &=& e^{i\gamma_2}\hbar[i \frac{1}{2 \sqrt{p_{E2}}} \frac{d}{dt}p_{E2}-\sqrt{p_{E2}}\frac{d}{dt}\gamma_2 ]
%\end{eqnarray}
%We obtain
\begin{eqnarray}
  \sqrt{p_{E1}}E_1+f_1\sqrt{p_{E2}}e^{i (\gamma_2-\gamma_1)}e^{i \xi} &=& \hbar[i \frac{1}{2 \sqrt{p_{E1}}} \frac{d}{dt}p_{E1}-\sqrt{p_{E1}}\frac{d}{dt}\gamma_1 ] \\
 f_1\sqrt{p_{E1}}e^{-i(\gamma_2-\gamma_1)}e^{-i\xi}+ \sqrt{p_{E2}}E_2 &=& \hbar[i \frac{1}{2 \sqrt{p_{E2}}} \frac{d}{dt}p_{E2}-\sqrt{p_{E2}}\frac{d}{dt}\gamma_2 ]
\end{eqnarray}
Consequently we have
%\begin{eqnarray}
$2f_1\sqrt{p_{E1}}\sqrt{p_{E2}}sin[(\gamma_2-\gamma_1)+ \xi] = \hbar \frac{d}{dt}p_{E1}=-\hbar \frac{d}{dt}p_{E2}$
% \\
%\end{eqnarray}
and by setting parametrization $[sin(\Theta(t))]^2=p_{E1}(t),[cos(\Theta(t))]^2=p_{E2}(t)$ we obtain
%%%that results in
\begin{eqnarray}
f_1(t) sin[(\gamma_2(t)-\gamma_1(t))+ \xi(t)] &=& \hbar \frac{d}{dt}\Theta(t).
\end{eqnarray}
Furthermore using $\sqrt{p_{E1}}E_1+f_1\sqrt{p_{E2}}cos[(\gamma_2-\gamma_1)+ \xi] = -\hbar \sqrt{p_{E1}} \frac{d}{dt}\gamma_{1}$  we obtain
\begin{eqnarray}
  %%%%\sqrt{p_{E1}}E_1+f_1\sqrt{p_{E2}}cos[(\gamma_2-\gamma_1)+ \xi] &=& -\hbar \sqrt{p_{E1}} \frac{d}{dt}\gamma_{1},\\
   E_1+f_1(t) ctan(\Theta(t))cos[(\gamma_2(t)-\gamma_1(t))+ \xi(t)] = -\hbar  \frac{d}{dt}\gamma_{1}(t), \nonumber \\
   E_2+f_1(t) tan(\Theta(t))cos[(\gamma_2(t)-\gamma_1(t))+ \xi(t)] = -\hbar  \frac{d}{dt}\gamma_{2}(t), \nonumber \\
  -\frac{(E_2-E_1)(t-t_0)}{\hbar} +\frac{1}{\hbar}\int_{t_0}^{t}dt'f_1(t')cos[(\gamma_2(t')-\gamma_1(t'))+ \xi(t')] (ctan(\Theta(t'))-tan(\Theta(t')))+(\gamma_{2}(t_0)-\gamma_{1}(t_0)) \nonumber \\ = \gamma_{2}(t)-\gamma_{1}(t).
%2f_1\sqrt{tan[\Theta(t)]}sin[(\gamma_2-\gamma_1)+ \xi] &=& \hbar \frac{d}{dt}\Theta(t).
\end{eqnarray}
%%We obtain
%%\begin{eqnarray}
%%\gamma_{2}(t)-\gamma_{1}(t)= -\frac{(E_2-E_1)t}{\hbar} +f_1(t)cos[(\gamma_2(t)-\gamma_1(t))+ \xi(t)] (ctan(\Theta(t))-tan(\Theta(t))).
%%\end{eqnarray}
Setting $(\gamma_2(t)-\gamma_1(t))+ \xi(t)=-\frac{(E_2-E_1)(t-t_0)}{\hbar}+(\gamma_2(t_0)-\gamma_1(t_0))+\xi(t)=\frac{\pi}{2}$ we have constant speed of change $\gamma_{2}(t)-\gamma_{1}(t)$ with time as well as constant value of $\hbar \frac{d}{dt}\Theta(t)$ under the condition that $f_1(t)$ is constant with time. In other case the situation is not so analytical and we need to deal with system of 3 coupled
ordinary differential equations given as
\begin{eqnarray} \label{gengenQ1}
  E_1+f_1(t) ctan(\Theta(t))cos[(\gamma_2(t)-\gamma_1(t))+ \xi(t)] &=& -\hbar  \frac{d}{dt}\gamma_{1}(t),\\
  \label{gengenQ2}
  E_2+f_1(t) tan(\Theta(t))cos[(\gamma_2(t)-\gamma_1(t))+ \xi(t)] &=& -\hbar  \frac{d}{dt}\gamma_{2}(t), \\
  \label{gengenQ3}
  f_1(t) sin[(\gamma_2(t)-\gamma_1(t))+ \xi(t)] &=& +\hbar \frac{d}{dt}\Theta(t),
\end{eqnarray}
where given parametric real value functions are $\xi(t)$ and $f_1(t)$.
In time-dependent case we can consider the existence of Rabi oscillations by assuming the effective Hamiltonian to be of the form
$H=E_1\ket{E_1}\bra{E_1}+E_2\ket{E_2}\bra{E_2}+f_1(t)e^{i\xi(t)}\ket{E_2}\bra{E_1}+f_1(t)e^{-i\xi(t)}\ket{E_1}\bra{E_2}$ with $f(t)=f_1(t)e^{i\xi(t)}$, where $f_1(t) \in R$ and $\xi(t)=\frac{(E_2-E_1)t}{\hbar}+\rho=\frac{(E_2-E_1)(t-t_0)}{\hbar}-(\gamma_2(t_0)-\gamma_1(t_0))+\frac{\pi}{2}$ and with $\rho=-(\gamma_2(t_0)-\gamma_1(t_0))+\frac{\pi}{2}$, so for certain classes of $V(x,t)$ potential we can write
\begin{eqnarray}
\hat{H}\ket{\psi}= (-\frac{\hbar^2}{2m}\frac{d^2}{dx^2}+V(x,t))
\begin{pmatrix}
e^{i\gamma_{E1}(t)}\sqrt{p_{E1}(t)}\psi_{E1}(x) \\
e^{i\gamma_{E2}(t)}\sqrt{p_{E2}(t)}\psi_{E2}(x)
\end{pmatrix}
=
\begin{pmatrix}
E_1 & f(t) \\
f(t)^{*}   & E_2
\end{pmatrix}
\begin{pmatrix}
e^{i\gamma_{E1}(t)}\sqrt{p_{E1}}\psi_{E1}(x) \\
e^{i\gamma_{E2}(t)}\sqrt{p_{E2}}\psi_{E2}(x)
\end{pmatrix}= \nonumber \\
= i \hbar \frac{d}{dt}
\begin{pmatrix}
e^{i\gamma_{E1}(t)}\sqrt{p_{E1}(t)}\psi_{E1}(x) \\
e^{i\gamma_{E2}(t)}\sqrt{p_{E2}(t)}\psi_{E2}(x)
\end{pmatrix}
%%%\sqrt{p_{E1}}|\psi>_{E1}+\sqrt{p_{E2}}|\psi>_{E2}
\end{eqnarray} that is equivalent to
\begin{eqnarray}
\begin{pmatrix}
E_1 & f(t) \\
f(t)^{*}   & E_2
\end{pmatrix}
\begin{pmatrix}
e^{i\gamma_{E1}(t)}\sqrt{p_{E1}(t)} \\
e^{i\gamma_{E2}(t)}\sqrt{p_{E2}(t)}
\end{pmatrix}
= i \hbar \frac{d}{dt}
\begin{pmatrix}
e^{i\gamma_{E1}(t)}\sqrt{p_{E1}(t)} \\
e^{i\gamma_{E2}(t)}\sqrt{p_{E2}(t)}
\end{pmatrix}
%%%\sqrt{p_{E1}}|\psi>_{E1}+\sqrt{p_{E2}}|\psi>_{E2}
\end{eqnarray}
what

\begin{eqnarray}
\Bigg[
\begin{pmatrix}
w_{1,1} & w_{1,2} \\
w_{2,1} & w_{2,2}
\end{pmatrix}
\begin{pmatrix}
E_1 & f(t) \\
f(t)^{*}   & E_2
\end{pmatrix}
\begin{pmatrix}
w_{2,2} & -w_{1,2} \\
-w_{2,1} & w_{1,1}
\end{pmatrix}
\Bigg]
\Bigg[
\begin{pmatrix}
w_{1,1} & w_{1,2} \\
w_{2,1} & w_{2,2}
\end{pmatrix}
\begin{pmatrix}
e^{i\gamma_{E1}(t)}\sqrt{p_{E1}(t)} \\
e^{i\gamma_{E2}(t)}\sqrt{p_{E2}(t)}
\end{pmatrix}
\Bigg]
= \nonumber \\
=i \hbar \frac{d}{dt}
\Bigg[
\begin{pmatrix}
w_{1,1} & w_{1,2} \\
w_{2,1} & w_{2,2}
\end{pmatrix}
\begin{pmatrix}
e^{i\gamma_{E1}(t)}\sqrt{p_{E1}(t)} \\
e^{i\gamma_{E2}(t)}\sqrt{p_{E2}(t)}
\end{pmatrix}
\Bigg]
%%%\sqrt{p_{E1}}|\psi>_{E1}+\sqrt{p_{E2}}|\psi>_{E2}
\end{eqnarray}
that implies
%\begin{eqnarray}
%\Bigg[
%\begin{pmatrix}
%w_{1,1} & w_{1,2} \\
%w_{2,1} & w_{2,2}
%\end{pmatrix}
%\begin{pmatrix}
%w_{2,2}E_1-f(t)w_{2,1} & -w_{1,2}E_1+f(t)w_{1,1} \\
%w_{2,2}f(t)^{*}-w_{2,1}E_2 & -w_{1,2}f(t)^{*}+w_{1,1}E_2
%\end{pmatrix}
%\Bigg]
%\begin{pmatrix}
%\alpha_q(t) \\
%\beta_q(t)
%\end{pmatrix}
%%%%= \nonumber \\
%=i \hbar \frac{d}{dt}
%\begin{pmatrix}
%\alpha_q(t) \\
%\beta_q(t)
%\end{pmatrix}
%%%%\sqrt{p_{E1}}|\psi>_{E1}+\sqrt{p_{E2}}|\psi>_{E2}
%\end{eqnarray}
\small
\begin{eqnarray}
\begin{pmatrix}
w_{1,1}w_{2,2}E_1-f(t)w_{2,1}w_{1,1}+w_{1,2}w_{2,2}f(t)^{*}-w_{1,2}w_{2,1}E_2 & -w_{1,2}w_{1,1}E_1+f(t)w_{1,1}w_{1,1}-w_{1,2}w_{1,2}f^{*}(t)+w_{1,1}w_{1,2}E_2 \\
w_{2,2}w_{2,2}f(t)^{*}-w_{2,2}w_{2,1}E_2+w_{2,1}w_{2,2}E_1-f(t)w_{2,1}w_{2,1} & -w_{2,1}w_{1,2}E_1+w_{2,1}f(t)w_{1,1}-w_{2,2}w_{1,2}f(t)^{*}+w_{2,2}w_{1,1}E_2
\end{pmatrix} %\times \nonumber \\
%\times
\begin{pmatrix}
\alpha_q(t) \\
\beta_q(t)
\end{pmatrix}
 \nonumber \\
=i \hbar \frac{d}{dt}
\begin{pmatrix}
\alpha_q(t) \\
\beta_q(t)
\end{pmatrix}
%%%\sqrt{p_{E1}}|\psi>_{E1}+\sqrt{p_{E2}}|\psi>_{E2}
\end{eqnarray}
\normalsize
%%%%We can rewrite the last equation in the form
\small
%\begin{eqnarray}
%\begin{pmatrix}
%cos(\frac{1}{2}ArcTan(r))^2 E_1+(f(t)+f(t)^{*})(sin(\frac{1}{2}ArcTan(r))cos(\frac{1}{2}ArcTan(r)))-E_2sin(\frac{1}{2}ArcTan(r))^2 & -w_{1,2}w_{1,1}E_1+f(t)w_{1,1}w_{1,1}-w_{1,2}w_{1,2}f^{*}(t)+w_{1,1}w_{1,2}E_2 \\
%cos(\frac{1}{2}ArcTan(r))^2f(t)^{*}+(E_2-E_1)(sin(\frac{1}{2}ArcTan(r))cos(\frac{1}{2}ArcTan(r)))-f(t)sin(\frac{1}{2}ArcTan(r))^2 & -w_{2,1}w_{1,2}E_1+w_{2,1}f(t)w_{1,1}-w_{2,2}w_{1,2}f(t)^{*}+w_{2,2}w_{1,1}E_2
%\end{pmatrix} %\times \nonumber \\
%%\times
%\begin{pmatrix}
%\alpha_q(t) \\
%\beta_q(t)
%\end{pmatrix}
% \nonumber \\
%=i \hbar \frac{d}{dt}
%\begin{pmatrix}
%\alpha_q(t) \\
%\beta_q(t)
%\end{pmatrix}.
%%%%\sqrt{p_{E1}}|\psi>_{E1}+\sqrt{p_{E2}}|\psi>_{E2}
%\end{eqnarray}
that can be written as
\begin{eqnarray}
\begin{pmatrix}
E_{p1}(t) & t_{s(1,2)}(t) \\
t_{s(2,1)}(t) & E_{p2}(t)
\end{pmatrix} %\times \nonumber \\
%\times
\begin{pmatrix}
\alpha_q(t) \\
\beta_q(t)
\end{pmatrix}
=
\begin{pmatrix}
a_{1,1}(t) & a_{1,2}(t) \\
a_{2,1}(t) & a_{2,2}(t)
\end{pmatrix} %\times \nonumber \\
%\times
\begin{pmatrix}
\alpha_q(t) \\
\beta_q(t)
\end{pmatrix}
%%%% \nonumber \\
=i \hbar \frac{d}{dt}
\begin{pmatrix}
\alpha_q(t) \\
\beta_q(t)
\end{pmatrix}.
%%%\sqrt{p_{E1}}|\psi>_{E1}+\sqrt{p_{E2}}|\psi>_{E2}
\end{eqnarray}
with 4 time-dependent coefficients

\begin{eqnarray} \label{generaleqns}
a_{1,1}=cos(\frac{1}{2}ArcTan(r))^2 E_1+(f(t)+f(t)^{*})(sin(\frac{1}{2}ArcTan(r))cos(\frac{1}{2}ArcTan(r)))+E_2sin(\frac{1}{2}ArcTan(r))^2= \nonumber \\
=cos(\frac{1}{2}ArcTan(r))^2 E_1+cos(\frac{(E_2-E_1)t}{\hbar}+\rho)f_1(t)sin(ArcTan(r)+E_2sin(\frac{1}{2}ArcTan(r))^2, \\
a_{2,1}=cos(\frac{1}{2}ArcTan(r))^2f(t)^{*}+(E_2-E_1)(sin(\frac{1}{2}ArcTan(r))cos(\frac{1}{2}ArcTan(r)))-f(t)sin(\frac{1}{2}ArcTan(r))^2 = \nonumber \\
=cos(\frac{1}{2}ArcTan(r))^22cos(\frac{(E_2-E_1)t}{\hbar}+\rho)f_1(t)+\frac{1}{2}(E_2-E_1)sin(\frac{1}{2}ArcTan(r)) \nonumber \\
-f_1(t)[cos(\frac{(E_2-E_1)t}{\hbar}+\rho)]+ i f_1(t)[ sin(\frac{(E_2-E_1)t}{\hbar}+\rho)], \\
a_{1,2}=sin(\frac{1}{2}ArcTan(r))cos(\frac{1}{2}ArcTan(r))(E_2-E_1)+f(t)(cos(\frac{1}{2}ArcTan(r)))^2-(sin(\frac{1}{2}ArcTan(r)))^2f^{*}(t)= \nonumber   \\
=\frac{1}{2}(E_2-E_1)sin(\frac{1}{2}ArcTan(r))+cos(\frac{(E_2-E_1)t}{\hbar}+\rho)f_1(t)[(cos(\frac{1}{2}ArcTan(r)))^2-1]-if_1(t)sin(\frac{(E_2-E_1)t}{\hbar}+\rho), \\
%%$a_{1,2}=-w_{1,2}w_{1,1}E_1+f(t)w_{1,1}w_{1,1}-w_{1,2}w_{1,2}f^{*}(t)+w_{1,1}w_{1,2}E_2,$  \\
%$a_{2,2}=-w_{2,1}w_{1,2}E_1+w_{2,1}f(t)w_{1,1}-w_{2,2}w_{1,2}f(t)^{*}+w_{2,2}w_{1,1}E_2$ \\
a_{2,2}=sin(\frac{1}{2}ArcTan(r))^2 E_1-[f(t)+f(t)^{*}]cos(\frac{1}{2}ArcTan(r))sin(\frac{1}{2}ArcTan(r))+E_2 cos(\frac{1}{2}ArcTan(r))^2= \nonumber \\
=sin(\frac{1}{2}ArcTan(r))^2 E_1-cos(\frac{(E_2-E_1)t}{\hbar}+\rho)f_1(t)sin(ArcTan(r))+E_2 cos(\frac{1}{2}ArcTan(r))^2
\end{eqnarray}
that gives
$f(t)+f^{*}(t)=2cos(\frac{(E_2-E_1)t}{\hbar}+\rho)f_1(t)$. We can always use
\begin{eqnarray}
a(t)f(t)-b(t)f(t)^{*}=a(t)f(t)-[-a+(b+a)]f(t)^{*}=a(t)[f(t)+f(t)^{*}]-(b+a)f(t)^{*}.
\end{eqnarray}
The obtained equations \ref{generaleqns} for tight-binding model coefficients can be generalized by using $(\gamma_2(t)-\gamma_1(t))+ \xi(t)$ obtained from \ref{gengenQ1}, \ref{gengenQ2}, \ref{gengenQ3} instead of used expression $\frac{(E_2-E_1)(t-t_0)}{\hbar}-(\gamma_2(t_0)-\gamma_1(t_0))+\frac{\pi}{2}$.
We can incorporate the dissipation to tight-binding model accompanied with Rabi oscillation and we obtain coeffcients
\begin{eqnarray} \label{generaleqnsD}
a_{(1,1)D}=cos(\frac{1}{2}ArcTan(r))^2 (E_{1r}+iE_{1i})+(f(t)+f(t)^{*})(sin(\frac{1}{2}ArcTan(r))cos(\frac{1}{2}ArcTan(r)))+(E_{2r}+iE_{2i})sin(\frac{1}{2}ArcTan(r))^2 \nonumber \\
=cos(\frac{1}{2}ArcTan(r))^2 (E_{1r}+iE_{1i})+cos(\frac{(E_{2r}+iE_{2i}-E_{1r}-iE_{1i})t}{\hbar}+\rho)f_1(t)sin(ArcTan(r)+(E_{2r}+iE_{2i})sin(\frac{1}{2}ArcTan(r))^2, \nonumber  \\ \\
a_{(2,1)D}=cos(\frac{1}{2}ArcTan(r))^2f(t)^{*}+(E_{2r}-E_{1r}+i(E_{2i}-E_{1i}))(sin(\frac{1}{2}ArcTan(r))cos(\frac{1}{2}ArcTan(r)))-f(t)sin(\frac{1}{2}ArcTan(r))^2 = \nonumber \\
=cos(\frac{1}{2}ArcTan(r))^22cos(\frac{(E_{2r}-E_{1r})t}{\hbar}+i\frac{(E_{2i}-E_{1i})t}{\hbar}+\rho)f_1(t)+\frac{1}{2}(E_{2r}-E_{1r})sin(\frac{1}{2}ArcTan(r))+ \nonumber \\
+i\frac{1}{2}(E_{2i}-E_{1i})sin(\frac{1}{2}ArcTan(r)) %\nonumber \\
-f_1(t)[cos(\frac{(E_{2r}-E_{1r})t}{\hbar}+i\frac{(E_{2i}-E_{1i})t}{\hbar}+\rho)]+ i f_1(t)[ sin(\frac{(E_{2r}-E_{1r})t}{\hbar}+i\frac{(E_{2i}-E_{1i})t}{\hbar}+\rho)], \\
a_{(1,2)D}=sin(\frac{1}{2}ArcTan(r))cos(\frac{1}{2}ArcTan(r))(E_{2r}-E_{1r})+i sin(\frac{1}{2}ArcTan(r))cos(\frac{1}{2}ArcTan(r))(E_{2i}-E_{1i})+ \nonumber \\
+f(t)(cos(\frac{1}{2}ArcTan(r)))^2-(sin(\frac{1}{2}ArcTan(r)))^2f^{*}(t)= \nonumber   \\
=\frac{1}{2}((E_{2r}-E_{1r})+i(E_{2i}-E_{1i}))sin(\frac{1}{2}ArcTan(r))+cos(\frac{(E_{2r}-E_{1r})t}{\hbar}+i\frac{(E_{2i}-E_{1i})t}{\hbar}+\rho)f_1(t)[(cos(\frac{1}{2}ArcTan(r)))^2-1] \nonumber \\
-if_1(t)sin(\frac{(E_{2r}-E_{1r})t}{\hbar}+i\frac{(E_{2i}-E_{1i})t}{\hbar}+\rho), %\\
%%$a_{1,2}=-w_{1,2}w_{1,1}E_1+f(t)w_{1,1}w_{1,1}-w_{1,2}w_{1,2}f^{*}(t)+w_{1,1}w_{1,2}E_2,$  \\
%$a_{2,2}=-w_{2,1}w_{1,2}E_1+w_{2,1}f(t)w_{1,1}-w_{2,2}w_{1,2}f(t)^{*}+w_{2,2}w_{1,1}E_2$ \\
\end{eqnarray}
\begin{eqnarray}
a_{(2,2)D}=sin(\frac{1}{2}ArcTan(r))^2 (E_{1r}+iE_{1i})-[f(t)+f(t)^{*}]cos(\frac{1}{2}ArcTan(r))sin(\frac{1}{2}ArcTan(r))+(E_{2r}+iE_{2i}) cos(\frac{1}{2}ArcTan(r))^2= \nonumber \\
sin(\frac{1}{2}ArcTan(r))^2 (E_{1r}+iE_{1i})-cos(\frac{(E_{2r}-E_{1r})t}{\hbar}+i\frac{(E_{2i}-E_{1i})t}{\hbar}+\rho)f_1(t)sin(ArcTan(r))+(E_{2r}+iE_{2i}) cos(\frac{1}{2}ArcTan(r))^2 , \nonumber \\
\end{eqnarray}
that imply non-Hermicity of dissipative tight-binding model with Rabi oscillations.
\section{Open curvy loops confining single electron in Cartesian coordinates in Schr\"{o}dinger formalism}
\subsection{Case of deformed curvy Wannier qubit }
\begin{figure}
    \centering
    \includegraphics[scale=0.8]{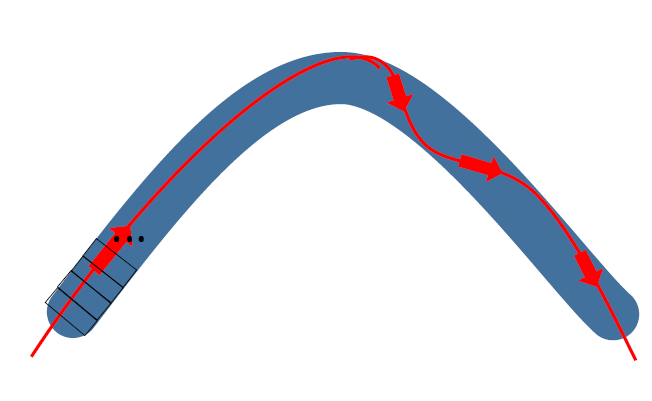}
    \caption{Schematic movement of wave-packet across cuvred nanowire that can be simplified as quasi-one dimensional object after proper transformation from 3D or 2D to 1D (Dimension) in visualization by Marcin Piontek. }
    \label{fiber}
 %   \end{figure}
 %   \begin{figure} %\label(centralfig)
  %  \centering
    \includegraphics[scale=0.7]{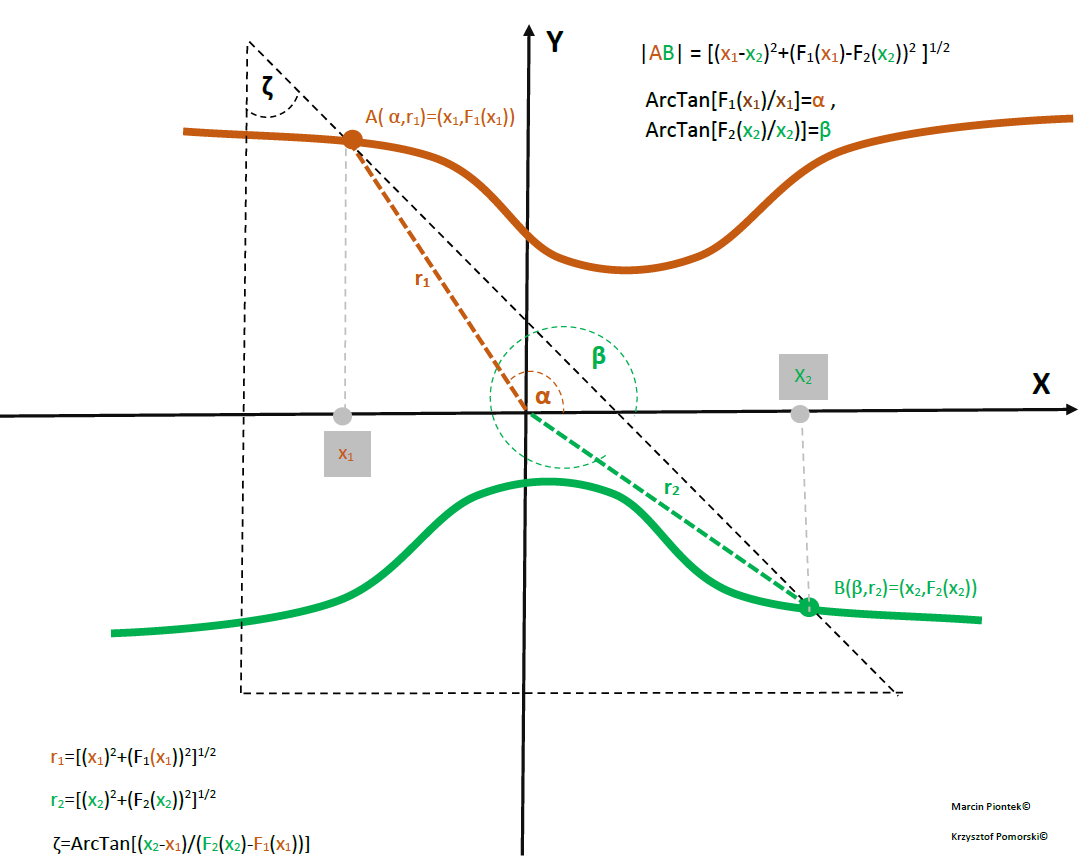}
    \caption{Generalized Q-Swap gate (Q-Inverter) for the case of 2 bended nanowires.}
    \label{2fibers}
   \end{figure}

We go beyond approach describing two straight interacting single-electron lines \cite{Szafran}, \cite{Nbodies}, \cite{qgates}.
We consider the set of open curvy quasi-one dimensional loops (that can be straight or curved smooth semiconductor nanowires with single electron) described by $x(s)$,$y(s)$ and $z(s)$, where s is the distance from beginning to the end of loop. We have Schr\"{o}dinger equation describing wave-packet movement in curvy nanowire
\begin{eqnarray} \label{eqnondiss1}
[-\frac{\hbar^2}{2m}(\frac{d^2}{dx^2}+\frac{d^2}{dy^2}+\frac{d^2}{dz^2})+V(x,y,z)] \psi(x,y,z)=E \psi(x,y,z).
 %%  &=&  \\
 %%  &=&  \\
 %%  &=& 
\end{eqnarray}
In such case the wave-packet moves in the way depicted Fig.\ref{fiber}. Moving along curvy cable trajectory

brings relation $\frac{d}{dx}=\frac{ds}{dx}\frac{d}{ds}$ and similarly with y and z.
We thus have $\frac{d^2}{dx^2}=\frac{ds}{dx}\frac{d}{ds}(\frac{ds}{dx}\frac{d}{ds})=(\frac{1}{x'(s)})^2\frac{d^2}{ds^2}-[\frac{x''(s)}{(x'(s))^3}]\frac{d}{ds}$.
Therefore we end up in equation in the form as
\begin{eqnarray}
\Bigg[-\frac{\hbar^2}{2m}\Bigg[([(\frac{1}{x'(s)})^2+(\frac{1}{y'(s)})^2+(\frac{1}{z'(s)})^2]\frac{d^2}{ds^2}-[\frac{x''(s)}{(x'(s))^3}+\frac{y''(s)}{(y'(s))^3}+\frac{z''(s)}{(z'(s))^3}]\frac{d}{ds})\Bigg]+V(s)\Bigg] \psi(s)=E \psi(s)
\end{eqnarray}
and it can be summarized as
\begin{eqnarray}
\Bigg[-\frac{\hbar^2}{2m}\Bigg[([(\frac{1}{x'(s)})^2+(\frac{1}{y'(s)})^2+(\frac{1}{z'(s)})^2]\frac{d^2}{ds^2}-[\frac{x''(s)}{(x'(s))^3}+\frac{y''(s)}{(y'(s))^3}+\frac{z''(s)}{(z'(s))^3}]\frac{d}{ds})\Bigg]+V(s)\Bigg] \psi(s)=E \psi(s),
\end{eqnarray}
so we have
\begin{eqnarray} \label{eqndiss1}
\Bigg[-\frac{\hbar^2}{2m}\Bigg[(f(s)\frac{d^2}{ds^2}-g(s)\frac{d}{ds})\Bigg]+V(s)\Bigg] \psi(s)=E \psi(s),
\end{eqnarray}
where $f(s)=[(\frac{1}{x'(s)})^2+(\frac{1}{y'(s)})^2+(\frac{1}{z'(s)})^2]$, $g(s)=[\frac{x''(s)}{(x'(s))^3}+\frac{y''(s)}{(y'(s))^3}+\frac{z''(s)}{(z'(s))^3}]$.
Similar reasoning with transformation from one coordinate system into another coordinate system is conducted by \cite{CurvedQM1}.
More detailed derivation of equation of motion for open loop curved nanocable in cylindrical, spherical coordines is given by \cite{Extended}.
The most prominent feature that can be observed from equation transformation from \ref{eqnondiss1} to \ref{eqndiss1} is the occurrence of dissipation term that is proportional to operator $\frac{d}{ds}$ as analogical to friction force (that is usually proportional to particle momentum). Indeed wave-packet travelling in semiconductor nanowire is being bent that corresponds to occurrence of force changing the
direction of wave-packet momentum. However from another perspective we can say that by bending straight trajectory of particle (flat space) we can generate dissipation. Thinking reversely we can say that fact of having certain type of dissipation in given system in given coordinates we can change by moving to space with another curvature so dissipation
is reduced or cancelled. The consequence of having dissipation in quantum system (or more precisely dissipation-like) will imply the fact of non-Hermicity of Hamiltonian matrix that shall imply the existence of complex value of eigenergies. What is interesting we can observe dissipation like term in the system of classical description of electron moving in nanowire what is given by equation \ref{cleqnodiss1} and \ref{CoupledODEs}. %%\ref{cleqdiss1}.

\begin{figure}
\centering
\includegraphics[scale=0.3]{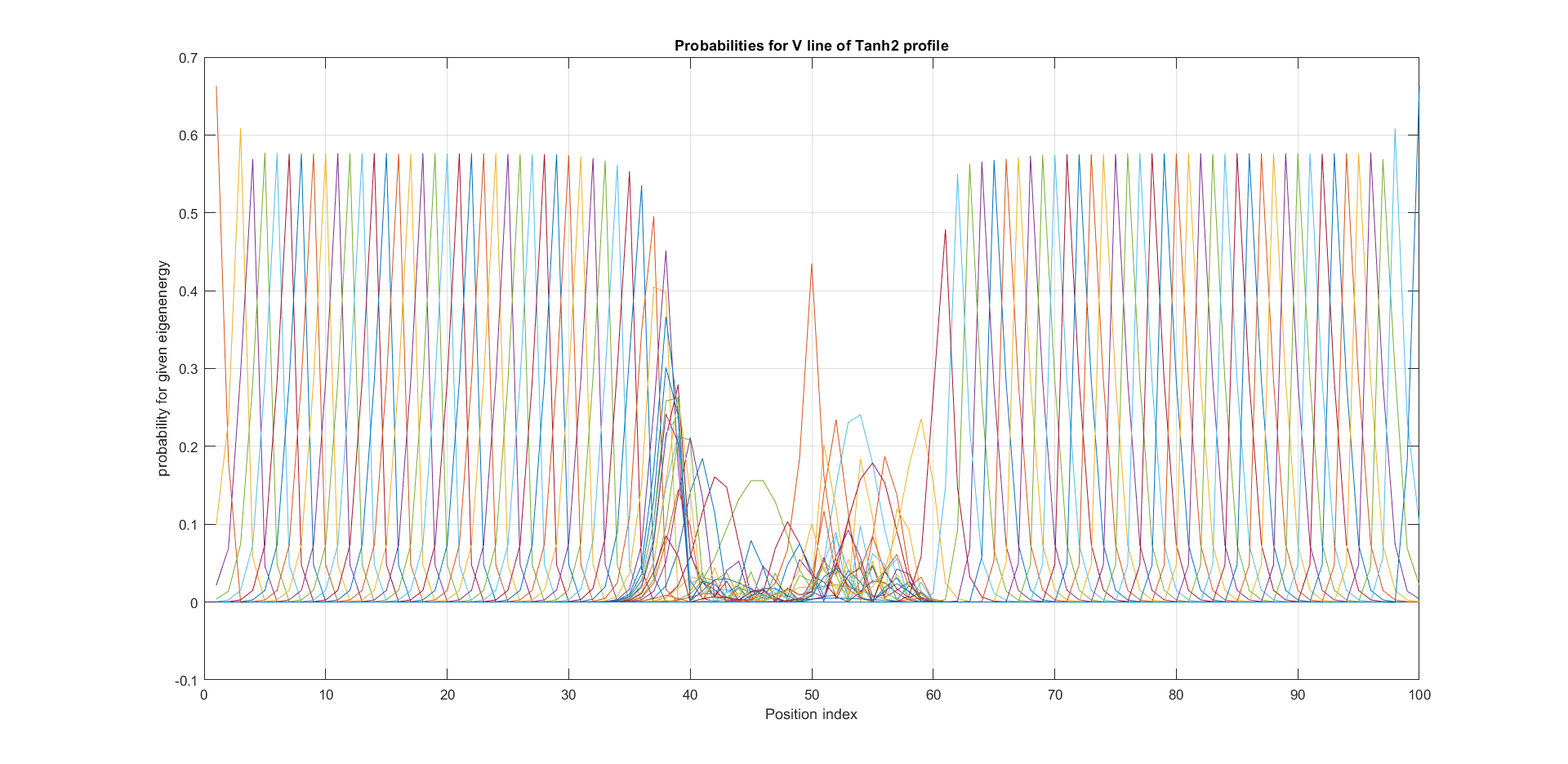} %{TwoWellsTanh.png}
\includegraphics[scale=0.3]{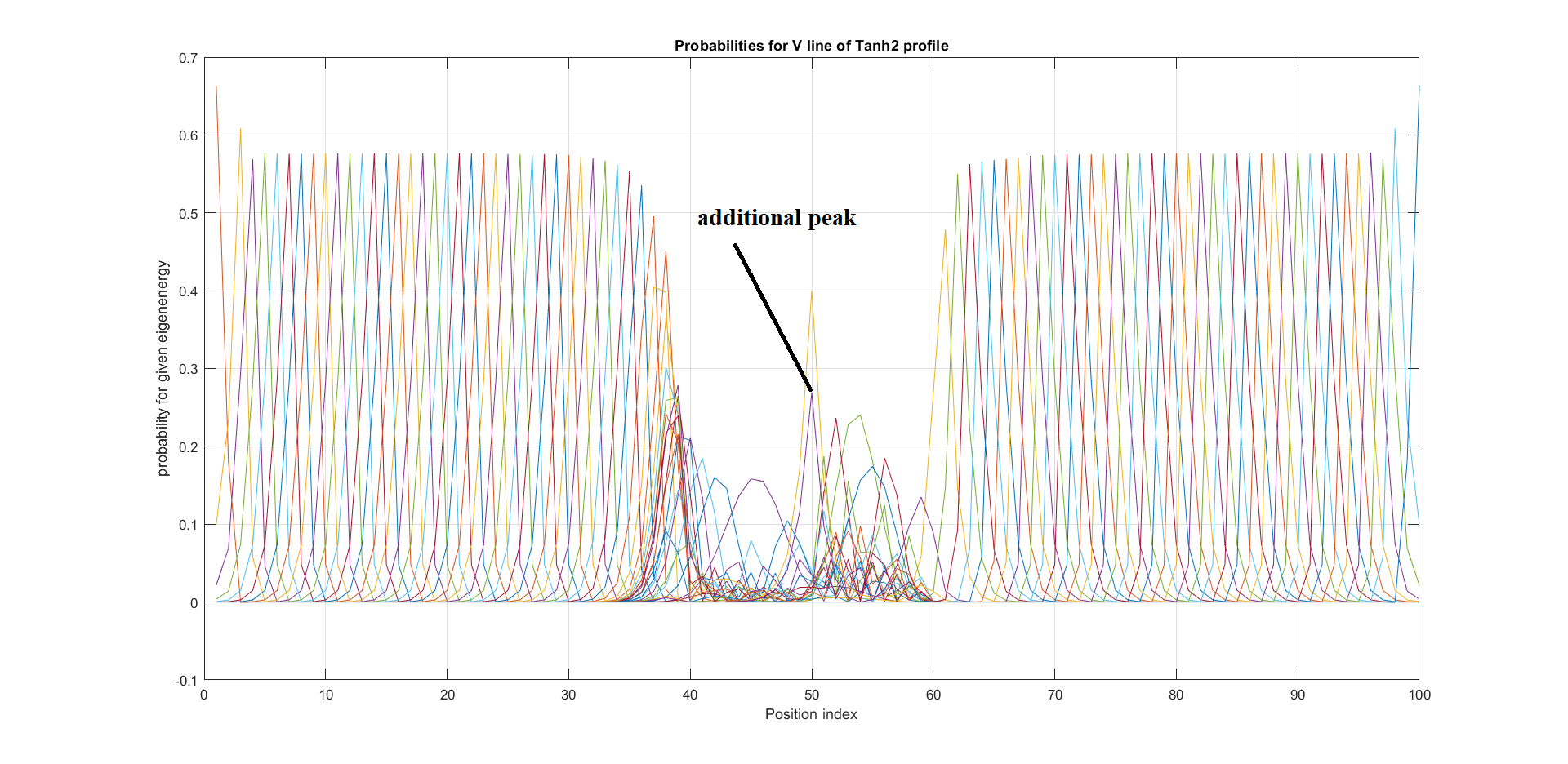} %{Probabilities50eigenenergiesNowellsTanh.png}
\includegraphics[scale=0.3]{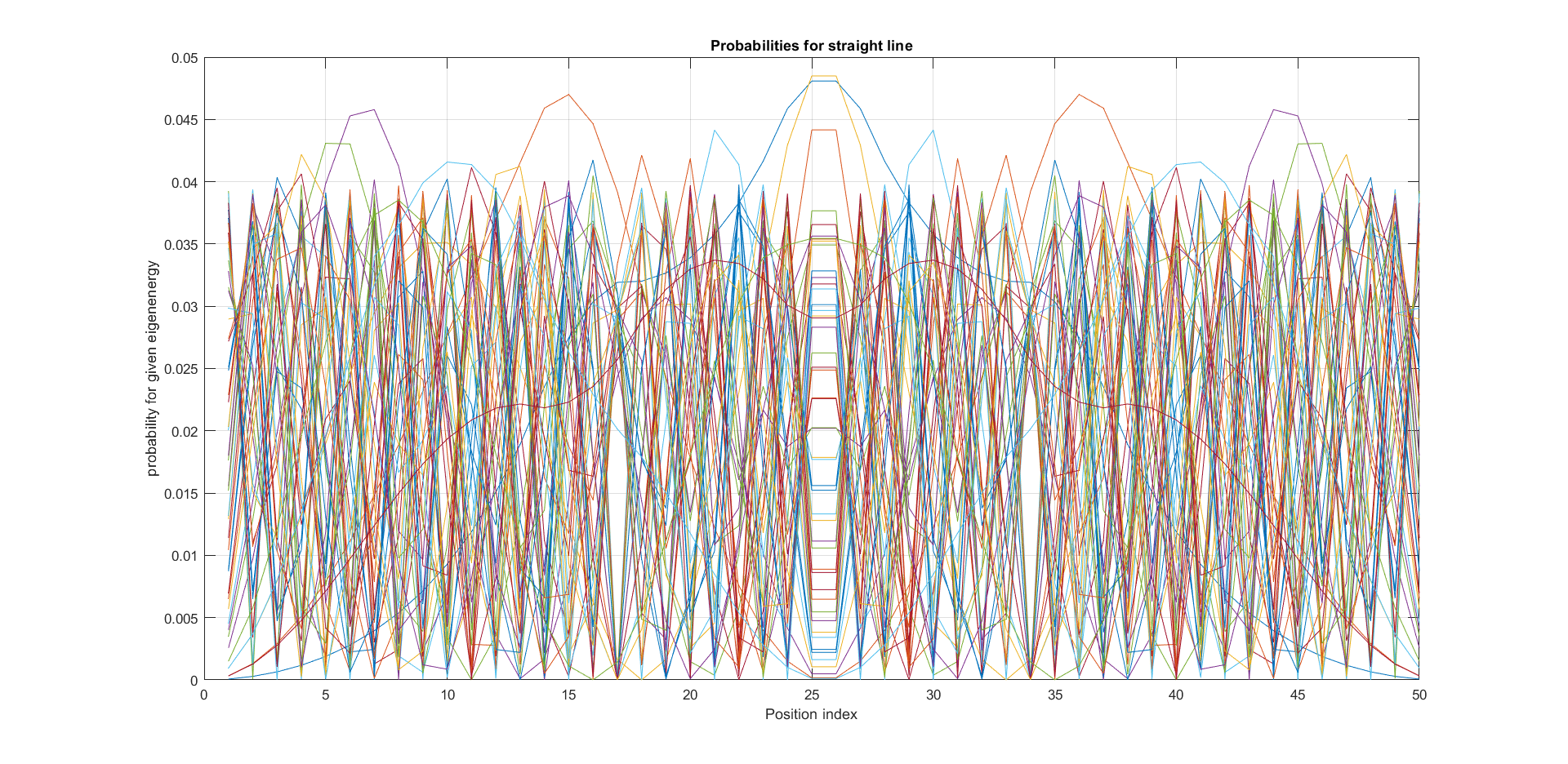}
\caption{Probability distributions corresponding to eigenenergy wavefunctions for tanh square nanowire ($\alpha=10$) with 3 built-in q-wells (UPPER), no q-wells (MIDDLE) and straight nanowire with 2 built-in q-wells (LOWER). }
\label{Alpha10QbitCurve}
\end{figure}

\begin{figure}
\includegraphics[scale=0.16]{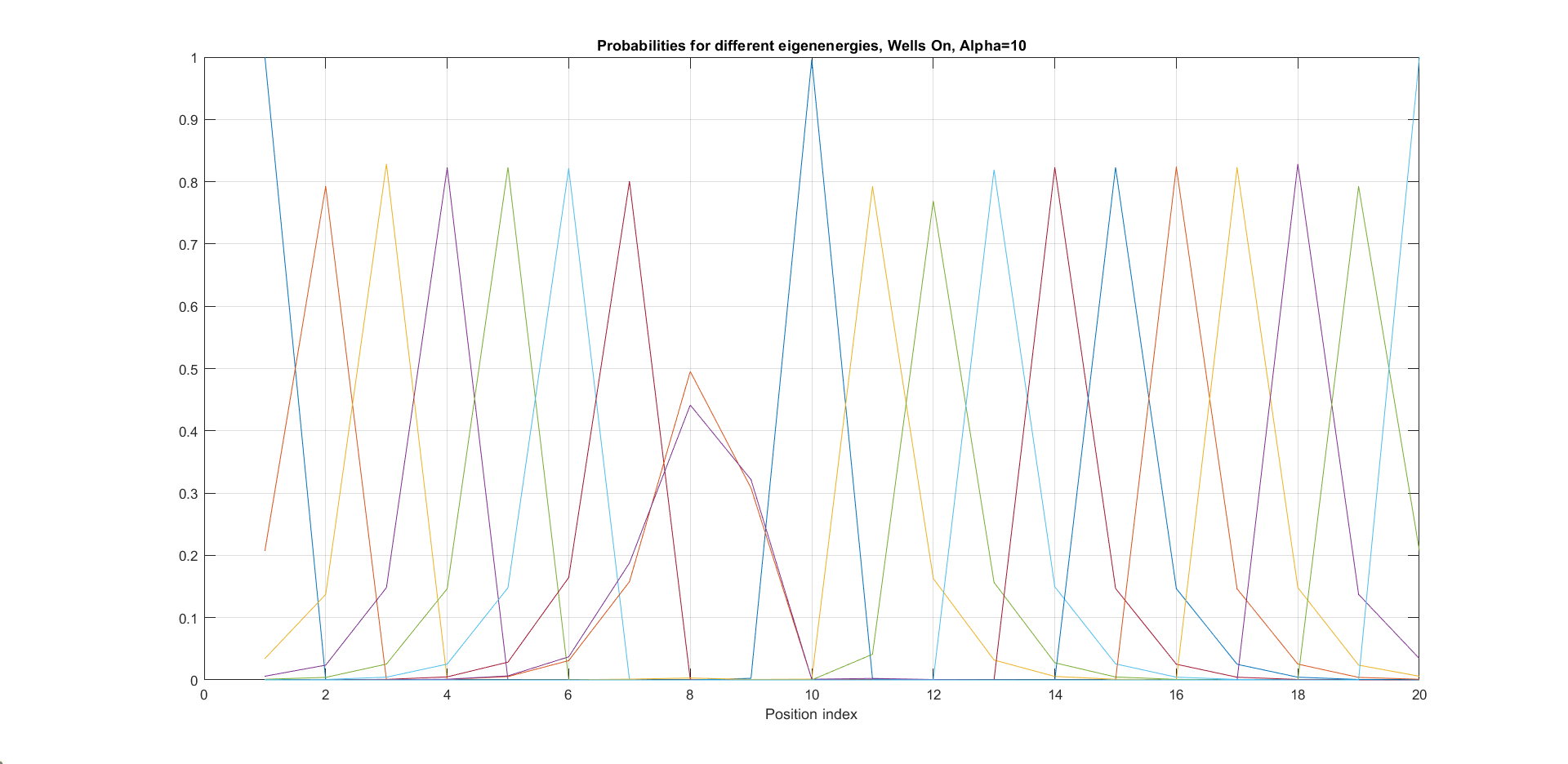}
\includegraphics[scale=0.16]{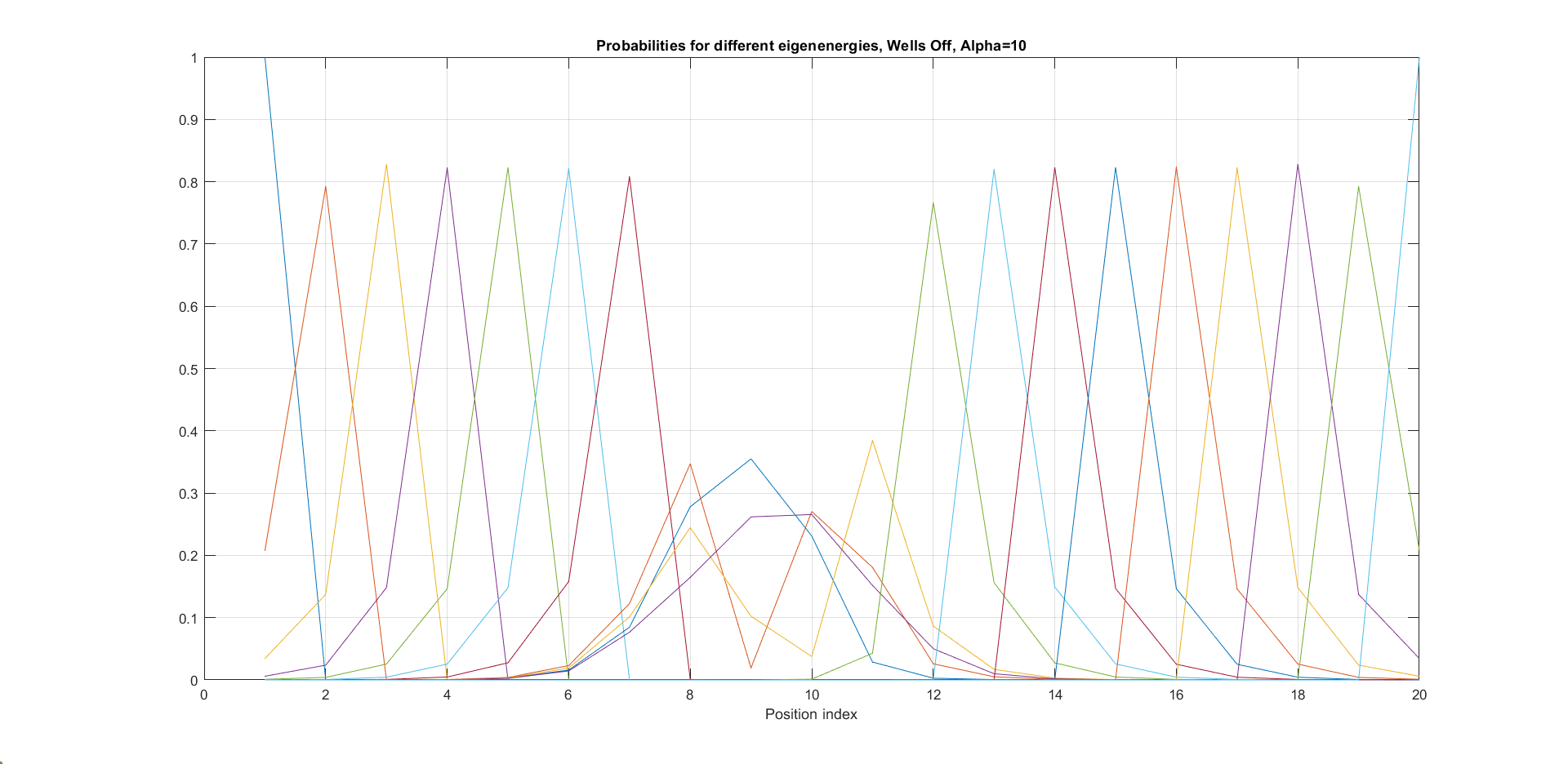}
\caption{Detailed analysis of probability distributions (for first 20 eigenenergy modes) for Tanh square nanowires in case of presence[LEFT]/no presence [RIGHT] of 3 q-wells with $\alpha=10$. Lack of built-in q-wells result in the gap of q-state presence in the middle of curved nanowire.}
\label{special1}
 % Analysis shows that eigenenergy wavefunctions are strongly localized due to the fact that nanowire has non-zero curvature (equivalent to condition that $\frac{\frac{d^2}{dx^2}}y(x){(\frac{d}{dx}y(x))^2} \neq 0$). }
%%%\end{figure}
%%%\begin{figure}
%%\includegraphics[scale=0.35]{EigenEnergy10noWellsPart1QQ.png}
%%\includegraphics[scale=0.35]{EigenEnergy10noWellsPart2QQ.png}
\includegraphics[scale=0.16]{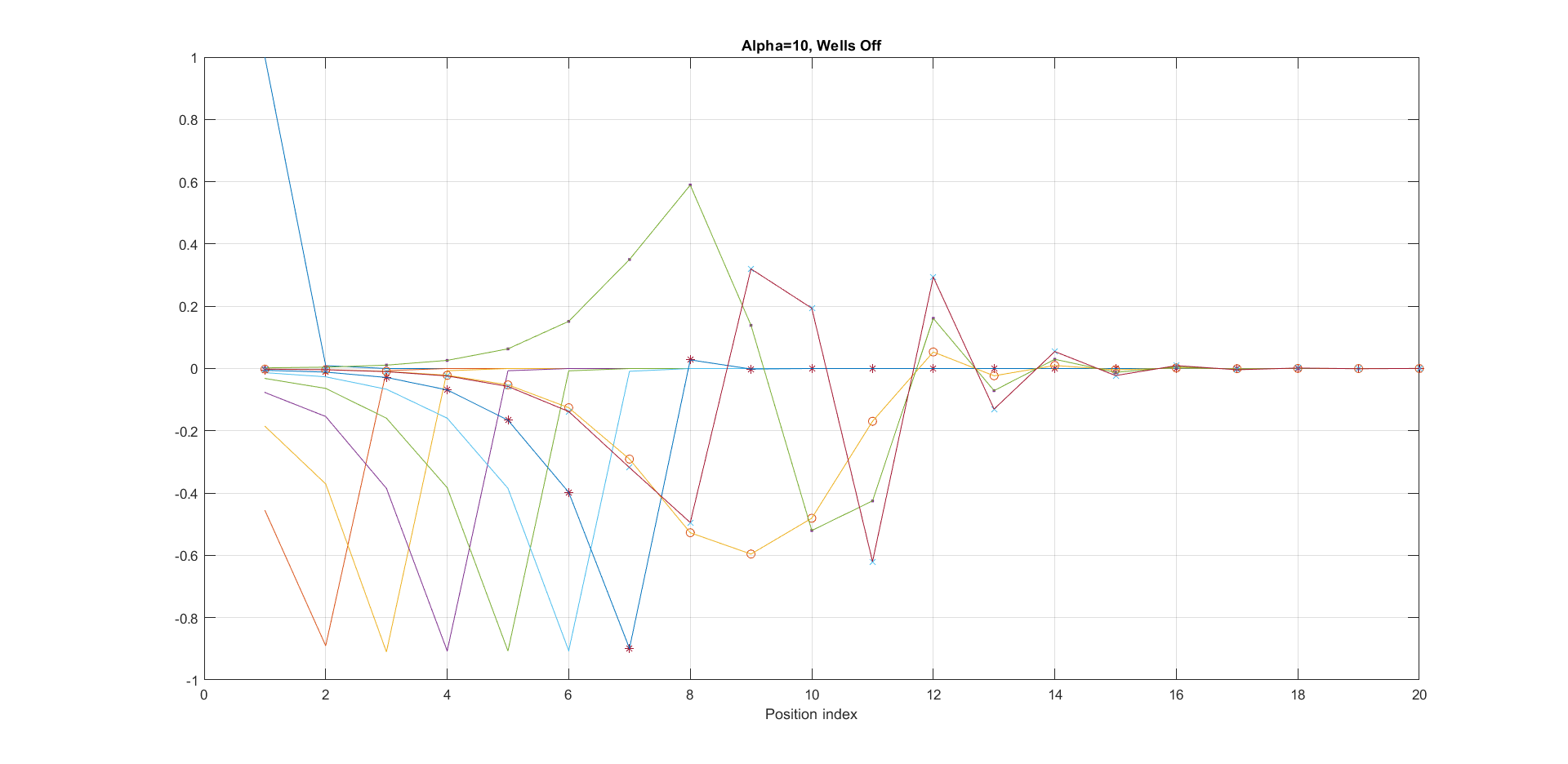}
\includegraphics[scale=0.16]{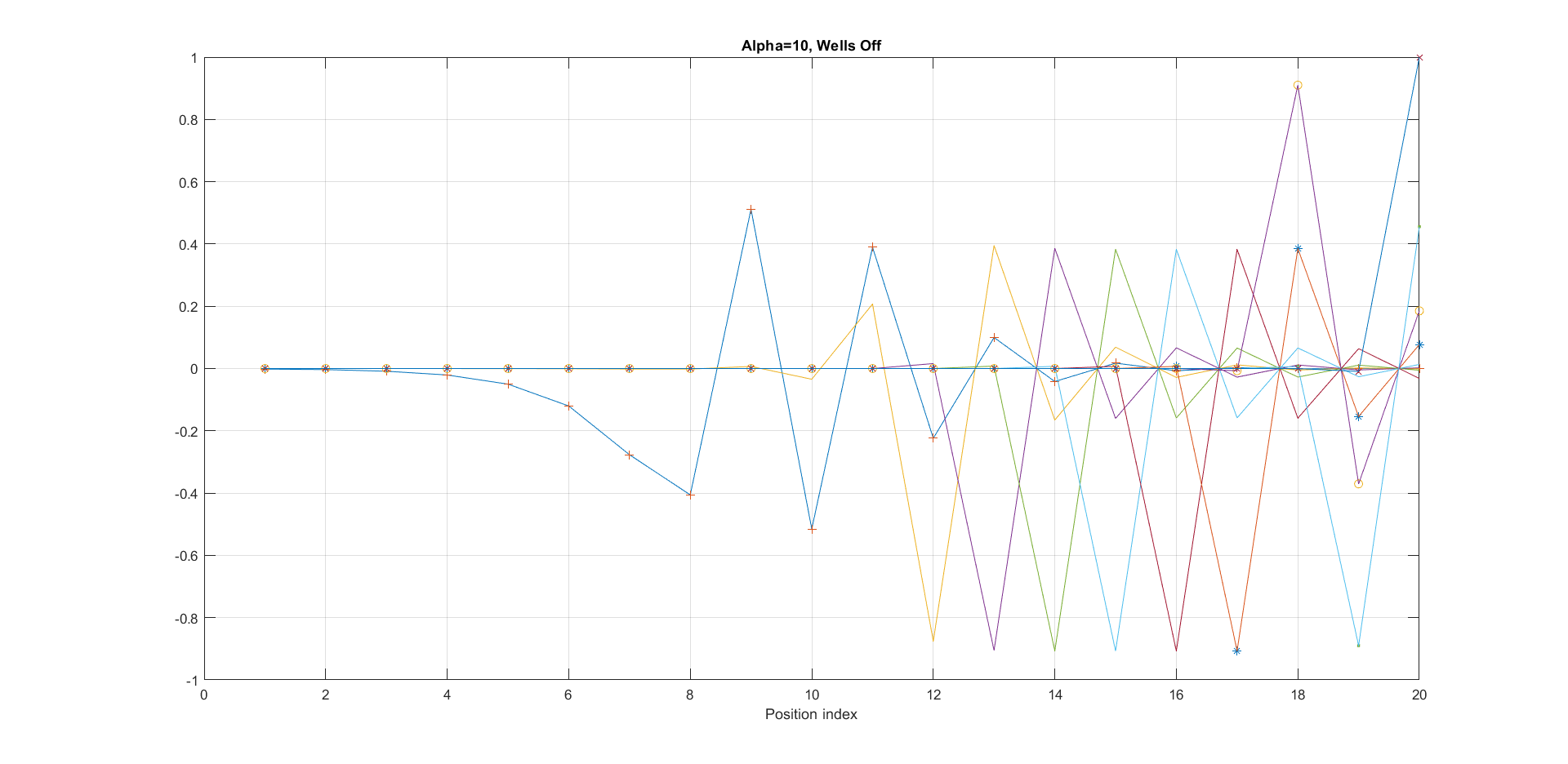}
\caption{Detailed analysis of eigen-energy wavefunctions (for first 20 eigen-energy modes) and with no-built in q-wells for Tanh square nanowires (with $\alpha=10$) shows that wave-functions are strongly localized due to the fact that nanowire has non-zero curvature (equivalent to condition that $\frac{\frac{d^2}{dx^2}}y(x){(\frac{d}{dx}y(x))^2} \neq 0$). }
\label{pq1}
%%\end{figure}
%%\begin{figure}
%%\includegraphics[scale=0.35]{EigenEnergy10noWellsPart1QQ.png}
%%\includegraphics[scale=0.35]{EigenEnergy10noWellsPart2QQ.png}
\includegraphics[scale=0.16]{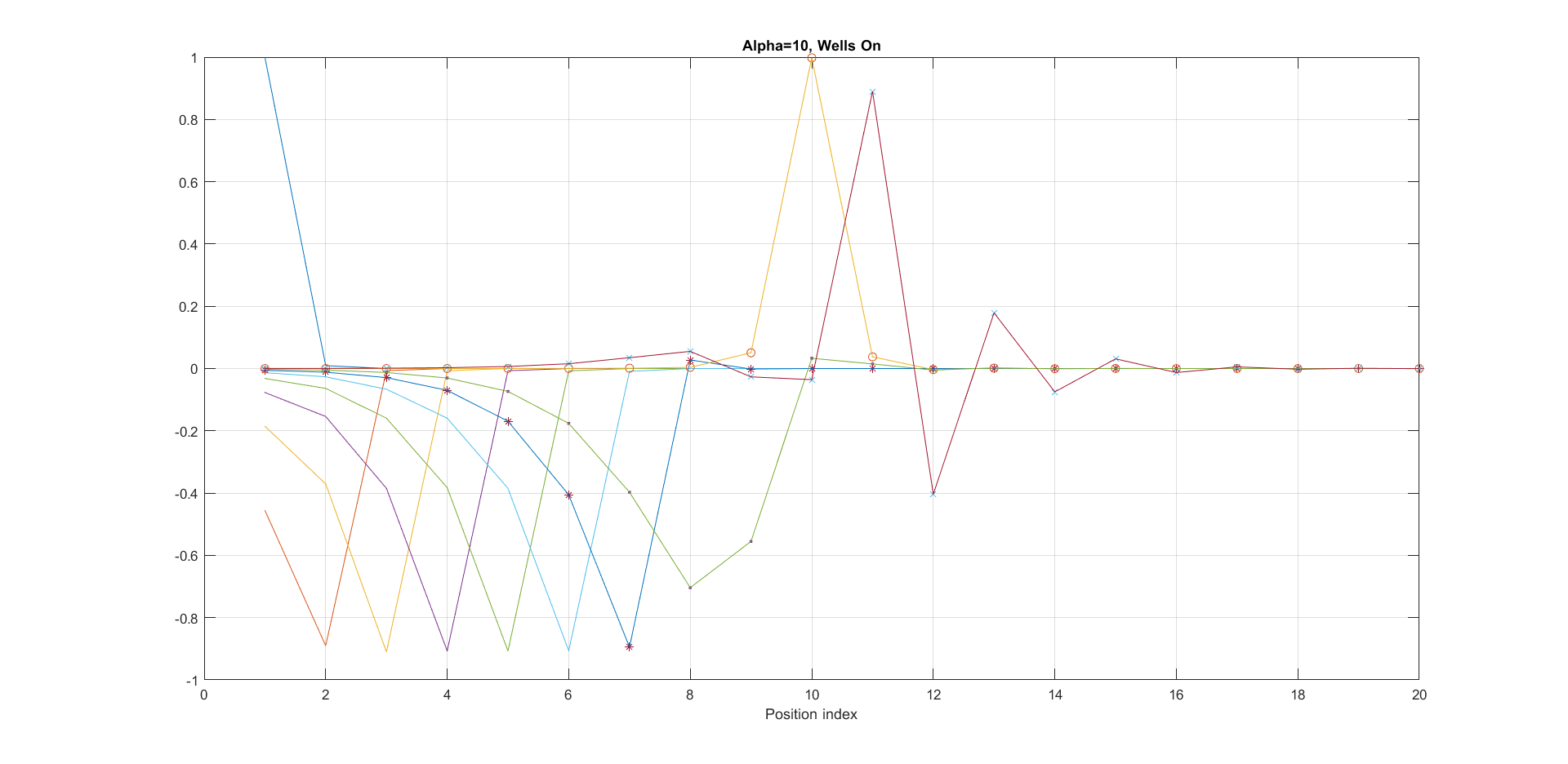}
\includegraphics[scale=0.16]{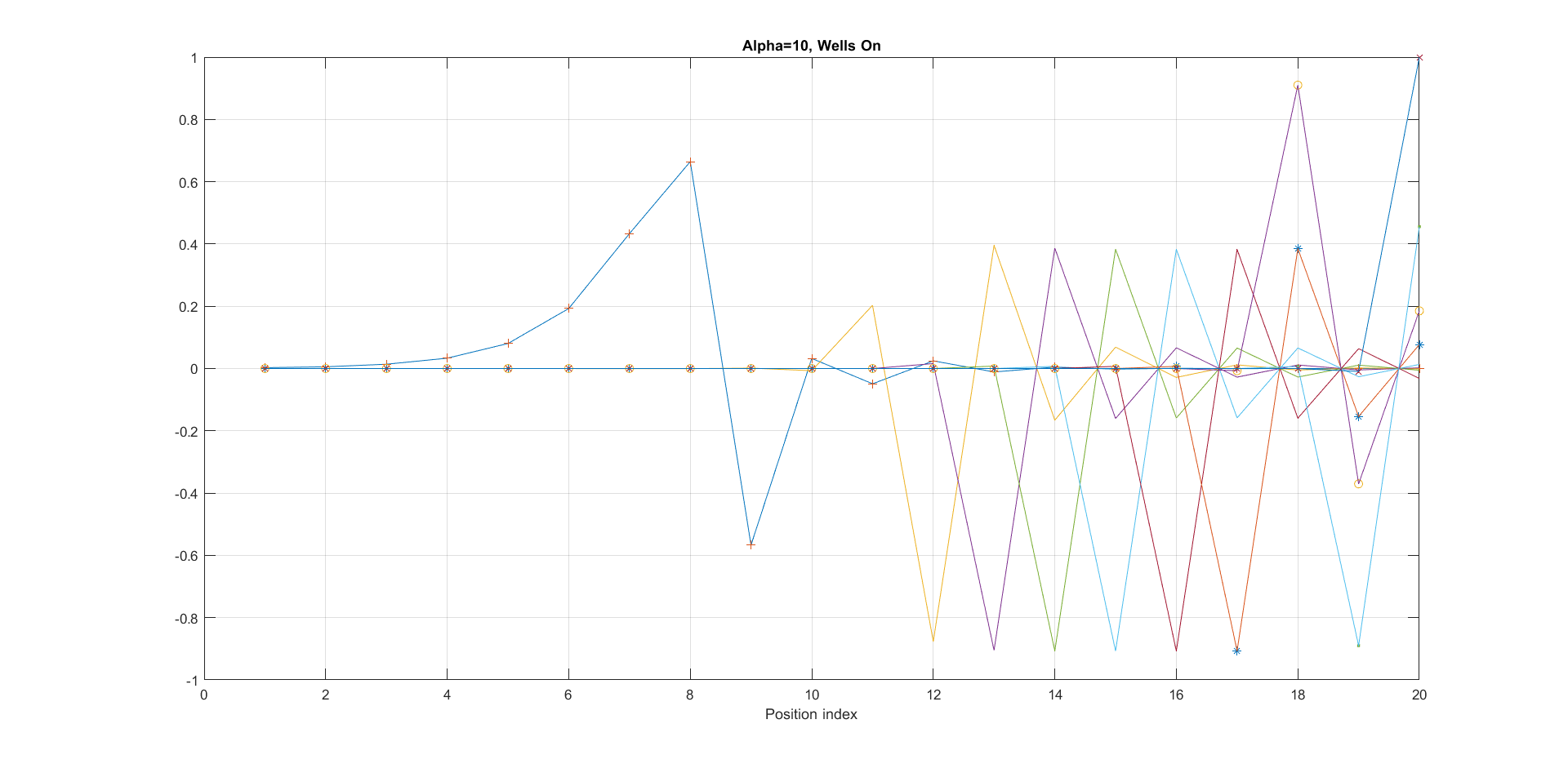}
\caption{Detailed analysis of eigenenergy wavefunctions (for first 20 eigenenergy modes) and with 3 built in q-wells for Tanh square nanowires (with $\alpha=10$) shows that wave-functions are strongly localized due to the fact that nanowire has non-zero curvature (equivalent to condition that $\frac{\frac{d^2}{dx^2}}y(x){(\frac{d}{dx}y(x))^2} \neq 0$). One can recognize the difference in quantum wave function behaviours when compared with the case of no-built in q-wells as given by Fig.\ref{pq1}.  }
\label{pq2}
\end{figure}

\begin{figure}
\centering
\includegraphics[scale=0.16]{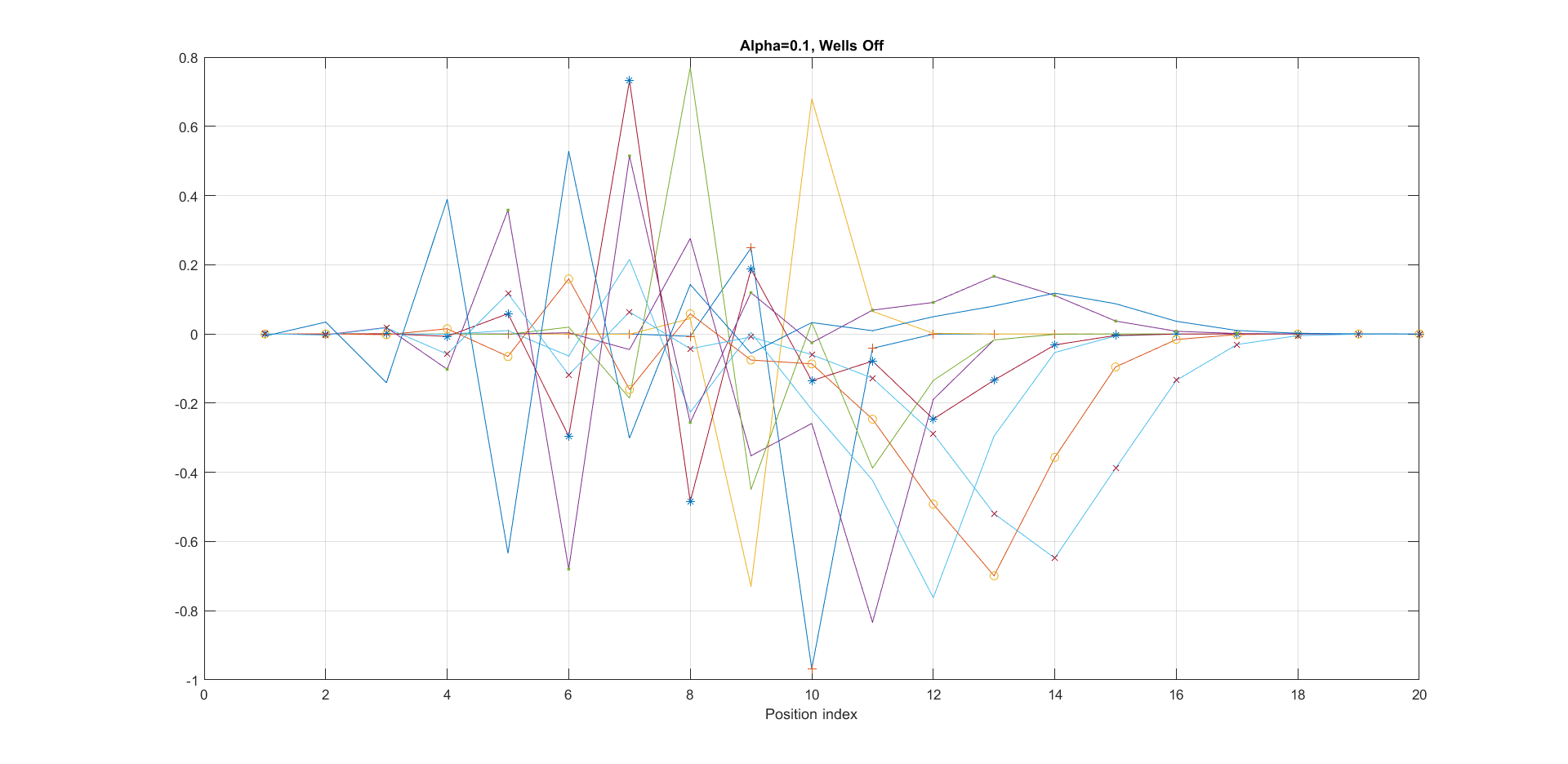}
\includegraphics[scale=0.16]{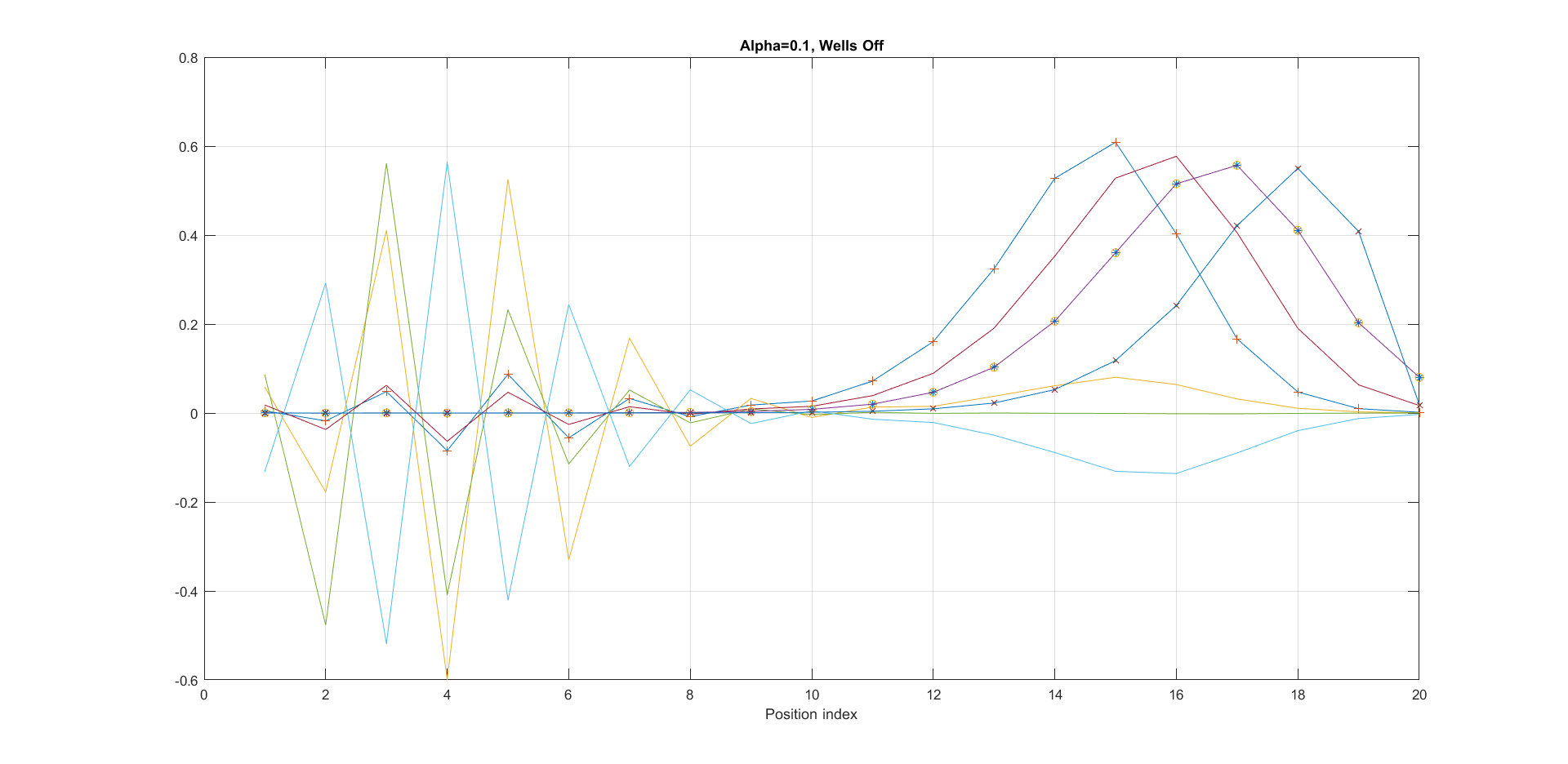}
\caption{Case of coefficient $\alpha=0.1$ reveals interesting eigenenergy wavefunction distitributions for cable with no built-in q-well. }
\label{pq4}
%%%\end{figure}
%%%\begin{figure}
\centering
\includegraphics[scale=0.16]{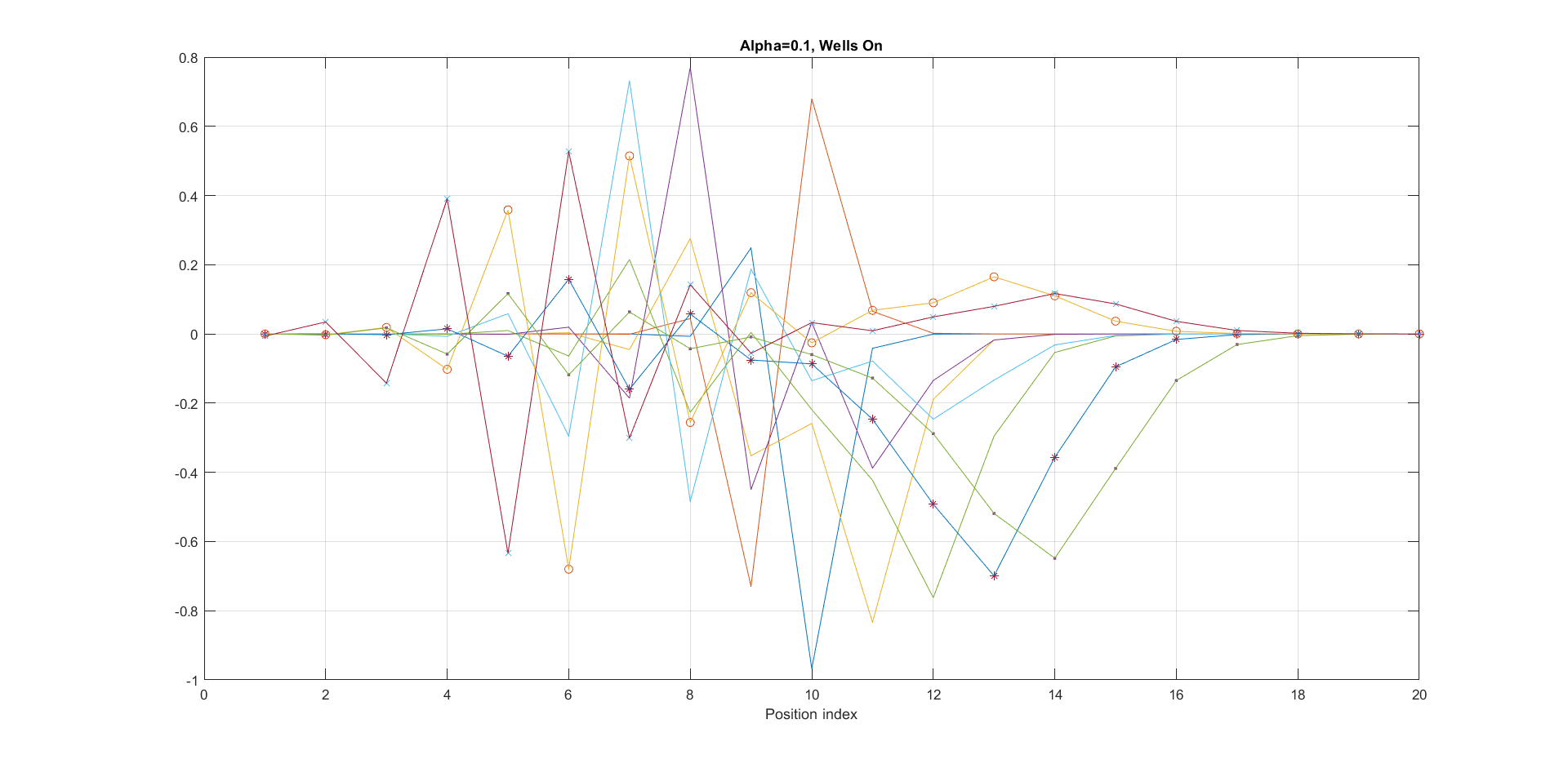} %{First10Alpha0p1_3WellsOn.png}
\includegraphics[scale=0.16]{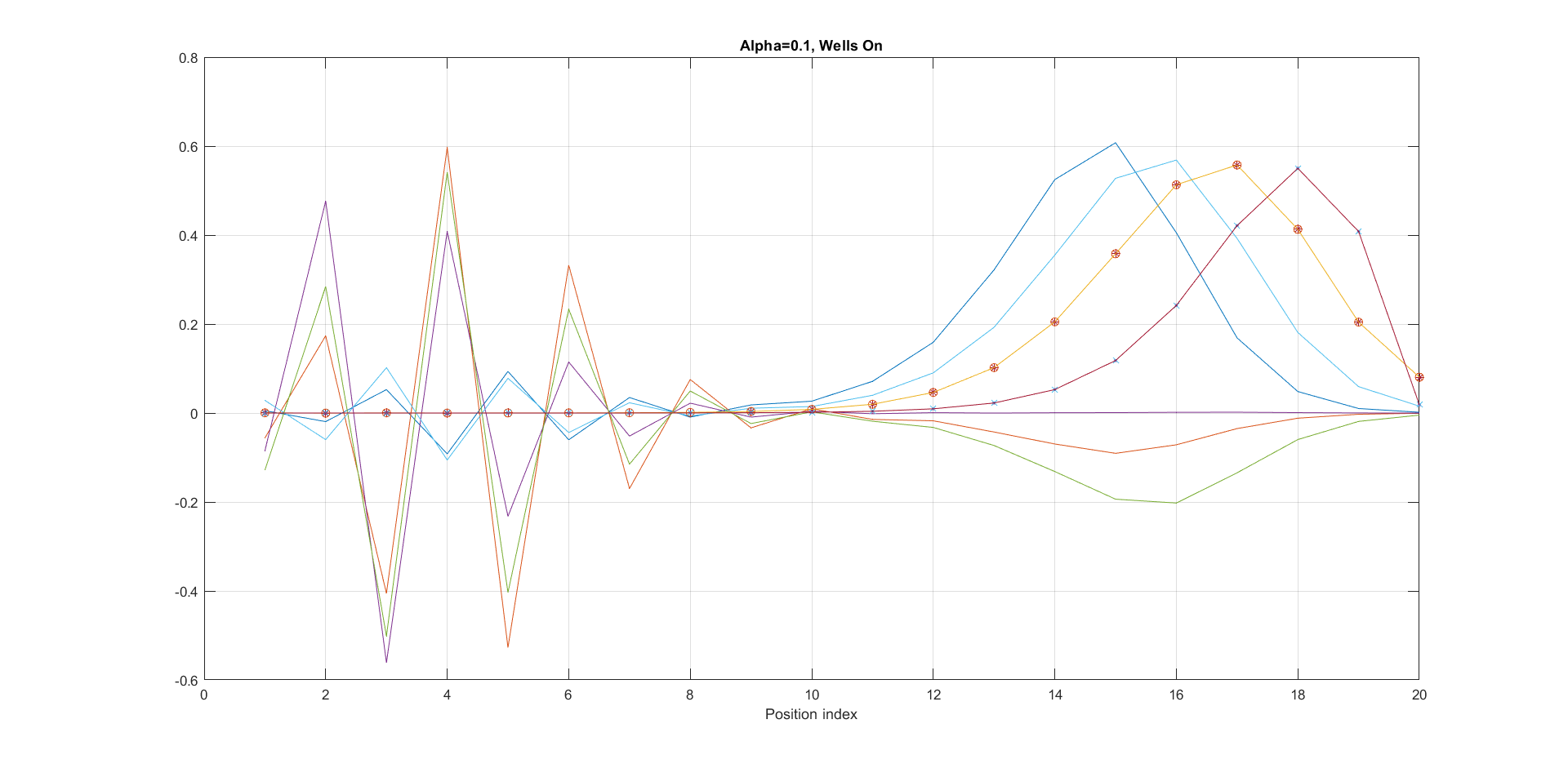} %{Last10Alpha0p1_3WellsOn.png}
\caption{Case of coefficient $\alpha=0.1$ reveals interesting eigenenergy wavefunction distitributions for nanowire cable with 3 built-in q-wells. }
\label{pq5}
%%\end{figure}
%%\begin{figure}
%%\centering
\includegraphics[scale=0.15]{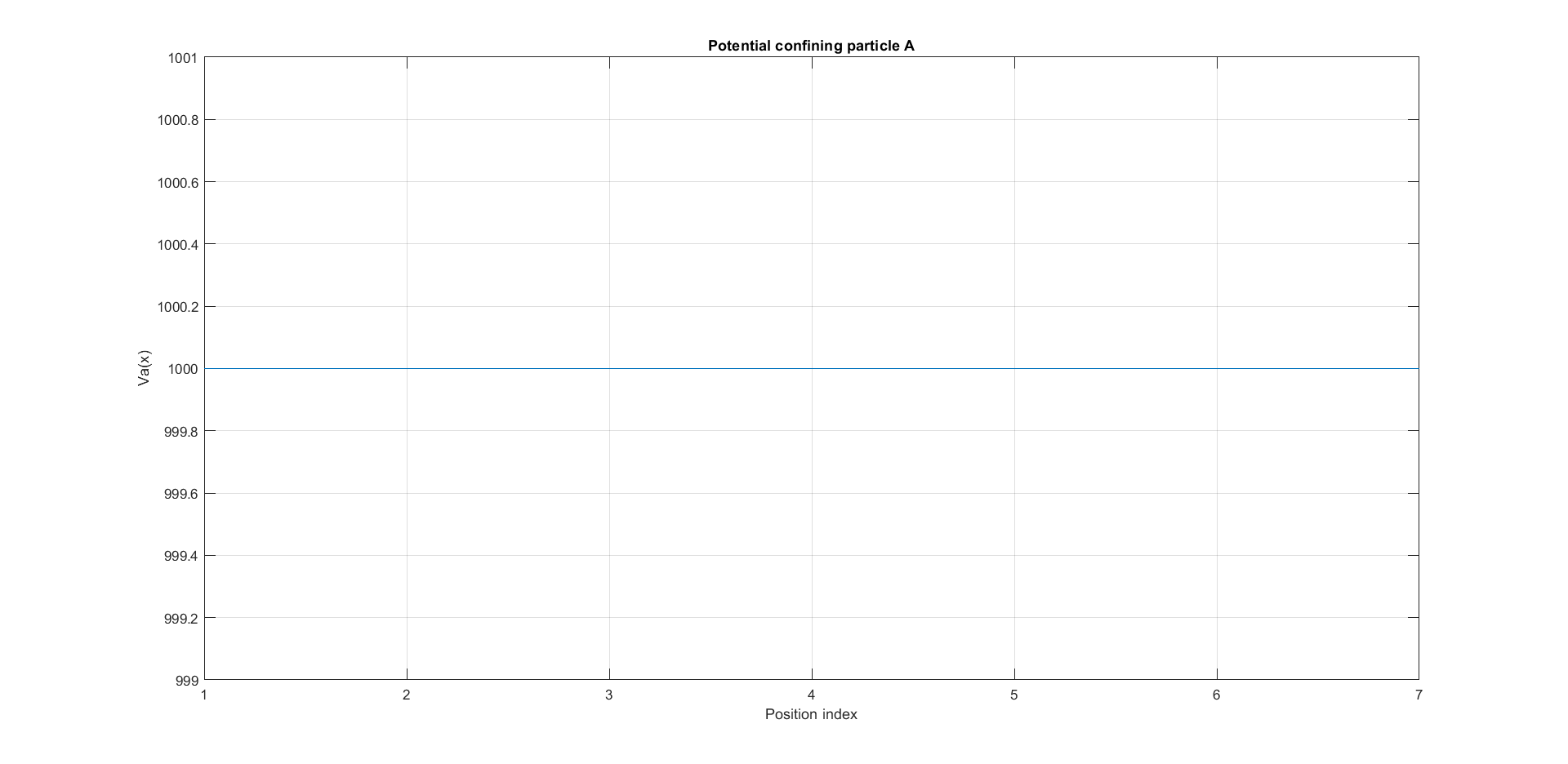} %{First10Alpha0p1_3WellsOn.png}
\includegraphics[scale=0.15]{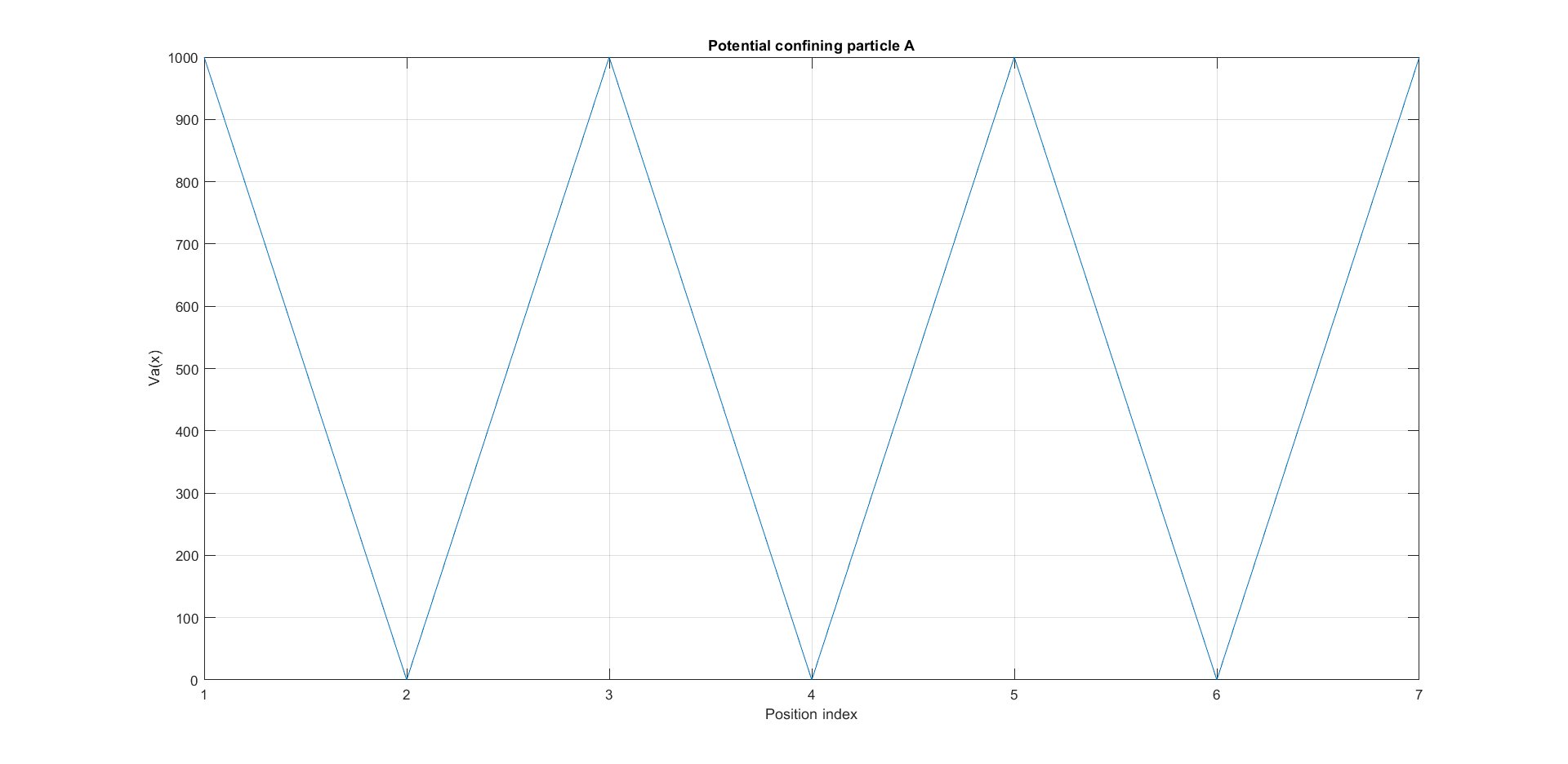} %{Last10Alpha0p1_3WellsOn.png}
\caption{Case of local confining potential $V_a(V_b)$ for Tanh Square interacting nanowire cables with no and built-in q-wells. }
\label{pq6}
\end{figure}

Using Cartesian coordinates we can formulate the following Schr\"{o}dinger equation of motion
\begin{eqnarray}
-\frac{\hbar^2}{2m}[(1+\frac{1}{(\frac{d}{dx}y(x)^2)})\frac{d^2}{dx^2}-\frac{\frac{d^2}{dx^2}y(x)}{(\frac{d}{dx}y(x))^2}\frac{d}{dx}]\psi(x,y(x))+V(x,y(x))\psi(x,y(x))=E \psi(x,y(x)).
\end{eqnarray}
Local confining potential is given by $V(x,y(x))$ and can simply take into account the existence of 1, 2, 3 and more quantum dots across semiconductor nanowire or can
omit the existence of quantum dots and external polarizing electric and magnetic fields by being constant.
Effectively we have obtained modified quasi-one dimensional Schr\"{o}dinger equation of the form
\begin{eqnarray}
-\frac{\hbar^2}{2m}[(1+\frac{1}{(\frac{d}{dx}y(x)^2)})\frac{d^2}{dx^2}-\frac{\frac{d^2}{dx^2}y(x)}{(\frac{d}{dx}y(x))^2}\frac{d}{dx}]\psi(x)+V(x)\psi(x)=E \psi(x).
\end{eqnarray}
Here the shape of open-loop nanowire is encoded in $y(x)$ function dependence (what is reflected in functions measuring cable curvature as by $(\frac{d}{dx}y(x)$ and by $(\frac{d^2}{dx^2}y(x)$). The last \textbf{CM-Schr\"{o}dinger} equation (\textbf{C}urvature \textbf{M}odified Schr\"{o}dinger equation) can easily be generalized to open-loop nanowire in 3 dimensions that brings quasi-one dimensional CM-Schr\"{o}dinger equation as well. We have obtained the following results for Tanh square nanocable as given by Fig.5-14. Fig.\ref{Alpha10QbitCurve} shows the case of presence of gap in quantum states presence due to nanowire cable bending. This effect is similar to the case of barrier existence due to nanocable bending as expressed by big $\alpha=10$ coefficient. Furthermore we can recognize that there is relatively very weak effect (but still noticable) of existence of local confining potential.Lack of built-in q-wells might result in the gap of q-state presence in the middle of curved nanowire as it is depicted in Fig.\ref{special1}. Now it is time for detailed analysis of each eigen-energy wavefunction analysis, so we separate  first 10 among 20 eigen-energy modes and last 10 modes  with no-built in q-wells for Tanh square nanowires (with $\alpha=10$). The analysis shows that wave-functions are strongly localized due to the fact that nanowire has non-zero curvature (equivalent to condition that $\frac{\frac{d^2}{dx^2}}y(x){(\frac{d}{dx}y(x))^2} \neq 0$) as it is depicted in Fig.\ref{pq1}. Now we conduct the same analysis but for 3 built-in q-wells for Tanh square nanowires (with $\alpha=10$) and we obtain wavefunction distribution as depicted in Fig.\ref{pq2}. Setting $\alpha=0.1$ and with the same detailed analysis of eigenenergy wavefunctions (for first 20 eigenenergy modes) we obtain two very similar probability distributions (as in contrast with case from Fig.\ref{pq2}.) for cable with 3 built-in q-wells [LEFT] no built-in q-wells [RIGHT] as depicted in Fig.\ref{pq4}. % Now we present the detailed eigenenergy wavefunctionies distribution in nanowire with coefficient $\alpha=0.1$ and with no built-in q-wells as depicted in Fig.\ref{pq4}.
The same as before the detailed eigenenergy wavefunctionies distribution in nanowire with coefficient $\alpha=0.1$ but with 3 built-in q-wells is depicted in Fig.\ref{pq5}. The case of effective potential distribution describing 3 built-in and no built-in q-wells potential is depicted in Fig.\ref{pq6}.

%We can also use detailed view of 200 eigenergy probability distributions with $\alpha=0.1$ and no built-in [UPPER] and built-in [LOWER] q-wells depicted in Fig.\ref{pq7}. Difference is very minor but it takes place and is bit similar as in case depicted by Fig.\ref{pq3}.
First 200 eigen-energy wavefunctions for Tanh Square V shape nanowire with no quantum wells for $\alpha=0.1$ are depicted in Fig. \ref{supsup}. Main conclusion is that bending nanowire brings the effect of separation of two reservoirs what means that there is no necessity of creation of separate material that is between two areas of nanowire.
%%\includegraphics[scale=0.16]{FirstRealImag200EigenEnergiesAlpha0p1_andNoWells.png}
%%\caption{First 200 eigen-energy wavefunctions for Tanh Square V shape nanowire with no built-in quantum wells. }
%%\end{figure}
%%\begin{figure}
%% \label{supsup}
%\label{pq5}

\subsection{Case of 2 interacting single-electron lines}
\begin{figure}
\centering
\includegraphics[scale=0.3]{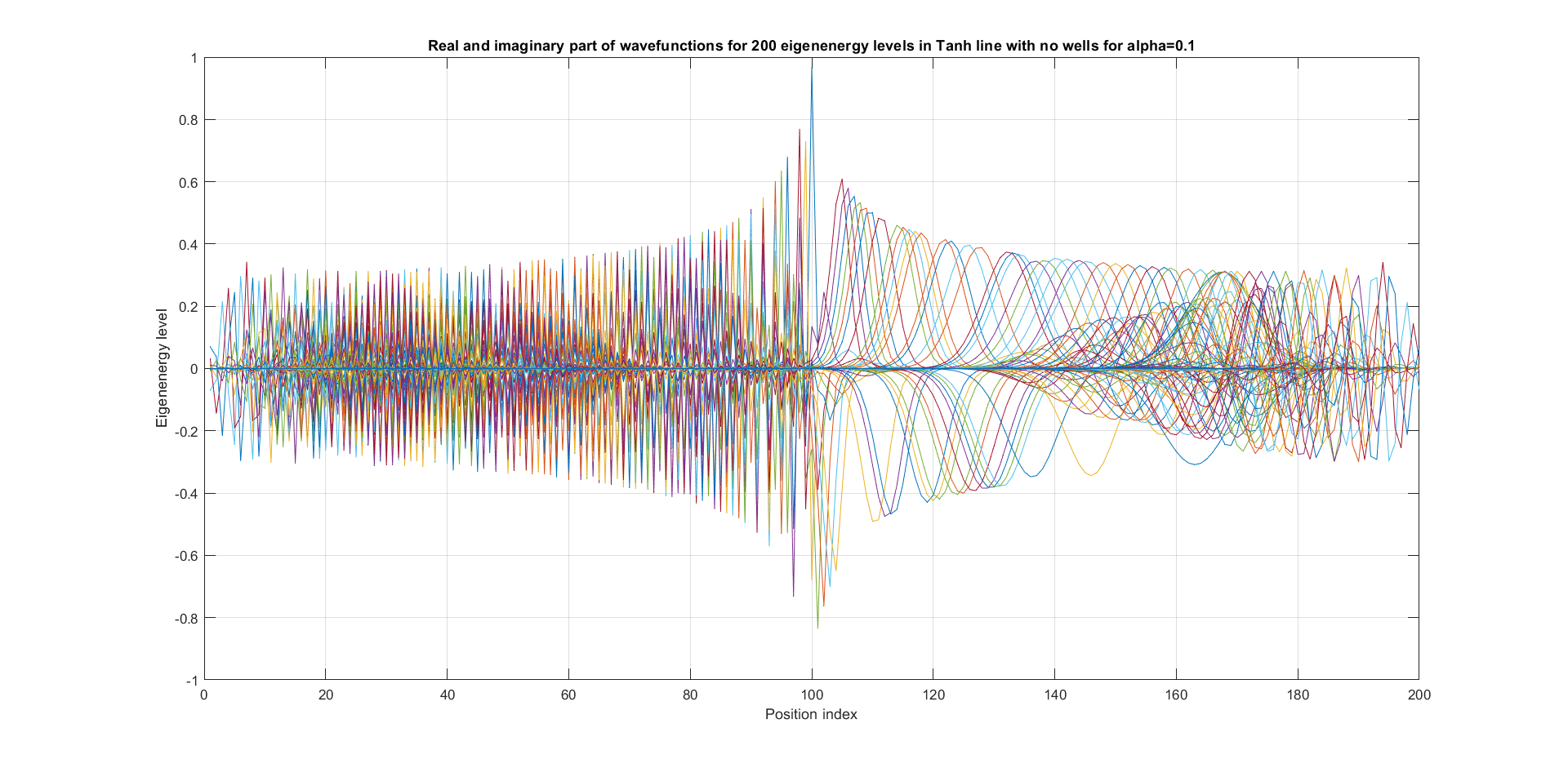}
%\caption{First 200 eigen-energy wavefunctions for Tanh Square V shape nanowire with 3 built-in [UPPER] and
\caption{First 200 eigen-energy wavefunctions for Tanh Square V shape nanowire with no built-in quantum wells for $\alpha=0.1$. }
%%\includegraphics[scale=0.16]{FirstRealImag200EigenEnergiesAlpha0p1_andNoWells.png}
%%\caption{First 200 eigen-energy wavefunctions for Tanh Square V shape nanowire with no built-in quantum wells. }
%%\end{figure}
%%\begin{figure}
 \label{supsup}
%%\end{figure}
%%\begin{figure}
\centering
\includegraphics[scale=0.17]{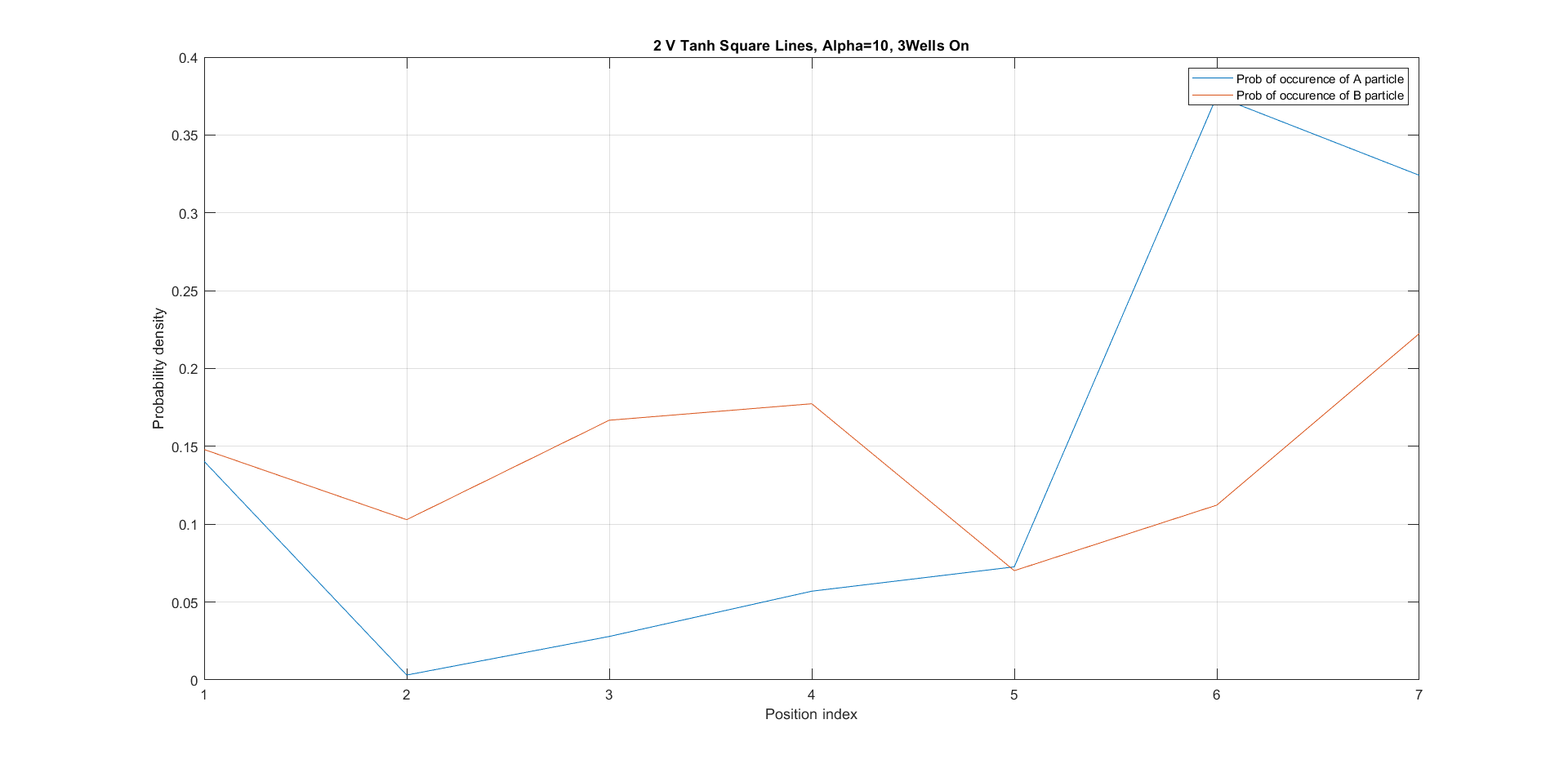}\includegraphics[scale=0.17]{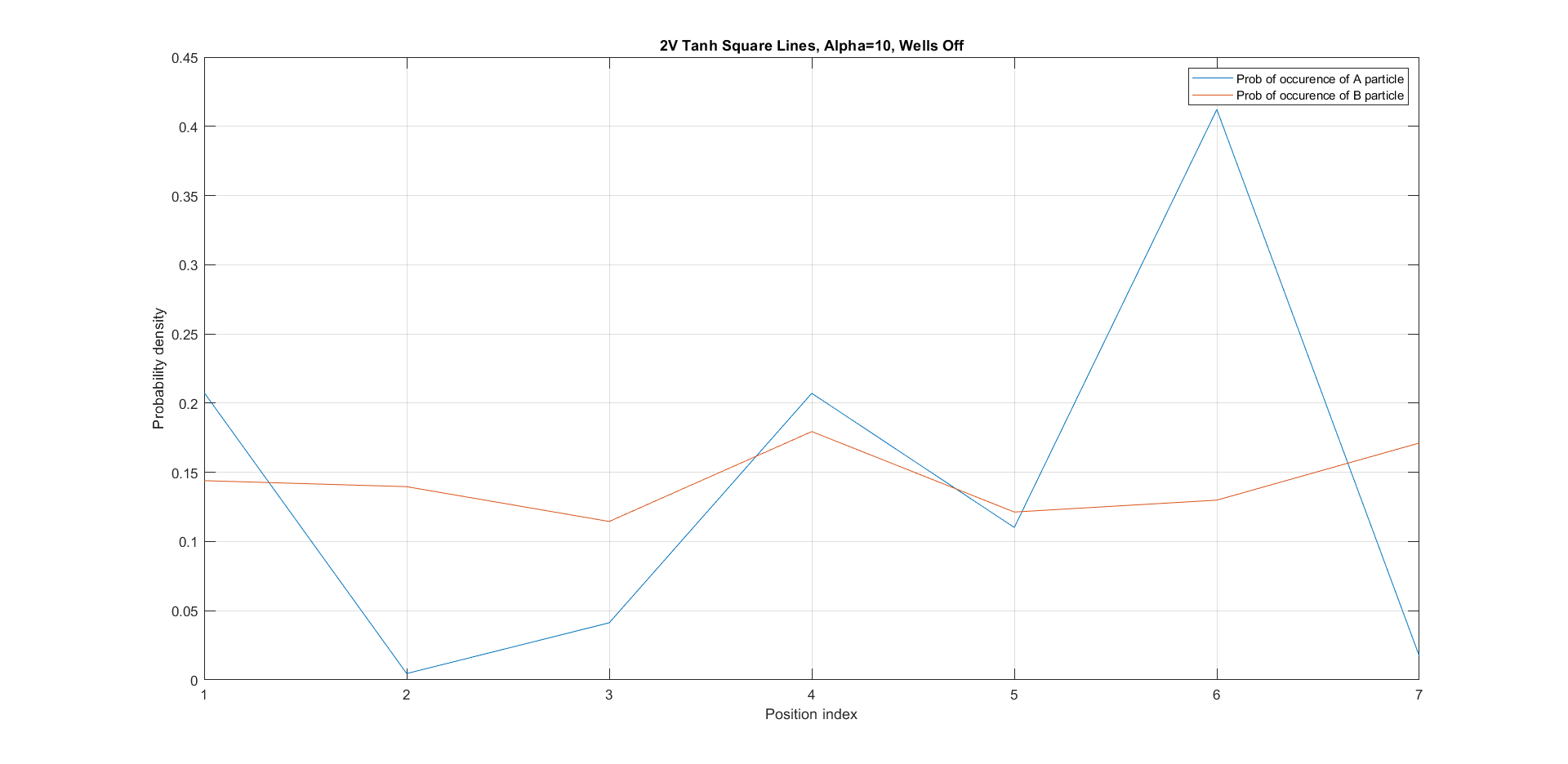}
\includegraphics[scale=0.17]{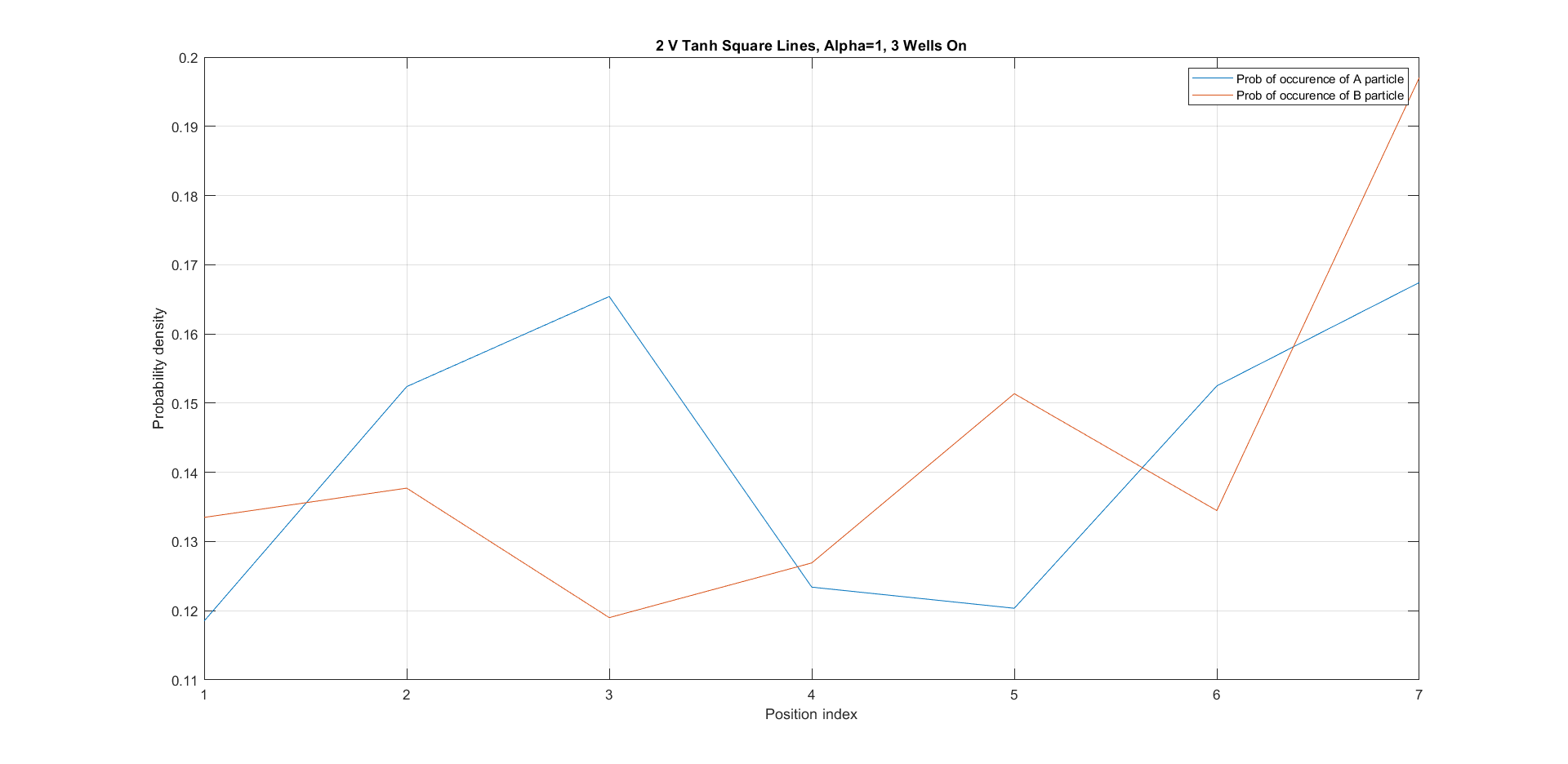}\includegraphics[scale=0.17]{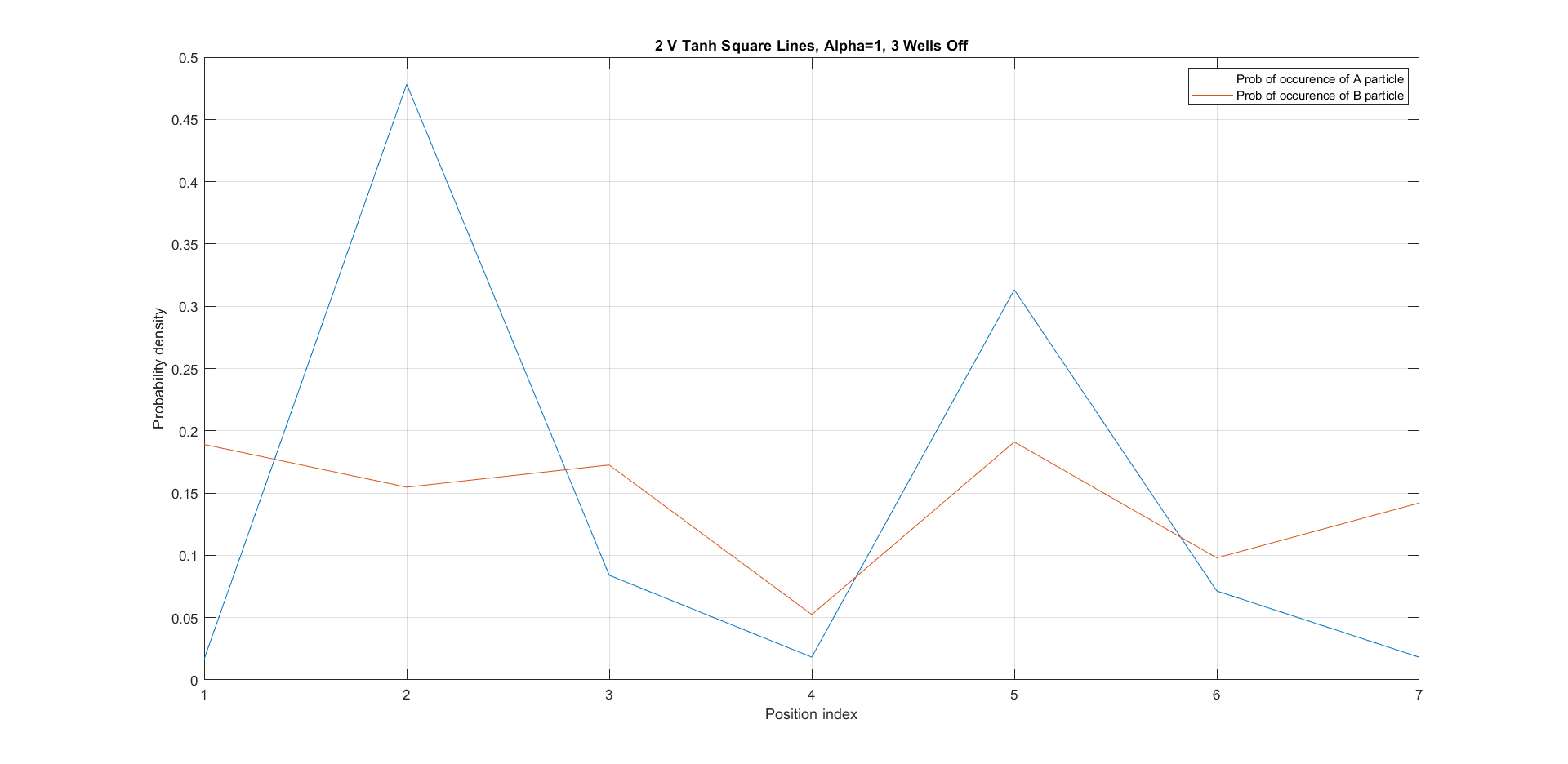}
\includegraphics[scale=0.17]{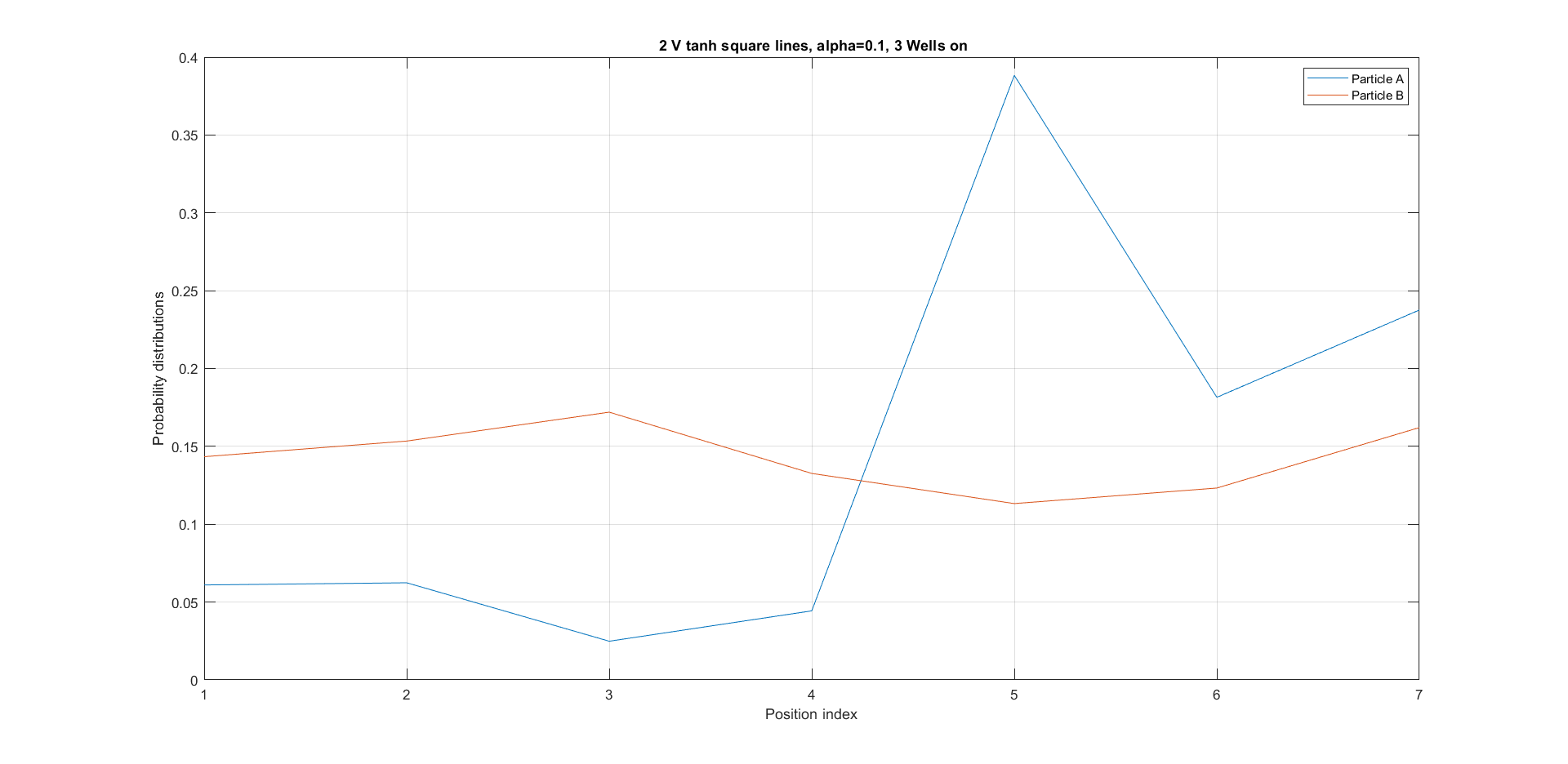}\includegraphics[scale=0.17]{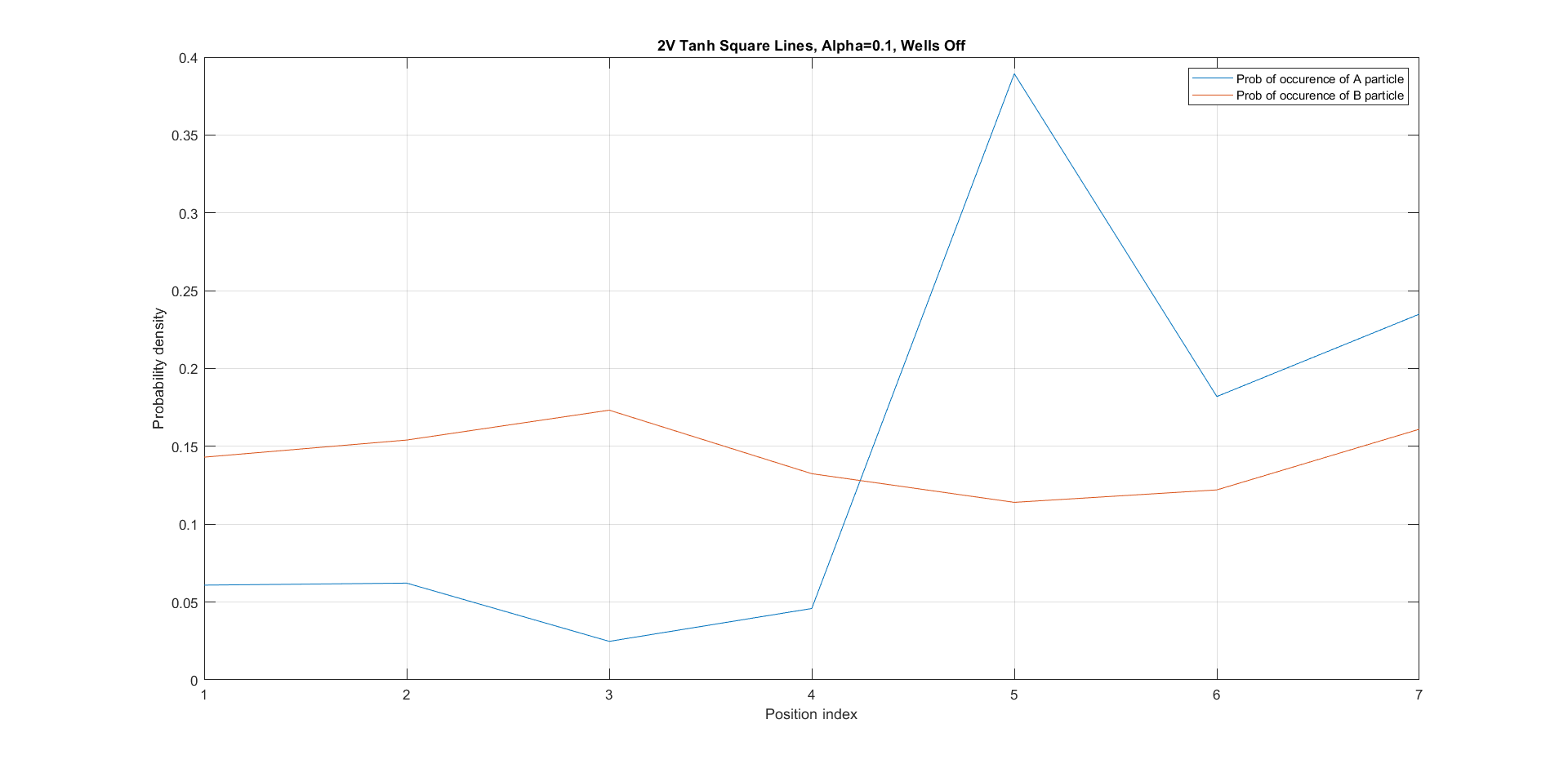}
\caption{Case of 2 V Tanh Squares Lines interacting and probability distributions around each line for electrons A and B with $\alpha=$ (10, 1, 0.1) for (UPPER, MIDDLE, LOWER) pictures with 3 quantum wells built-in [LEFT] and no quantum wells built-in [RIGHT].  }
%%\end{figure}
%%\begin{figure}
\label{supsup1}
%%\centering
%%\includegraphics[scale=0.20]{TwoVInt3WellsOffAlpha10.png}\includegraphics[scale=0.20]{TwoVInt3WellsOffAlpha1.png}\includegraphics[scale=0.20]{TwoVInt3WellsOffAlpha0p1.png}
%%\caption{Case of 2 V Tanh Squares Lines interacting and probability distributions around each line for electrons A and B with $\alpha=$ (10, 1, 0.1) for (UPPER, MIDDLE, LOWER) pictures with no q-wells built-in.  }
%%\label{supsup2}
\end{figure}

Furthermore two parallel lines in $><$ configuration with single-electron distributed at each line are expressed by 2 body Schr\"{o}dinger modified equations in the form as
\begin{eqnarray}
-\frac{\hbar^2}{2m_A}[(1+\frac{1}{(\frac{d}{dx_A}y_A(x_A)^2)})\frac{d^2}{dx_A^2}-\frac{\frac{d^2}{dx_A^2}y_A(x_A)}{(\frac{d}{dx_A}y(x_A))^2}\frac{d}{dx_A}]\psi(x_A,y_A,x_B,y_B) \nonumber \\
-\frac{\hbar^2}{2m_B}[(1+\frac{1}{(\frac{d}{dx_B}y_B(x_B)^2)})\frac{d^2}{dx_B^2}-\frac{\frac{d^2}{dx_B^2}y_B(x_B)}{(\frac{d}{dx_B}y(x_B))^2}\frac{d}{dx_B}]\psi(x_A,y_A,x_B,y_B)+ \nonumber \\
+[V_A(x_A)+V_B(x_B)+V_{A-B}(x_A,y_A,x_B,y_B)]\psi(x_A,y_A)=E \psi(x_A,y_A,x_B,y_B)=E \psi(s_A,s_B), \nonumber \\
%%-\frac{\hbar^2}{2m}[(1+\frac{1}{(\frac{d}{dx_B}y_B(x_B)^2)})\frac{d^2}{dx_B^2}-\frac{\frac{d^2}{dx_B^2}y_B(x_B)}{(\frac{d}{dx_B}y(x_B))^2}]\psi(x_A,y_A)+[V_B(x_B)+V_{A-B}(x_A,y_A,x_B,y_B)]\psi(x_A,y_A)=E %%\psi(x_A,y_A).
\end{eqnarray}
where $V_A$ and $V_B$ are local confining potentials for electron A and B, while Coulomb interaction between electrons $V_{A-B}(x_A,y_A,x_B,y_B)=\frac{q^2}{d((x_A,y_A),(x_B,y_B))}$ and $(s_A,s_B)$ is the pair of variables parametrising two-dimensional 2-body wavefunction.
This is the case of 2 body interaction that is considered with omission spin degrees of freedom. If each of nanowires (bent or straight) is divided into m pieces we deal with Hamiltonian matrix of size $M^{p}$ by $M^{p}$ that has $M^{p=2}$ energy eigenfunctions and eigenvalues. Here we set number of particles p to 2 (for q-swap gate p=2 and p=3 for CNOT gate). Those particles are interacting and represented by number of electrons placed at different open loop semiconductor nanowires. We set M=7 for two symmetric around OX axes V lines and assumed $\alpha_A=\alpha_B=c$ with formulas for A and B cables given by $F_{A(B)}(x)=a+b*(Tanh(c*x+d))^2$, so nanolines are given by $(x,F_{1}(x))$ and $(x,F_{2}(x))$ and depicted in Fig.\ref{2fibers}.
We have obtained the following results of probability distribution for the case of 3 built-in q-wells (dots) as depicted in left part of Fig.\ref{supsup1} and with case of no-built in q-wells as depicted in right part of Fig.\ref{supsup1}. We can trace electron anticorrelation under different nanowire cable bending. We can identify anticorrelation and correlation factors as occurring in the case of electrostatically interacting electrons placed at different nanocables as it was indicated already in \cite{SEL}. For the purpose of computations in this work all eigenenergies were set with equal probability of occupancy. Quite clearly the presented results
go beyond tight-binding model expressed by \cite{SEL} and \cite{Cryogenics}.
%%%$2^{}$.
%\ref(fig1q}.

%%\begin{figure}
%%\centering
%%\includegraphics[scale=0.16]{Prob20Alpha0p1_3WellsOnQ.png}
%%\includegraphics[scale=0.16]{Prob20Alpha0p1_3WellsOff.png}
%%\caption{Detailed analysis of eigenenergy wavefunctions (for first 20 eigenenergy modes) with coefficient $\alpha=0.1$ reveals very similar probability distributions (as in contrast with case from Fig.\ref{pq2}.) for cable with 3 built-in q-wells [LEFT] no built-in q-wells [RIGHT]. }
%%\label{pq3}
%%\end{figure}

%\begin{figure}
%\centering
%\includegraphics[scale=0.4]{Prob200Alpha0p1_WellsOff.png} %%{Probabilies_VTanh_alpha0p1.png}
%\includegraphics[scale=0.4]{Prob200Alpha0p1_WellsOnS.png}
%\caption{Probability distributions for 200 different eigenenergies in function of position index for $\alpha=0.1$ with no built-in (upper) and built-in q-wells (lower). Difference is very minor but it takes place and is bit similar as in case depicted by Fig.\ref{pq3}. }
%\label{pq7}
%\end{figure}

\section{Physical system implementing Wannier qubit swap gate in classical description}

    \begin{figure} %\label(centralfig)
    \centering
    \includegraphics[scale=0.8]{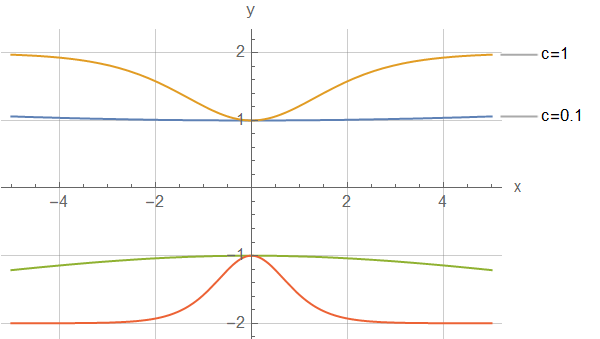}
    \includegraphics[scale=0.6]{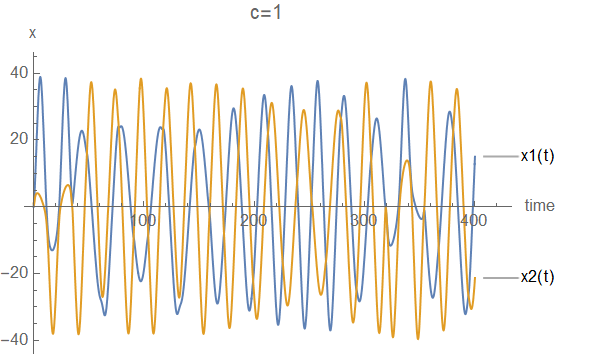}\includegraphics[scale=0.6]{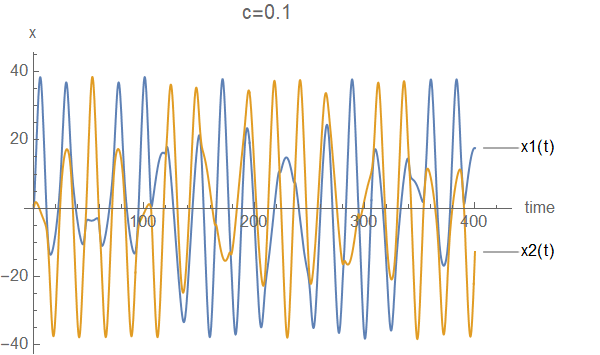}
    \caption{Two particles (electrons) placed at semiconductor nanowires interacting electrostatically and family of V shape lines parametrized by $F_{1(2)}(x)=a+b*(Tanh(c*x+d))^2$, so nanolines are given by $(x,F_{1}(x))$ and $(x,F_{2}(x))$ points. }
    \label{fig1q}
    \end{figure}
%We assume movement of 2 particles at trajectories $(x_1(s_1),y_1(s_1))$ and $(x_2(s_2),y_2(s_2))$ as given by Fig.\ref{fig1q}.
%    \begin{eqnarray}
%      -\frac{d}{dx_1}H=\frac{d}{dt}p_{1,x}, -\frac{d}{dy_1}H=\frac{d}{dt}p_{1,y}, \frac{d}{dp_{1,x}}H=\frac{d}{dt}x_{1}, \frac{d}{dp_{1,y}}H=\frac{d}{dt}y_{1},   \nonumber \\
%      -\frac{d}{dx_2}H=\frac{d}{dt}p_{2,x}, -\frac{d}{dy_2}H=\frac{d}{dt}p_{2,y}, \frac{d}{dp_{2,x}}H=\frac{d}{dt}x_{2}, \frac{d}{dp_{2,y}}H=\frac{d}{dt}y_{2}.
%    \end{eqnarray}

    We refer to Fig.\ref{2fibers} and we have the Hamiltonian describing interaction of electrons confined in different nanowires of the structure
    \begin{eqnarray}
      H =H_1+H_2+H_{1-2}= \nonumber \\
      =\frac{(p_{1,x})^2}{2m_1}+\frac{(p_{1,y})^2}{2m_1}+\frac{(p_{2,x})^2}{2m_2}+\frac{(p_{2,y})^2}{2m_2}+V(x_1(s_1),y_1(s_1))+V(x_2(s_2),y_2(s_2))+H_C(s_1,s_2), \nonumber
      \\
    %%   &=&  \\
    %%   &=& 
    \end{eqnarray}
    and we have
     \begin{eqnarray} \label{cleqnodiss1}
     \frac{d}{dx_1}=\frac{ds_1}{dx_1}\frac{d}{ds_1},  \frac{d}{dy_1}=\frac{ds_1}{dy_1}\frac{d}{ds_1}, \nonumber \\
     \frac{d}{dx_2}=\frac{ds_2}{dx_2}\frac{d}{ds_2},  \frac{d}{dy_2}=\frac{ds_2}{dy_2}\frac{d}{ds_2}, \nonumber \\
     \frac{d}{dt}x_1=\frac{dx_1}{ds_1}\frac{d}{dt}s_1,  \frac{d}{dt}y_1=\frac{dy_1}{ds_1}\frac{d}{dt}s_1, \nonumber \\
     \frac{d}{dt}x_2=\frac{dx_1}{dx_1}\frac{d}{dt}s_2,  \frac{d}{dt}y_2=\frac{dy_2}{ds_2}\frac{d}{dt}s_2, \nonumber \\
    \frac{1}{m_1}\frac{d}{dt}p_{1,x}=\frac{d^2}{dt^2}x_1=\frac{dx_1}{ds_1}\frac{d}{dt}[\frac{dx_1}{ds_1}\frac{d}{dt}s_1]=[(\frac{dx_1}{ds_1})^2\frac{d^2}{dt^2}s_1+\frac{d^2x_1}{ds_1^2}\frac{dx_1}{ds_1}\frac{d}{dt}s_1], \nonumber \\
%     \frac{1}{m_2}\frac{d}{dt}p_{2,x}=\frac{d^2}{dt^2}x_2=\frac{dx_2}{ds_2}\frac{d}{dt}s_2[\frac{dx_2}{ds_2}\frac{d}{dt}s_2]=[(\frac{dx_2}{ds_2})^2\frac{d^2}{dt^2}s_2+\frac{d^2x_2}{ds_2^2}\frac{dx_2}{ds_2}\frac{d}{dt}s_2], \nonumber \\
%     \frac{1}{m_1}\frac{d}{dt}p_{1,y}=\frac{d^2}{dt^2}y_1=\frac{dy_1}{ds_1}\frac{d}{dt}s_1[\frac{dy_1}{ds_1}\frac{d}{dt}s_1]=[(\frac{dy_1}{ds_1})^2\frac{d^2}{dt^2}s_1+\frac{d^2y_1}{ds_1^2}\frac{dy_1}{ds_1}\frac{d}{dt}s_1], \nonumber \\
%     \frac{1}{m_2}\frac{d}{dt}p_{2,y}=\frac{d^2}{dt^2}y_2=\frac{dy_2}{ds_2}\frac{d}{dt}s_2[\frac{dy_2}{ds_2}\frac{d}{dt}s_2]=[(\frac{dy_2}{ds_2})^2\frac{d^2}{dt^2}s_2+\frac{d^2y_2}{ds_2^2}\frac{dy_2}{ds_2}\frac{d}{dt}s_2], \nonumber \\
    %%   &=&  \\
    %%   &=& 
    \end{eqnarray}
    We set $V(s_1,s_2)=q^2/((x_1(s_1)-x_2(s_2))^2+(f_1(x_1(s_1))-f_2(x_2(s_2)))^2)^{1/2}$ and thus obtain $\frac{d}{dx1}H=\frac{ds_1}{dx_1}\frac{d}{ds_1}H$
    and $\frac{d}{dx2}H=\frac{ds_2}{dx_1}\frac{d}{ds_2}H$.
Let us solve the practical set of coupled non-linear ODE equations by setting $s_1=x_1$ and $s_2=x_2$, so we have
\begin{eqnarray}
  m_1\frac{d^2}{dt^2}x_1(t) &=& -m_1(\frac{d}{dx_1}f_1(x_1))(\frac{d^2}{dx_1^2}f_1(x_1))(\frac{d}{dt}x_1(t))^2-\frac{d}{dx_1}V_1(x_1,f_1(x_1))-\frac{d}{dx_1}\frac{q^2}{((x_1-x_2)^2+(f_1(x_1)-f_2(x_2)))^{\frac{1}{2}}}, \nonumber  \\
  m_2\frac{d^2}{dt^2}x_2(t) &=& -m_2(\frac{d}{dx_2}f_2(x_2))(\frac{d^2}{dx_2^2}f_2(x_2))(\frac{d}{dt}x_2(t))^2-\frac{d}{dx_2}V_2(x_2,f_2(x_2))-\frac{d}{dx_2}\frac{q^2}{((x_1-x_2)^2+(f_1(x_1)-f_2(x_2)))^{\frac{1}{2}}}, \nonumber  \\
  \label{CoupledODEs}
  \end{eqnarray}
  where $V_1(x_1(t),y_1(t))$ and $V_2(x_2(t),y_2(t))$ are local confining potentials in nanowires. In particular we have set $V_1(x_1)= e^{0.1 \sqrt{x_1^2 +(f_1(x_1))^2 }}$ and
 $V_2(x_2)= e^{0.1 \sqrt{x_2^2 +(f_2(x_2))^2 }}$ with $f_1(x_1)=1 + Tanh(c 0.5 x_1)^2$ and $f_2(x_2)=1 + Tanh(c x_2)^2$. There is occurrence of additional "dissipative" terms in equations of motions as by expressions \newline $-m_1(\frac{d}{dx_1}f_1(x_1))(\frac{d^2}{dx_1^2}f_1(x_1))(\frac{d}{dt}x_1(t))^2$ and $ -m_2(\frac{d}{dx_2}f_2(x_2))(\frac{d^2}{dx_2^2}f_2(x_2))(\frac{d}{dt}x_2(t))^2$ due to non-zero nanocable's curvature.  One can observe the emergence of deterministic chaos as depicted in Fig. \ref{fig1q}. The wider considerations for various coordinate systems and curved semiconductor nanowires is given by \cite{Extended}.

\section{Conclusions}
The usage of hopping terms in tight-binding model \ref{eqn:fundamentalbasic} was justified by Schr\"{o}dinger formalism with formulas \ref{MainFormula},
\ref{formts12},
\ref{formts21},
\ref{formts11},
\ref{formts22} for static electric and magnetic field as well as for the case of Rabi oscillations and non-static electric and magnetic field expressed by formulas \ref{generaleqns} . The prescription for exact computation of localized energy terms as $E_{p1}$ and $E_{p2}$ was given by formulas \ref{basicexp}.

 %%\ref{given}.
The origin of tight-binding model dissipation \cite{epidemic},\cite{dissipation} was identified in the framework of Schr\"{o}dinger formalism, since real value eigenergies of 2 quantum dot system ($E_1$, $E_2$) can be replaced with $(E_{1r}+iE_{1i},E_{2r}+iE_{2i})$, where ($E_{1i}$,$E_{2i}$) are dissipative terms and ($E_{1r}$,$E_{2r}$) are real-valued non-dissipative terms (originally given by Schr\"{o}dinger equation). The mathematical structure of dissipative tight-binding model in static case with constant electric and magnetic field was given by \ref{MainFormulaD} and in case of Rabi oscillation was given by \ref{generaleqnsD}. The concept of Wannier functions in system of 2 coupled semiconductor qubits (Wannier qubit) can be applied not only for single-electron occupancy, but also for many electron occupancy. In such case the fermion Hubbard model can be used as given by \cite{Spalek} and \cite{InteractingWF}. Various correlation functions using Schr\"{o}dinger formalism and justifying the tight-binding model has been proposed and are presented in the extended version of this work \cite{Extended} using \cite{HB},\cite{HB1}.
The case of 2 Wannier qubit interaction was formulated in Schr\"{o}dniger formalism using the straight and curvy semiconductor nanowires and basic preliminary numerical results were presented as depicted in Fig.\ref{supsup1}.
It consequently points to Hermitian and non-Hermitian Hamiltonian matrix formula. The effect or curvature of semiconductor quasi-one dimensional nanowires was expressed by proper equations of motion both in case of classical and quantum picture. Main conclusion is that bending nanowire brings the effect of separation of two reservoirs as depicted in Fig.\ref{pq1}. It has certain importance for future photonic technologies \cite{InegratedQPH}.
It allows for modeling of quantum and classical SWAP gates using electron-electron interaction.
The effects of topology of open loop semiconductor nanowires can be studied by usage of Toeplitz matrix approach in different coordinate systems as by : Cartesian, Cylindrical and Spherical coordinates.
The conducted considerations allows for description of Wannier position based qubits with single and many electrons injected into source and drain of field effect transistor. They also give the base for modeling the quantum neural networks implemented by the chain of coupled quantum dots. The presented fundamental approach is useful in enhancement of tight-binding scheme as used in the design of quantum gates \cite{qgates},\cite{Nbodies},\cite{QInternet}, \cite{epidemic}. The presented work is the extension of methodology given by \cite{Xu} as well as by \cite{Panos},\cite{Nbodies},\cite{Szafran}. The results obtained in Fig.5-13 for the case of Tanh Square cables shall be tested using Local Density of States observed in STM for different $\alpha$ coefficient. Various physical phenomena observed in condensed matter systems \cite{Spalek} can be simulated with the concept of quantum programmable matter demonstrated by position-dependent qubits controlled by electric signals \cite{QInternet},\cite{Jaynes}. Furthermore quantum machine learning can mimic any stochastic finite state machine by usage of tight-binding model as it was shown in \cite{epidemic}. Using the concept of reconfigurable q-graph of quantum dot \cite{Cryogenics} we can simulate the behaviour of quantum particle in curved space what is the subject of future work \cite{CurvedQM1}, \cite{GeneralRelativityQM}, \cite{RotatingQM}. Another future direction is the investigation of holonomic quantum computation \cite{holonomic},\cite{TaoChen} with qubits constructed from curvy semiconductor nanowires described in this work.

\end{document}